\documentclass[preprint]{aastex}

\usepackage{graphicx}

\usepackage{caption,subcaption}
\usepackage{amsmath, amssymb}
\usepackage{calligra}
\usepackage{url}
\usepackage{natbib}
\usepackage{color}
\usepackage{pdflscape}
\usepackage{hyperref}
\usepackage{tikz} 
\usepackage[T1]{fontenc}
\usetikzlibrary{shapes,arrows}
\let\new=\newcommand
\new{\diff}{{\rm d}}

\slugcomment{The Astrophysical Journal}

\shorttitle{Tidal disruption by SMBHs}
\shortauthors{Mageshwaran \& Mangalam}

\begin{document}

\title{Stellar and gas dynamical model for tidal disruption events in a quiescent galaxy}

\author{T.~Mageshwaran$^{\dagger}$}
\author{A.~Mangalam$^{\ddagger}$} 

\affil{Indian Institute of Astrophysics, Bangalore-560034, India}

\email{ $^{\dagger}$mageshwaran@iiap.res.in, $^{\ddagger}$mangalam@iiap.res.in}

\begin{abstract}
A detailed model of the tidal disruption events (TDEs) has been constructed using stellar dynamical and gas dynamical inputs that include black hole (BH) mass $M_{\bullet}$, specific orbital energy $E$ and angular momentum $J$, star mass $M_{\star}$ and radius $R_{\star}$, and the pericenter of the star orbit $r_{p}(E,\hspace{1mm}J,\hspace{1mm}M_{\bullet})$. We solved the steady state Fokker--Planck equation using the standard loss cone theory for the galactic density profile $\rho (r) \propto r^{-\gamma}$ and stellar mass function $\xi(m) $ where $m=M_{\star}/M_{\odot}$ and obtained the feeding rate of stars to the BH integrated over the phase space as $\dot{N}_{t} \propto M_{\bullet}^\beta$, where $\beta= -0.3\pm 0.01$ for $M_{\bullet}>10^7 M_{\odot}$ and $\sim  6.8 \hspace{1mm} \times 10^{-5}$ Yr$^{-1}$ for $\gamma=0.7$. We use this to model the in-fall rate of the disrupted debris, $\dot{M}(E,\hspace{1mm}J,\hspace{1mm}m,\hspace{1mm}t)$, and discuss the conditions for the disk formation, finding that the accretion disk is almost always formed for the fiduciary range of the physical parameters. We also find the conditions under which the disk formed from the tidal debris of a given star with a super Eddington accretion phase. We have simulated the light curve profiles in the relevant optical g band and soft X-rays for both super and sub-Eddington accretion disks as a function of $\dot{M}(E,\hspace{1mm}J,\hspace{1mm}t)$. Using this, standard cosmological parameters, and mission instrument details, we predict the detectable TDE rates for various forthcoming surveys finally as a function of $\gamma$.

\end{abstract}


\section{Introduction}

It is well known from observations that supermassive black holes (SMBHs) reside at the center of galactic nuclei \citep{1995ARA&A..33..581K,2001eaa..bookE2635K}. If a star passes within the radius $\displaystyle{r_{t}\sim R_{\star} (M_{\bullet}/ M_{\star})^{1/3}}$ of the galaxy's central black hole (BH), the BH's tidal gravity exceeds the star's self-gravity and it is tidally disrupted \citep{1975Natur.254..295H,1988Natur.333..523R}. Stars in the galactic center move in the combined potential field of the  SMBH and other stars in the galactic center. The star with orbital energy $E$ is tidally captured if the orbital angular momentum is $J \leq J_{lc}=\sqrt{2 r^2_t (\Phi(r_t)-E})$, where $J_{lc}$ is the angular momentum of the loss cone \citep{1976MNRAS.176..633F} and $\Phi(r)$ is the potential by the BH and other stars in the galactic center. The stellar interactions result in the diffusion of the stars into the loss cone. The stellar distribution function (DF) $f(E,\hspace{1mm}J)$ follows the Fokker--Planck (FP) equation \citep{1976ApJ...209..214B,1977ApJ...211..244L} and the rate of feeding of stars into the loss cone gives the theoretical tidal disruption event (TDE) rate $\dot{N}_{t}$. The rate of TDE per galaxy depends on the stellar distribution in the galactic center, the SMBH mass $M_{\bullet}$ and the structure of galactic nuclei that could be symmetric, axis-symmetric, or triaxial nuclei \citep{2013CQGra..30x4005M}. \citet{1978ApJ...226.1087C} obtained the numerical solution to the FP equation for spherical nuclei by means of a detailed boundary layer analysis and applied it to globular clusters. \citet{2004ApJ...600..149W} solved the steady state FP equation for the 51 galaxies with the Nuker profile by assuming a single mass star distribution and obtained the $\dot{N}_{t}\sim 10^{-4}-10^{-5} \hspace{1mm} \rm{Yr^{-1}}$. They further predicted that $\dot{N}_{t}\propto M^{-0.25}_{\bullet}$ for the isothermal case (also see \citet{2013degn.book.....M}). \citet{2014arXiv1410.7772S} employed a stellar mass function, $\xi(m)$, in their DF, and applied it to a sample of 200 galaxies, and obtained  $\dot{N}_{t}\propto M^{-0.4}_{\bullet}$. \citet{1999MNRAS.309..447M} solved the steady state FP equation for an axis-symmetric nuclei with stars on a centrophobic orbits and obtained $\dot{N}_{t}\sim 10^{-4} M^{-0.19}_{\bullet} \hspace{1mm}\rm{Yr^{-1}}$. The non-spherical mass distribution in the galactic center provides an additional torque to the star's orbit, which results in additional diffusion of stars in the loss cone, and such orbits are called as drain orbits. \citet{2013ApJ...774...87V} solved the full FP equation numerically for an axis-symmetric nuclei with a slight deviation from the sphericity and found that the $\dot{N}_{t}$ is two to three times higher than that for spherical geometry. The orbital evolution in the galactic nuclei is more complicated in the presence of triaxial potentials, which support two distinct families of tube orbits circulating about the long and short axes of the triaxial figure \citep{2011ApJ...726...61M}. In addition, there is a new class of centrophilic orbits called the pyramids, and the defining feature of the pyramid orbit is that the minimum of $J$=0 and a star on such an orbit will eventually find its way into the SBH even without the assistance of collisional relaxation \citep{1999AJ....118.1177M}. Feeding rates due to collisional loss cone refilling are very large in such galaxies compared with the spherical galaxies \citep{2004ApJ...606..788M}. Due to the complexity of the orbits in the non-spherical galaxies, we assume that the galaxy is spherical in this paper and plan to study the more general ellipsoidal or triaxial case later.

Approximately two dozen TDE candidates have been observed and found with diverse mixture of optical \citep{2011ApJ...741...73V,2012MNRAS.420.2684C,2012Natur.485..217G,2014ApJ...780...44C,2014ApJ...793...38A}, UV \citep{2008ApJ...676..944G,2009ApJ...698.1367G}, and X-ray \citep{2002luml.conf..436K,2002AJ....124.1308D,2013MNRAS.435.1904M} detection. The TDE rates are highly uncertain from observations due to low sample size, which are typically a value of $10^{-5} \hspace{1mm} \rm{Yr^{-1}}$ per galaxy is inferred from X-ray \citep{2002AJ....124.1308D}, UV \citep{2008ApJ...676..944G}, and optically \citep{2014ApJ...792...53V} selected events. Although uncertain, the values obtained are below the theoretical estimates.

The second theoretical aspect of TDE is the accretion dynamics of the disrupted debris and the resulting luminosity and spectral energy distribution. A star tidally captured by the BH is disrupted in a dynamical timescale. \citet{1988Natur.333..523R} calculated the energy of the disrupted debris for the initial parabolic orbit of the star. The debris is assumed to follow a Keplerian orbit around the BH and the mass in-fall rate is inferred to be $\dot{M}\propto t^{-5/3}$. In general, the mass in-fall rate depends on the internal structure and properties of the star, and follows the $t^{-5/3}$ law only at the late stages of its evolution \citep{1989IAUS..136..543P,2009MNRAS.392..332L,2013ApJ...767...25G}. The debris experience stream collision either due to incoming stream that intersects with the outflowing stream at the pericenter \citep{1994ApJ...422..508K} or due to relativistic precession at the pericenter \citep{2013MNRAS.434..909H}. These interactions result in circularization of the debris to form an accretion disk \citep{2013MNRAS.434..909H,2015arXiv150104635B,2015ApJ...804...85S}. \citet{2009MNRAS.400.2070S} proposed a simplistic steady accretion model, with the fraction of mass outflow caused by the strong radiative pressure in the super Eddington phase constant. We follow the work of \citet{2009MNRAS.400.2070S} to obtain the light curve profile and duration of the flare detection $\delta t_{f}$ for  given instrument parameters. 

The energy of the debris $E_d$ depends on the pericenter of the star orbit $r_{p}(E,\hspace{1mm}J,\hspace{1mm}M_{\bullet},\hspace{1mm} m)$. Hence, the mass accretion rate $\dot{M}(E,\hspace{1mm}J,\hspace{1mm}M_{\bullet},\hspace{1mm} m)$, the flux, and  $\delta t_{f}$ depends on $E$ and $J$. We build an accretion model based on the initial stellar system parameters and simulate light curve profiles as a function of $E$ and $J$ in the optical and X-ray bands. For a DF that depends on the $E$ only, the stars are diffused into the loss cone through  star-star interactions, which leads to the change in orbital angular momentum of the star \citep{1977ApJ...211..244L}; thus the DF of stars within the loss cone depends on both $E$ and $J$ and we calculate $\dot{N}_{t}(E,\hspace{1mm}J,\hspace{1mm} m,\hspace{1mm}\gamma)$ and the detectable rates of TDE $\dot{N}_{D}$ for the various optical and X-ray surveys.

The crucial point is that the $J$ plays an important role in the stellar dynamical process through $\dot{N}_{t}(E,\hspace{1mm}J,\hspace{1mm} m,\hspace{1mm}\gamma)$ and the accretion process through $r_{p}(E,\hspace{1mm}J,\hspace{1mm}M_{\bullet},\hspace{1mm} m)$ which impacts the detectable TDE rates; this has not been taken into account in previous calculations. 

The observed sample of candidate TDE is expanding rapidly, mainly at the optical frequencies due to the advent of the highly sensitive wide field surveys such as Panoramic Survey Telescope and Rapid Response System (Pan-STARRS), observing in both the medium deep survey (MDS) mode and 3$\pi$ survey mode \citep{2002SPIE.4836..154K}. The study of TDE will be further revolutionized in the next decade by the Large Synoptic Survey Telescope (LSST) with high sensitivity in optical frequencies \citep{2009arXiv0912.0201L} and the extended Roentgen Survey with an Imaging Telescope Array (eROSITA) in the X-ray band, which performs an all sky survey (ASS) twice a year \citep{2012arXiv1209.3114M} and can detect hundreds or thousands of TDE per year \citep{2008ApJ...676..944G,2014ApJ...792...53V,2014MNRAS.437..327K}. We calculate the detectable rate $\dot{N}_{D}$ for Pan-STARRS 3$\pi$, Pan-STARRS MDS, and LSST in the optical g band and eROSITA in soft X-ray band; whose instrument details are given in Table \ref{survey}. 

\begin{figure}
\begin{center}
\scalebox{0.85}
{
\tikzstyle{decision} = [diamond, draw,  
    text width=6em, text badly centered, node distance=5cm, inner sep=0pt]
\tikzstyle{block} = [rectangle, draw, 
    text width=12em, text centered, rounded corners, minimum height=2em]
\tikzstyle{line} = [draw, -latex']
\tikzstyle{cloud} = [draw, ellipse,node distance=3cm,
    minimum height=2em]

\begin{tikzpicture}[node distance = 1.4cm, auto]
\node [block] (para) {Required parameters \\ $M_{\bullet}$, $M_{\star}, R_{\star}$,  $E$, $J$ and $z$};
\node [block, right of=para, node distance=6cm] (df) {Distribution function\\ $F(E,\hspace{1mm}m)$ \\(Equation ((\ref{fem}), \ref{dff}))};
\node [block, left of=para, node distance=6cm] (ed) {Debris energy $E_{d}$\\ (Equation (\ref{ed}))};
\node [block, below of=para, node distance=1.5cm] (macc) {Accretion rate $\dot{M}$\\(Equation (\ref{acc}))};
\node [block, below of=macc, node distance=1.6cm] (time) {Timescales: $t_{a}\hspace{1mm} \& \hspace{1mm}t_{r}  $\\ (Equation (\ref{acctime}) \& (\ref{trring}))};
\node [block, below of=time, node distance=2.0cm] (mcheck) {Critical black hole mass $M_{c}$ such that $\dot{M} = \dot{M}_E$ };
\node [decision, below of=ed, node distance=5.1cm] (tcir) {Is \\ $t_{r} \leq t_{a}$ ?};
\node [block, below of=tcir, node distance=3.5cm] (nd) {No disk is formed\\ $L=\eta \dot{M}c^2$; $\dot{M}=M/t_{d}$};
\node [decision, below of=mcheck, node distance=2.8cm] (se) {Is \\ $M_{\bullet} \leq M_{c}$ ?};
\node [block, below of=se, node distance=3.0cm] (seh) {Super Eddington phase \\$\dot{M} > \dot{M}_E$};
\node [block, right of=seh, node distance=6cm] (seg) {Sub Eddington phase \\$\dot{M} < \dot{M}_E$};
\node [block, below of=seh, node distance=1.5cm] (lum) {Flux observed $f_{obs}$ \\ (Equation (\ref{cond}))};
\node [block, below of=lum, node distance=1.3cm] (tdur) {Flare duration $\delta t_f$};
\node [block, below of=tdur, node distance=1.4cm] (pro) {Probability of detection \\ $P(M_{\bullet},\hspace{1mm}z)$ (Equation (\ref{proba}))};
\node [block, below of=pro,fill=yellow!20, node distance=1.8cm] (tnet) {Net Event Rate $\dot{N}_{D}$\\ (Equation (\ref{det}))};
\node [block, right of=tdur, node distance=6cm](vol) {Luminosity distance and galaxy density};
\node [block, left of=tdur, node distance=6cm ](ins) {Mission Instrument Detail\\
$f_{s},\hspace{1mm} t_{cad},\hspace{1mm} t_{int}, \hspace{1mm} f_{l}$};
\node [block, right of=time, node distance=9.5cm](FP) {Fokker-Planck (FP) equation\\ (Equation \ref{fty})};
\node [block, below of=FP, node distance=2.5cm](ss) {Distribution of orbital elements \\ (Equation (\ref{flr}))};
\node [block, right of=tnet, node distance=9.5cm](cr1) {Theoretical rate $\dot{N}_{t}$\\ (Equation (\ref{tdnth1})) };
\path [line] (para) -- (ed);
\path [line] (df) -| (FP);
\path [line] (FP) -- (ss);
\path [line] (ss) -- (cr1);
\path [line] (para) -- (df);
\path [line] (ed) |- (macc);
\path [line] (time) -| (tcir);
\path [line] (macc) -- (time);
\path [line] (tcir) -- node {No} (nd);
\path [line] (tcir) -- node {Yes} (mcheck);
\path [line] (mcheck) -- (se);
\path [line] (se) -| node {No} (seg);
\path [line] (se) -- node {Yes} (seh);
\path [line] (seh) -- (lum);
\path [line] (lum) -- (tdur);
\path [line] (nd) |- (lum);
\path [line] (seg) |- (lum);
\path [line] (tdur) -- (pro);
\path [line] (vol) -- (tdur);
\path [line] (vol) -- (lum);
\path [line] (pro) -- (tnet);
\path [line] (vol) |- (pro);
\path [line] (ins) |- (tdur);
\path [line] (ins) |- (pro);
\path [line] (ins) |- (tnet);
\path [line] (cr1) -- (tnet);
\end{tikzpicture}
}
\end{center}
\caption{ The flow chart of the procedure we have adopted in the calculation of event rates. The stellar dynamics and gas dynamics are connected by the parameters of specific energy $E$ and specific angular momentum $J$ of the star's initial orbit. The flux $f_{\rm obs}$ is compared with the sensitivity $f_{l}$ of the detector to obtain flare duration. For the given instrument details, such as cadence $t_{\rm cad}$, integration time $t_{\rm int}$ and fraction of sky observed $f_{s}$, we calculate the net detectable TDE rate for the detector. The $t_d$ is the dynamical time of the in-fall of the debris to the black hole.}
\label{fc}
\end{figure}
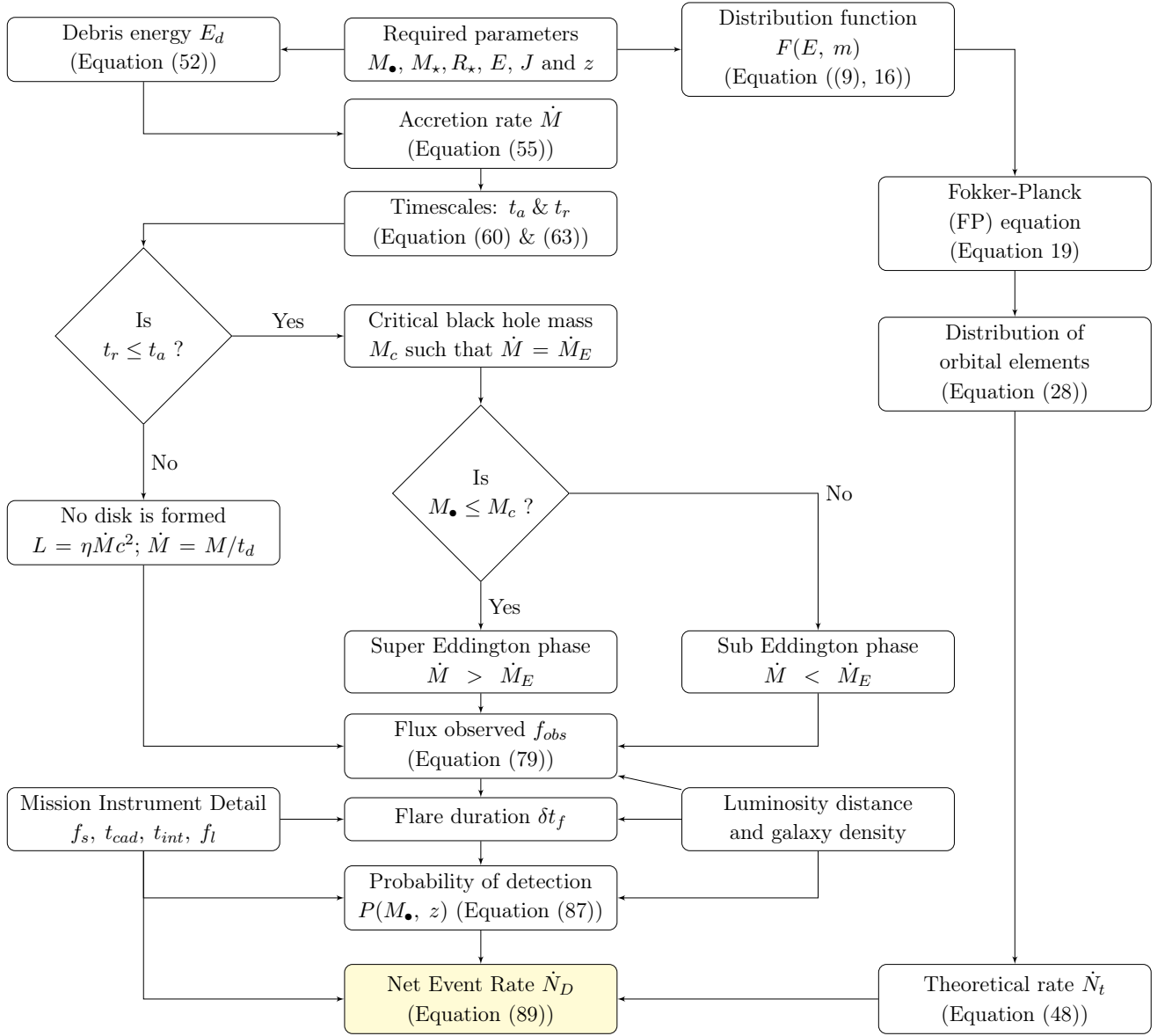

The Figure \ref{fc} shows the methodology we have adopted to calculate the detectable TDE with the initial parameters $M_{\bullet}$, $M_{\star}, R_{\star}$, $E$, $J$ and redshift $z$. In Section 2, we solve the steady state FP equation for a power law density profile $\rho (r)\propto r^{-\gamma}$ and stellar mass function $\xi(m)$. We obtain $\dot{N}_{t}(E,\hspace{1mm}J,\hspace{1mm} m,\hspace{1mm}\gamma)$ for the typical parametric range of density profiles ($\gamma= 0.6$ \hspace{1mm}--\hspace{1mm} 1.4), energy, and angular momentum ($J\leq J_{lc}$), which we use it later to calculate the detection rate, $\dot{N}_{D}$. In Section 3, we have calculate the energy $E_d$ of the disrupted debris and the maximum radius $R_{l}(E,\hspace{1mm}J)$ from the star center to the point where the debris is bound to the BH. We then simulate $\dot{M}(E,\hspace{1mm}J,\hspace{1mm}t)$. In Section 4, we compare the accretion $t_{a}$, ring formation $t_{r}$, viscous $t_{v}$ and radiation $t_{R}$ timescales and discuss the conditions for the formation of an accretion disk. In Section 5, by equating the $\dot{M}$ and Eddington mass accretion rate $\dot{M}_{E}$, we obtain the critical BH mass $M_{c}(E,\hspace{1mm}J)$, such that for $M_{\bullet}<M_{c}$, the accretion disk formed has a super Eddington phase. We then simulate the light curve profiles in the optical and X-ray as a function of $E$, $J$, and $M_{\bullet}$, depending on whether the accretion disk is super Eddington or sub Eddington. The flux from the source at a redshift $z$ is compared with the sensitivity of the mission instrument to obtain the $\delta t_{f}$. In Section 6, we calculate $\dot{N}_{D}$ for optical and X-ray missions by assuming the standard cosmological parameters and BH mass function to obtain the galaxy density. The Table \ref{gloss} shows a glossary of symbols we use in this paper.


\begin{table}
\scriptsize
\scalebox{0.9}{
\begin{tabular}{|c l | c l|}
\hline
{\bf Common Parameters} & && \\
\hline
&&&\\
$M_{\bullet}$ & Black hole mass & $J$ & Orbital angular momentum \\
$M_{6}$ & $M_{\bullet}/(10^{6}M_{\odot})$ & $J_{lc}$ & Loss cone angular momentum\\
$M_{\star}$ & Stellar mass  & $J_c$ & Angular momentum of circular orbit \\
$E$ & Orbital energy & $\ell$ & $J/J_{lc}$ \\
$\xi(m)$ & Stellar mass function &$m$& $M_{\star}/M_{\odot}$\\
\hline
{\bf Stellar Dynamical Parameters} &&& \\
\hline
$j$ & $J^2/J^2_{c}$ & $\mathcal{E}$ & $E/\sigma^2$ \\
$r_t$ & Tidal radius & $\rho$ & Galactic density  \\
$r_h$ & Black hole influence radius & $\gamma$ & Galaxy density power law \\
$s$ & $r/r_h$ & $s_t$ & $r_t/r_h$ \\
$R_s$ & Schwarzschild radius & $\Phi_{\star}$ & Stellar potential \\
$r_{b}$ & Break radius of Nuker profile & $\Phi_{\bullet}$ & Black hole potential \\
$\sigma$ & stellar velocity dispersion & $\Phi $ & Total potential= $\Phi_{\bullet}+\Phi_{\star}$\\
$q$ & Diffusion parameter & $t_{\rm MS}$ & Main sequence lifetime \\
$\mathcal{E}_{c}$ & Critical energy for $q=1$ & $T_{r}$ & Radial period of orbit\\
$\dot{N}_{t}$ & Theoretical TDE rate & $f_{\star}$ & Probability of main sequence star capture\\
\hline
{\bf Accretion Dynamical Parameters } &&& \\
\hline
$r_p$ & Pericenter of the orbit & $t_d$ & dynamical time \\
$\bar{e}$ & $E/(GM_{\bullet}/r_{t})$= $(r_{t}/r_{h})\mathcal{E}$ & $\Gamma$ & Adiabatic index  \\
$E_{d}$ & Energy of disrupted debris & $k$ & Spin factor \\
$R_{l}$ & Maximum radius from star center to bound debris & $t_{m}$ & Orbital period of inner-most debris\\
$x_l$ & $R_l/R_{\star}$ & $t_{a}$ & Accretion timescale\\ 
$x$ & $\Delta R/R_{\star}$ &$\Delta R$ & Debris radius from star center \\
$\varepsilon$ & $E_d/E_{dm}$ &$E_{\rm dm}$ &  Energy of inner-most bound debris   \\
$\mu$ & $M/M_{\star}$ & $M$ &  Debris mass with energy $E_d$ \\
$\dot{M}$ & Mass accretion rate & $\tau$ & $t/t_m$ \\
$\dot{M}_{E}$ & Eddington mass accretion rate &$t_{r}$ & Ring formation timescale\\
$f_{r}$ & Fraction of star mass bound to black hole & $t_{v}$ & Viscous timescale  \\
$M_{c}$ & Critical black hole mass & $t_{R}$ & Radiation timescale\\
$\kappa$ & Opacity of the medium & $\mathcal{T}_{r}$ & $t_{r}/t_{a}$ \\
$r_c$ & Circularization radius  &$\mathcal{T}_{v}$ & $t_{v}/t_{R}$ \\
$r_{L}$ & Outflowing wind launch radius &$T_{\rm ph}$ & Temperature of photosphere \\
$r_{\rm ph}$ & Radius of photosphere of outflowing wind & $T_{e}$ & Effective temperature of disk  \\
$L_e$ & Luminosity emitted from the source &  $L$ & Luminosity  \\
$\psi(M_{\bullet})$ & Black hole mass function & $L_E$ & Eddington luminosity \\
$P(M_{\bullet},\hspace{1mm}z)$ & Probability of detection  & $z$ & Redshift  \\
$\Upsilon$ & Detection efficiency of a detector & $\dot{N}_{D}$ & Detectable rate \\
\hline
{\bf Instrumental Parameters} & &&\\
\hline
$f_{l}$ & sensitivity of the detector & $t_{\rm cad}$ & Cadence of instrument \\
$t_{\rm int}$ & Integration time of detector & $f_{s}$ & Fraction  of sky survey \\
\hline
\end{tabular}
}
\caption{Glossary of symbols}
\label{gloss}
\end{table}

\section{Theoretical capture rate}
\label{tcr}

The strength of tidal encounter is expressed in terms of parameter $\eta_t$, the ratio of surface gravity of the star $GM_{\star}/R^2_{\star}$ and tidal gravity due to the BH $GM_{\bullet}R_{\star}/r^3_p$ at the pericenter $r_p$, given by \citep{2013CQGra..30x4005M}

\begin{equation}
\eta_t=\left(\frac{GM_{\star}}{R^2_{\star}}\frac{r^3_p}{GM_{\bullet}R_{\star}}\right)^{\frac{1}{2}}.
\end{equation}

The tidal radius $r_t$ is defined as the value of $r_p$ that satisfies

\begin{subequations}
\begin{align}
r_t &=\left(\eta^2_t\frac{M_{\bullet}}{M_{\star}}\right)^{\frac{1}{3}}R_{\star} \\
 &= 2.25\times 10^{-6}\eta^{\frac{2}{3}}_t\left(\frac{M_{\bullet}}{10^6M_{\odot}}\right)^{\frac{1}{3}}\left(\frac{M_{\star}}{M_{\odot}}\right)^{-\frac{1}{3}}\left(\frac{R_{\star}}{R_{\odot}}\right)~{\rm pc}
\end{align}
\label{tidrad}
\end{subequations} 

\begin{figure}
\begin{center}
\includegraphics[scale=0.50]{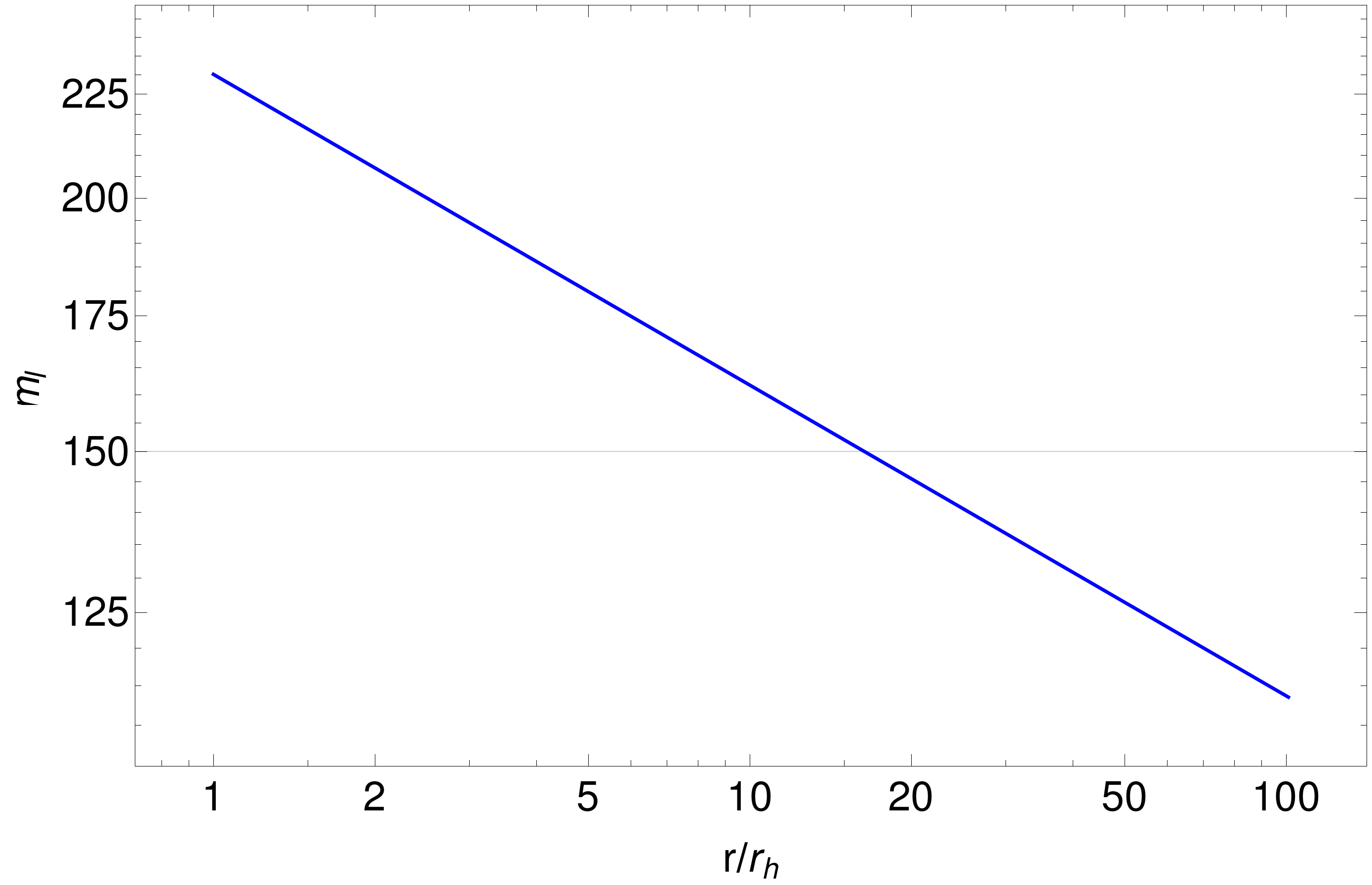}
\end{center}
\caption{ The mass limit of star $m_{l}$ as a function of black hole mass $M_{\bullet}=10^{6}M_{\odot}M_{6}$ for $a=r_h$. It is the maximum mass of star in the cluster that can be captured during its main sequence lifetime. The thin gray line shows the maximum mass in the \citet{2001MNRAS.322..231K} sample of stars.  }
\label{mlim}
\end{figure}

The quantity $\eta_t$ is the form factor of order unity and we have taken $\eta_t=1$. For a main sequence star with the mass-radius relation $R_{\star}=R_{\odot} m^n$, where $m=M_{\star}/M_{\odot}$, the tidal radius given by Equation (\ref{tidrad}b) reduces to 

\begin{equation}
r_{t}(M_{\bullet},~m)\approx 2.25\times 10^{-6}\left(\frac{M_{\bullet}}{10^6M_{\odot}}\right)^{\frac{1}{3}}m^{n-\frac{1}{3}}~{\rm pc}
\end{equation}

and for $n>1/3$, $r_{t}(M_{\bullet},\hspace{1mm}m)$ increases with $m$. We take $n=0.8$ for the entire range of stellar masses in our calculations \citep{1994sse..book.....K}. We take the lifetime of the main sequence star $t_{MS} \propto M^{-2.5}_{\star}$ and the dynamical time of the star to fall into the BH $t_{\rm dyn}=\sqrt{a^3/GM_{\bullet}}$, where $a$ is the semimajor axis of the star to the BH. For a star to be captured during its main sequence lifetime, $t_{\rm dyn}<t_{MS}$, which gives 

\begin{equation}
m_{l}=\left(t_{\odot}\sqrt{\frac{GM_{\bullet}}{a^3}}\right)^{0.4},
\end{equation}

where $t_{\odot}$ is the life time of the sun and is shown in Figure \ref{mlim} for $a=r_{h}$.

Stars in the galactic center move in the potential field of both SMBH and other stars in the galaxy. The DF is assumed to be a function of energy $E=\Phi(r)-v^2/2$ only and is given by $f(E)\propto E^p$ for $r \leq r_h$ and $\Phi (r)=GM_{\bullet}/r$, where $r_h=GM_{\bullet}/\sigma^2$ is the radius of influence and $\sigma$ is the stellar velocity dispersion and is related to the $M_{\bullet}$ through the $M_{\bullet}-\sigma$ relation given by \citet{2005SSRv..116..523F}

\begin{equation}
M_{\bullet}=1.66\times 10^8 M_{\odot}\left(\frac{\sigma}{200~{\rm Km ~ sec^{-1}}}\right)^{4.86}
\label{msigma}
\end{equation}

\citet{1976ApJ...209..214B} introduced the stellar scattering and diffusion and found that $p=3/4$ \citep{1972ApJ...178..371P} gives a negatively divergent flux; they also obtained $p=1/4$ for the steady state distribution, which gives a constant energy flux. The stars are tidally captured if the angular momentum is $J\leq J_{lc}(E,r_{t})$ where $J_{lc}(E,r_{t})=\sqrt{2r^2_t(\Phi (r_t)-E)}$ is the loss cone angular momentum  \citep{1976MNRAS.176..633F}. The maximum value of $J$ is $J_{lc}(E,r_{t})$. As $J_{lc}(E,r_{t})\geq 0$, the maximum value of energy is $E_{\boldsymbol{m}}= \Phi (r_t)$.

Because $J_{lc}(E,r_{t})$ depends on $r_{t}$, which varies with $M_{\star}$, so we consider a DF that depends on the stellar mass function $\xi(m)$ given by \citep{2001MNRAS.322..231K}

\begin{equation}
\xi (m) \approx \left\{
\begin{array}{ll}
H m^{-1.3}& 0.08<m<0.5 \\
B m^{-2.3}& 0.5<m<150  
\end{array}
\right. 
\label{smf}
\end{equation}

where 

\begin{equation*}
H=2B \hspace{5mm},  \hspace{5mm} B=\frac{1}{7.91-0.77m^{-1.3}_m}\hspace{5mm} {\rm and} \hspace{5mm} m=\frac{M}{M_\ast},
\end{equation*}

where $m_m$ is the maximum mass of a main sequence star in the stellar distribution, taken to be 150.

\begin{figure}
\begin{center}
\includegraphics[scale=0.70]{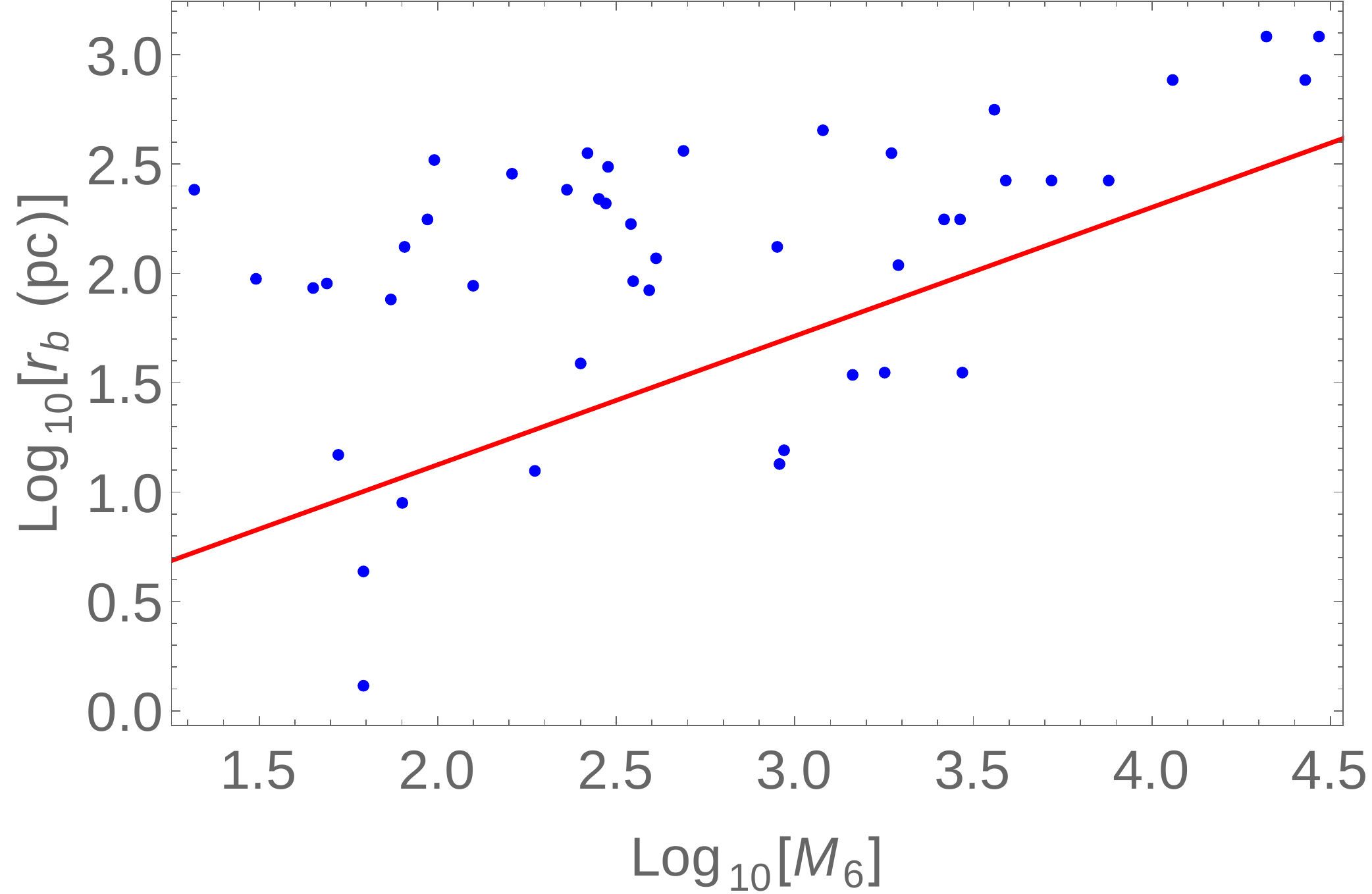}
\end{center}
\caption{ The blue points show the break radius $r_{b}$ for the sample of galaxies given in \citet{2004ApJ...600..149W}. The red line shows the radius of influence $r_{h}$. The plot indicates that for most of galaxies $r_{b}>r_{h}$, which implies that for $r\leq r_{h}$, the density $\rho(r)$ can be taken to be a single power law profile. }
\label{rb}
\end{figure}

\citet{2004ApJ...600..149W} and \citet{2014arXiv1410.7772S} have taken the sample of galaxies with a Nuker law, which is basically a double power law profile with break radius $r_{b}$ that separates inner and outer slopes. In Figure \ref{rb}, we have shown the range of $r_{b}$ and $r_{h}$ for the sample of galaxies given in \citet{2004ApJ...600..149W}. For most of the galaxies $r_{b}>r_{h}$, which is also true for the sample of galaxies given in \citet{2014arXiv1410.7772S}. Because the stellar dynamics in the galactic center is influenced by the BH for $r\leq r_h$, we consider a single power law density profile $\rho(r)=\rho_{0} (r/r_{0})^{-\gamma}$ for $r \leq r_{h}$, where $\gamma$ is the inner slope of the Nuker law. We define the $r_{h}$, where $M_{\star}(r_{h})=2M_{\bullet}$ \citep{2004ApJ...600..149W}, such that

\begin{equation}
\rho_{0}r^{\gamma}_{0}=\frac{3-\gamma}{2\pi}M_{\bullet}r^{\gamma-3}_h
\label{rono}
\end{equation} 

The potential due to the stellar distribution is obtained from the Poisson equation and  given by

\begin{equation}
\Phi_{\star}(r) = 2\sigma^2 \left\{
\begin{array}{ll}
\frac{1}{2-\gamma}  \left[1-\left(\frac{r}{r_h}\right)^{2-\gamma}\right]& \hspace{2mm} \gamma \neq 2 \\
\ln (\frac{r_h}{r})& \hspace{2mm}  \gamma =2
\end{array}
\right. 
\label{poten}
\end{equation}

The total potential is given by $\Phi(r)=\displaystyle{\Phi_{\bullet}(r)+\Phi_{\star}(r)}$, where $\Phi_{\bullet}(r)=GM_{\bullet}/r$ is the potential due to the BH. We consider the DF as

\begin{equation}
F(E,\hspace{1mm}m)=f(E)\xi(m)
\label{fem}
\end{equation}

The density of stars for $F(E,\hspace{1mm}m)$ is given by

\begin{equation}
\rho(r)= \int \diff^3 v \hspace{1mm} M_{\star} \hspace{1mm} F(E,\hspace{1mm}m) =\int \diff^3 v \hspace{1mm} f(E) \int \diff m \hspace{1mm} \xi(m) \hspace{1mm} M_{\star},
\label{den}
\end{equation}

where $\diff^3 v=2\pi v_{t} \diff v_{t} \diff v_{r}$ with radial velocity $v_{r}$ and tangential velocity $v_{t}$ is given by

\begin{equation}
v_{t}=\frac{J}{r}; \hspace{4mm} v_{r}=\sqrt{2(\Phi(r)-E)-\frac{J^2}{r^2}}; \hspace{4mm} \diff^3 v=\frac{2\pi}{r^2 v_{r}} J \diff E \diff J
\end{equation}

For a spherically isotropic galactic center, the function $f(E)$ is obtained through the inverse transform of Equation (\ref{den}), and is known as the Eddington formula, given by \citep{2008gady.book.....B}

\begin{equation}
f(E)=\frac{1}{\sqrt{8}\pi^2 \left<M_{\star}\right>}\frac{\diff}{\diff E} \int_{E_{min}}^E \frac{\diff \rho}{\diff \Phi} \frac{1}{\sqrt{E-\Phi}} \,\diff \Phi,
\label{edf}
\end{equation}

where

\begin{equation}
\left<M_{\star}\right> =\int_{0.08}^{150} \hspace{1mm} M_{\star} \hspace{1mm} \xi(m) \,\diff m
\end{equation}

and $E_{\rm min}$ is the minimum of $E$ taken to be -100. The number of stars in the cluster for a given $F(E,\hspace{1mm}m)$ is 

\begin{equation}
N=\int \diff^3 r \int \diff^3 v \int \diff m \hspace{1mm} F(E,\hspace{1mm}m), 
\end{equation}

where $\diff^3 r=4\pi r^2 \diff r$ for spherical galaxy. In terms of dimensionless variables $\ell=J/J_{lc}$ and $\mathcal{E}=E/\sigma^2$, the number of stars is given by

\begin{equation}
N(\mathcal{E},\hspace{1mm}\ell)\hspace{1mm}\diff \mathcal{E} \hspace{1mm}\diff \ell=\int \diff m~8\pi^2\hspace{1mm} J^2_{lc} \hspace{1mm}\sigma^2 \hspace{1mm}\ell \hspace{1mm}T_r(\mathcal{E},\hspace{1mm}\ell)\hspace{1mm} F(\mathcal{E},\hspace{1mm}m) \hspace{1mm}\diff \mathcal{E} \hspace{1mm}\diff \ell
\label{nel}
\end{equation}

where $\displaystyle T_r(\mathcal{E},\hspace{1mm}\ell)=\oint \frac{\diff r}{v_{r}}$ is the radial period. For a spherical geometry, $T_r(\mathcal{E},\hspace{1mm}\ell)$ is a function of $\mathcal{E}$ only, and the $N(\mathcal{E},\hspace{1mm}\ell)$ increases with $\ell$.  

From the given stellar density profile and potential $\Phi(r)$, the DF $f(E)$ is given by 

\begin{equation}
f(\mathcal{E})=\frac{M_{\bullet}}{\left<M_{\star}\right>}\frac{1}{r^3_h \sigma^3} g(\mathcal{E})
\label{dff}
\end{equation}

where 

\begin{equation}
\displaystyle{g(\mathcal{E})=\frac{1}{4\sqrt{2}\pi^3}\left\{
\begin{array}{ll}
\gamma(3-\gamma)\frac{\diff}{\diff \mathcal{E}} \int_{s_1}^{s_2} \frac{s^{\gamma-1}}{\sqrt{\mathcal{E}-s-\frac{2}{2-\gamma}(1-s^{\gamma-2})}} \,\diff s & \hspace{2mm} \gamma \neq 2 \\
\int_{\mathcal{E}_{\rm min}}^{\mathcal{E}} \frac{\mathcal{L}(2+\mathcal{L})}{(1+\mathcal{L})^3} \frac{1}{\sqrt{\mathcal{E}-\Psi}} \,\diff \Psi & \hspace{2mm} \gamma =2
\end{array}
\right. }
\label{ge}
\end{equation}

and $\mathcal{L}$ is the Lambert function given by $\mathcal{L}e^{\mathcal{L}}=(1/2) e^{\Psi/2}$ where $\Psi=\Phi/\sigma^2$, $s=r_h/r$, $\mathcal{E}=E/\sigma^2$ \citep{2004ApJ...600..149W} and $s_1$ and $s_2$ are obtained by solving

\begin{subequations}
\begin{align}
s_1+\frac{2}{2-\gamma}(1-s_{1}^{\gamma-2}) &=\mathcal{E}_{\rm min} \\
s_2+\frac{2}{2-\gamma}(1-s_{2}^{\gamma-2}) &=\mathcal{E} 
\end{align}
\end{subequations} 

\begin{figure}
\begin{center}
\includegraphics[scale=0.38]{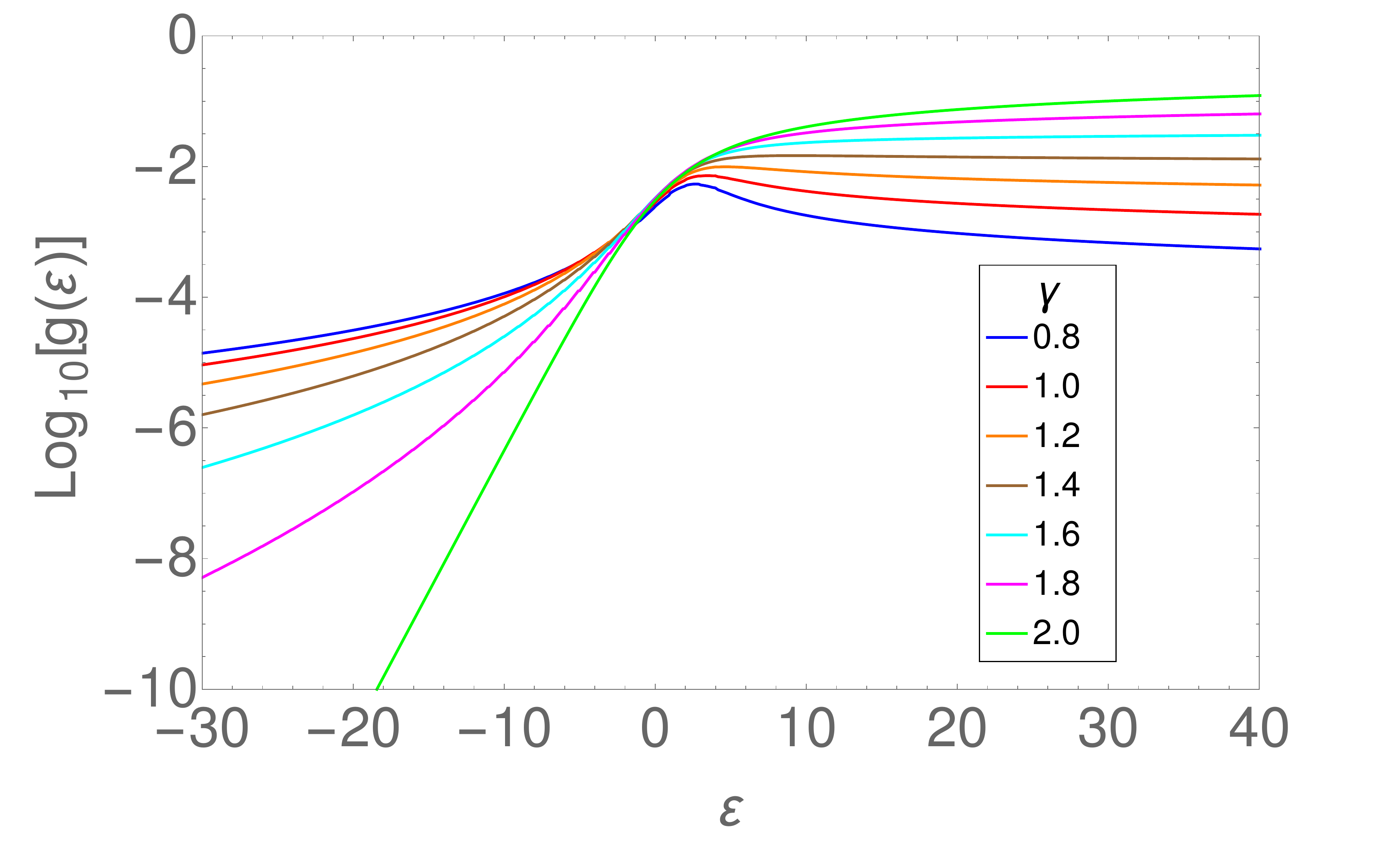}
\end{center}
\caption{The dimensionless $g(\mathcal{E})$ is shown for various $\gamma$. For $\gamma=2$, $g(\mathcal{E})$ corresponds to Equation (17c) of \citet{2004ApJ...600..149W}. }
\label{gec}
\end{figure}

\begin{figure}
\begin{center}
\includegraphics[scale=0.5]{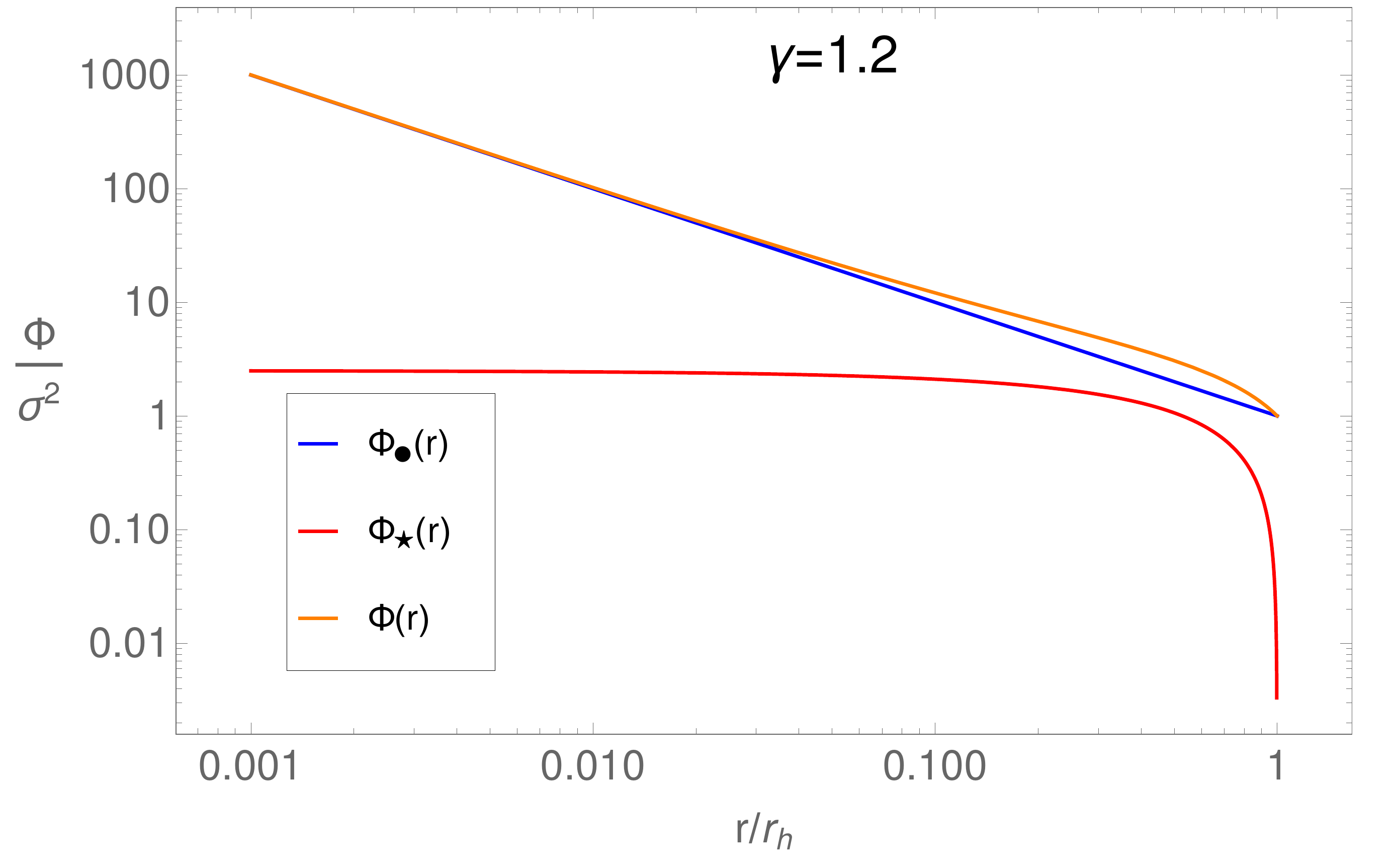}
\caption{The stellar potential $\Phi_{\star}(r)$, black hole potential $\Phi_{\bullet}(r)$ and total potential $\Phi(r)$ are shown for $\gamma=1.2$. For $r\ll r_{h}$, the black hole potential $\Phi_{\bullet}(r)=GM_{\bullet}/r$ dominates over the star potential.}   
\label{pot}
\end{center}
\end{figure}

The Figure \ref{gec} shows the plot of $g(\mathcal{E})$ for various $\gamma$. For $\gamma=2$, $g(\mathcal{E})$ corresponds to Equation (17c) of \citet{2004ApJ...600..149W}. The BH potential dominates over the star potential for $r \ll r_{h}$ as shown in Figure \ref{pot} and thus $g(\mathcal{E})\propto \mathcal{E}^{\gamma-3/2}$ for $\mathcal{E} \gg 1$.
 
\subsection{Loss cone theory}
\label{lct}

The loss cone is a geometrical region in phase space for which $r_{p} \leq r_{t}$. We adopt the \citet{1978ApJ...226.1087C} formalism for computing the flux of stars into the loss cone in $\mathcal{E}$ and the $j=J^2/J^2_{c}(\mathcal{E})$ phase space, where $J_{c}(\mathcal{E})$ is the angular momentum of circular orbit with energy $\mathcal{E}$. We consider the diffusion in the $j$ space only and in the limit $j\rightarrow 0$, the steady state FP equation reduces to \citep{2013CQGra..30x4005M}

\begin{equation}
\frac{\diff \mathcal{F}(\chi,\hspace{1mm}y)}{\diff \chi}=\frac{\diff }{\diff y}\left(y\frac{\diff \mathcal{F}(\chi,\hspace{1mm}y)}{\diff y}\right)
\label{fty}
\end{equation}

with the boundary condition

\begin{subequations}
\begin{align}
\mathcal{F}(0,\hspace{1mm}y) &=0 \hspace{14mm} \forall \hspace{2mm} y< y_{lc} \\
\mathcal{F}(0,\hspace{1mm}y) &=\mathcal{F}(1,\hspace{1mm}y)\hspace{4mm} \forall \hspace{2mm} y \geq y_{lc}  
\end{align}
\label{bc}
\end{subequations} 

where 

\begin{equation}
\chi=\frac{1}{\left<D(\mathcal{E})\right>}\int_{r_{p}}^{r} \lim_{j\rightarrow 0}\frac{\left<(\Delta j)^2\right>}{2j } \,\frac{\diff r}{v_{r}}\hspace{5mm} {\rm and} \hspace{5mm} y=\frac{j}{\left<D(\mathcal{E})\right>}
\end{equation}

and $\displaystyle{\left<D(\mathcal{E})\right>=\oint \lim_{j\rightarrow 0}\frac{\left<(\Delta j)^2\right>}{2j} \,\frac{\diff r}{v_{r}}} $ is the orbit averaged angular momentum diffusion coefficient and $y_{lc}$ is $y$ at $j=j_{lc}$. \citet{2015ApJ...804..128M} has expressed $\mathcal{F}(\chi,\hspace{1mm}y)$ in terms of the distribution of the pericenters $r_{p}$ and apocenters $r_{a}$, and calculated the capture rate $\dot{N}_{t}$ in terms of $r_{a}$ and $r_{p}$ while we use appropriately scaled values of $E$ and $J^2$ for calculating $r_p$ and various other parameters required in the gas dynamical calculation in the following sections. Note that in the calculation we use the total potential inclusive of the stars and the BH. 

The local diffusion coefficient in the limit $j\rightarrow 0$ is given by \citep{1999MNRAS.309..447M}

\begin{equation}
\lim_{j\rightarrow 0}\frac{\left<(\Delta j)^2\right>}{2j}=\frac{r^2}{J^2_{c}(\mathcal{E})}\left<(\Delta v_{\perp})^2\right>
\end{equation}

where $\left<(\Delta v_{\perp})^2\right>$ is given in Appendix L of \citet{2008gady.book.....B}. Thus, the orbit averaged diffusion coefficient is given by

\begin{equation}
\left<D(\mathcal{E})\right>=\frac{32\sqrt{2}}{3}\frac{\pi^2 G^2 \left<M^2_{f}\right>\log\Lambda}{J^2_{c}(\mathcal{E})}\frac{M_{\bullet}}{\left<M_{\star}\right>}\frac{1}{\sigma^2}(2h_{1}(\mathcal{E})+3h_2(\mathcal{E})-h_{3}(\mathcal{E}))
\end{equation}

where $M_{f}$ is the mass of field star with the maximum mass taken to be 150 $M_{\odot}$, $\Lambda\approx M_{\bullet}/M_{\star}$ and

\begin{subequations}
\begin{align}
\left<M^2_{f}\right> &=M^2_{\odot}\int_{0.08}^{150} m^2_{f}\xi(m_f) \,\diff m_f \\
h_{1}(\mathcal{E}) &=\int_{0}^{s(\mathcal{E})} \diff s' \frac{s^{'^2}}{\sqrt{\Psi(s^{'})-\mathcal{E}}} \int^{\mathcal{E}}_{-\infty} \diff \mathcal{E}' \hspace{1mm} g(\mathcal{E}')\\
h_{2}(\mathcal{E}) &=\int_{0}^{s(\mathcal{E})} \diff s^{'} \frac{s^{'^2}}{\Psi(s^{'})-\mathcal{E}} \int^{\Psi(s^{'s})}_{\mathcal{E}} \diff \mathcal{E}' \sqrt{\Psi(s^{'})-\mathcal{E}'}\hspace{1mm} g(\mathcal{E}') \\
h_{3}(\mathcal{E}) &=\int_{0}^{s(\mathcal{E})} \diff s^{'} \frac{s^{'^2}}{(\Psi(s^{'})-\mathcal{E})^2} \int^{\Psi(s^{'})}_{\mathcal{E}} \diff \mathcal{E}' (\Psi(s^{'})-\mathcal{E}')^{\frac{3}{2}}\hspace{1mm} g(\mathcal{E}') 
\end{align}
\end{subequations} 

where $s(\mathcal{E})$ is obtained by solving $\Psi(s)=\mathcal{E}$, $s=r/r_{h}$ and 

\begin{equation}
\Psi(s)=\frac{\Phi}{\sigma^2}=\frac{1}{s}+\frac{2}{2-\gamma}(1-s^{{2-\gamma}}).
\label{psis}
\end{equation}

Now $J^2_{c}(\mathcal{E})$ is given by

\begin{equation}
J^2_{c}(\mathcal{E})=\sigma^2 r^2_{h} \hspace{1mm}[s_{c}(\mathcal{E})+2s^{{4-\gamma}}_{c}(\mathcal{E})]
\label{angm}
\end{equation}

where $s_{c}(\mathcal{E})$ is given by

\begin{equation}
\frac{1}{2s_{c}}+\frac{2}{2-\gamma}(1-s^{{2-\gamma}}_{c})-s^{{2-\gamma}}_{c}=\mathcal{E}.
\label{sce}
\end{equation}

\begin{figure}
\begin{center}
\includegraphics[scale=0.6]{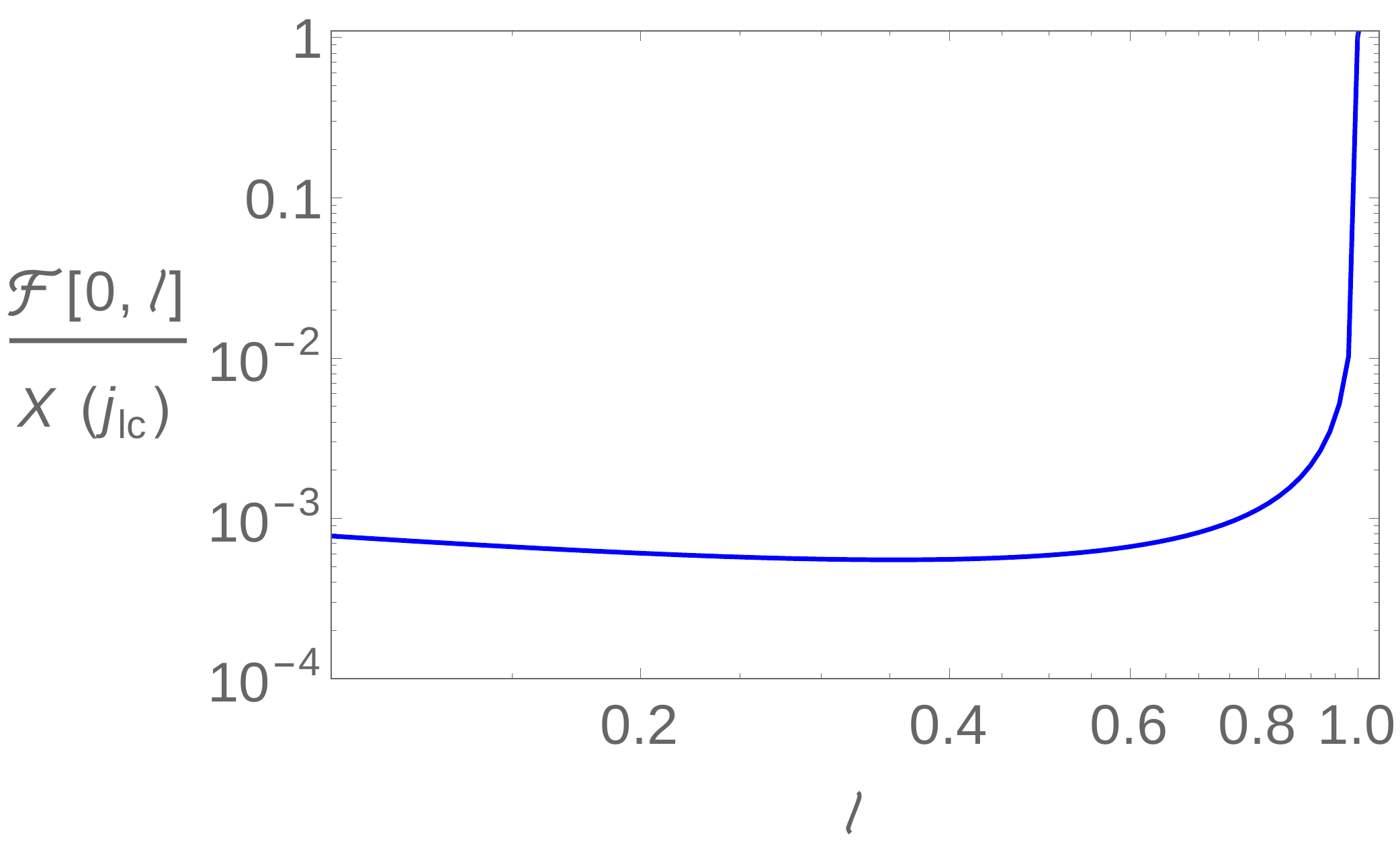}
\end{center}
\caption{The $\mathcal{F}(0,\hspace{1mm}\ell)$ is obtained by transforming Equation (\ref{flr}) from $y$ to $\ell=J/J_{lc}$ space such that $y/y_{lc}=\ell^2$, and calculating for 1000 terms in summation of Equation (\ref{flr}). The boundary condition given by Equation (\ref{bc}a) shows that $\mathcal{F}(0,\hspace{1mm}\ell)$ is a step function with $\ell$ for $\ell\leq 1$. Thus, taking 1000 terms in summation satisfies the boundary condition within a fraction of about $10^{-3}$. }
\label{foy}
\end{figure}

The solution of Equation (\ref{fty}) with the boundary conditions (\ref{bc}) is given by \citep{2013CQGra..30x4005M}

\begin{equation}
\mathcal{F}(\chi,~y)=X(j_{lc})\left(1-2\sum\limits_{n=1}^{\infty}\frac{e^{\frac{-\alpha^2_{n}q}{4}\chi}}{\alpha_{n}}\frac{J_{0}\left(\alpha_{n} \sqrt{y/y_{lc}}\right)}{J_{1}(\alpha_{n})}\right)
\label{flr}
\end{equation}

where $\alpha_{n}$ are the consecutive zeros of the Bessel function $J_{0}(\alpha)$, $q=1/y_{lc}$, and $X(j_{lc})$ is given by

\begin{equation}
X(j_{lc})=\frac{f(\mathcal{E})}{1+q^{-1}\zeta(q)\log\left(\frac{1}{j_{lc}}\right)}; \hspace{5mm} \zeta(q)=1-4\sum\limits_{n=1}^{\infty}\frac{e^{\frac{-\alpha^2_{n}q}{4}}}{\alpha^2_{n}}
\label{xlc}
\end{equation}

where $f(\mathcal{E})$ is given by Equation (\ref{dff}). The Equation (\ref{flr}) gives the DF of stars in the loss cone in terms of energy $\mathcal{E}$ and angular momentum $\ell=J/J_{lc}=\sqrt{y/y_{lc}}$. Figure \ref{foy} shows the plot of $\mathcal{F}(0,\hspace{1mm}\ell)$ for 1000 terms in the summation in Equation (\ref{flr}), which matches with the boundary condition given by Equation (\ref{bc}a) with an accuracy of $\sim 10^{-3}$. With increase in the number of terms in summation, the order of accuracy of $\mathcal{F}(0,\hspace{1mm}\ell)$ to Equation (\ref{bc}a) increases; however, we use 1000 terms in summation to calculate $\mathcal{F}(\chi,\hspace{1mm}\ell)$ beacuse it is in close agreement with the boundary condition. 

The approximation to $\zeta(q)$ given by \cite{1978ApJ...226.1087C} is

\begin{equation}
\zeta(q) \approx \zeta_{CK}(q)=  \left\{
\begin{array}{ll}
1&q \gg 1 \\
\displaystyle{\frac{q}{0.186q+0.824\sqrt{q}}}&q \ll 1  
\end{array}
\right. 
\label{ckzeta}
\end{equation}

\begin{figure}
\begin{subfigure}[]{0.45\textwidth}
\centering
\includegraphics[scale=0.38]{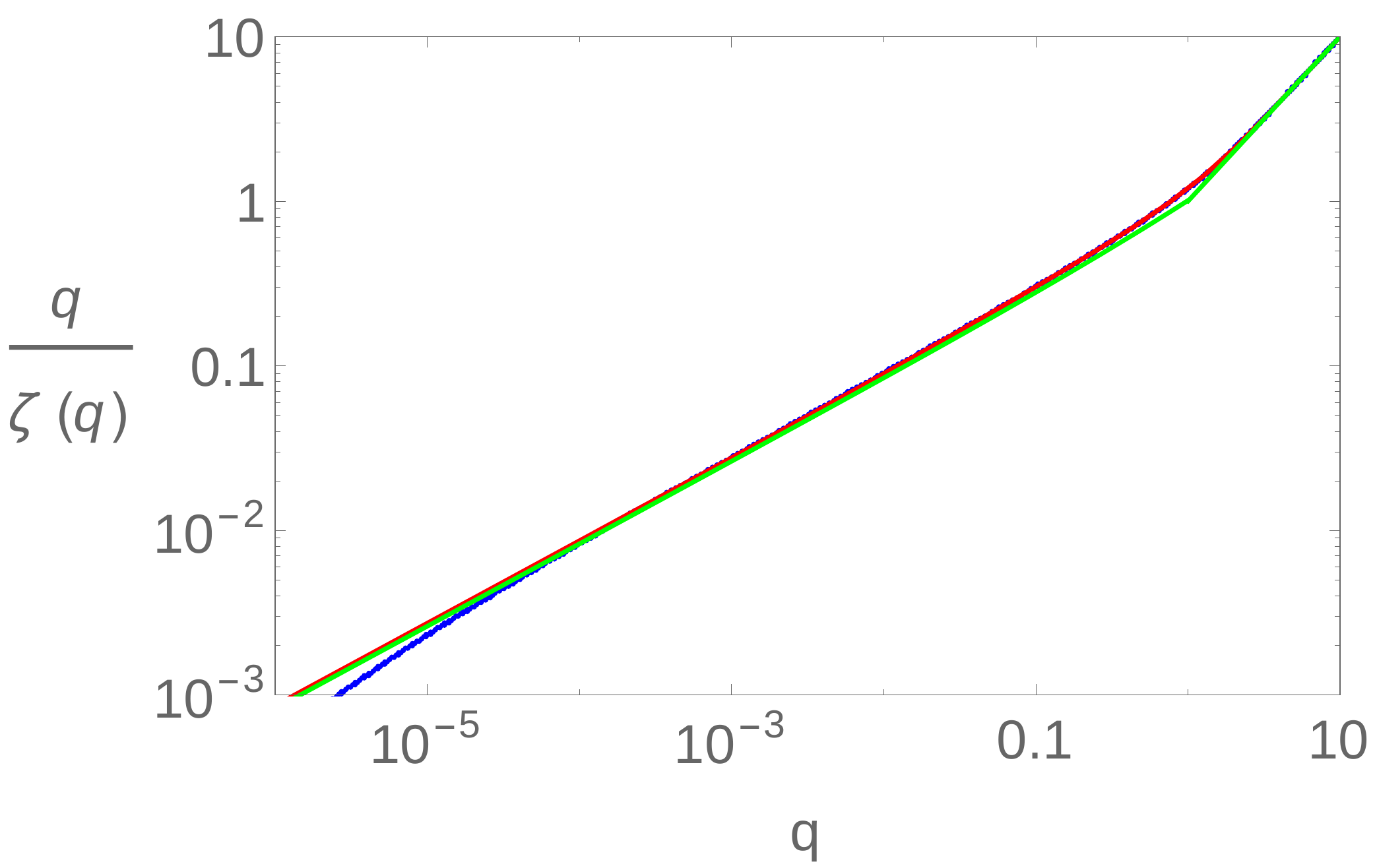}
\caption{}
\end{subfigure}
\quad
\begin{subfigure}[]{0.45\textwidth}
\centering
\includegraphics[scale=0.38]{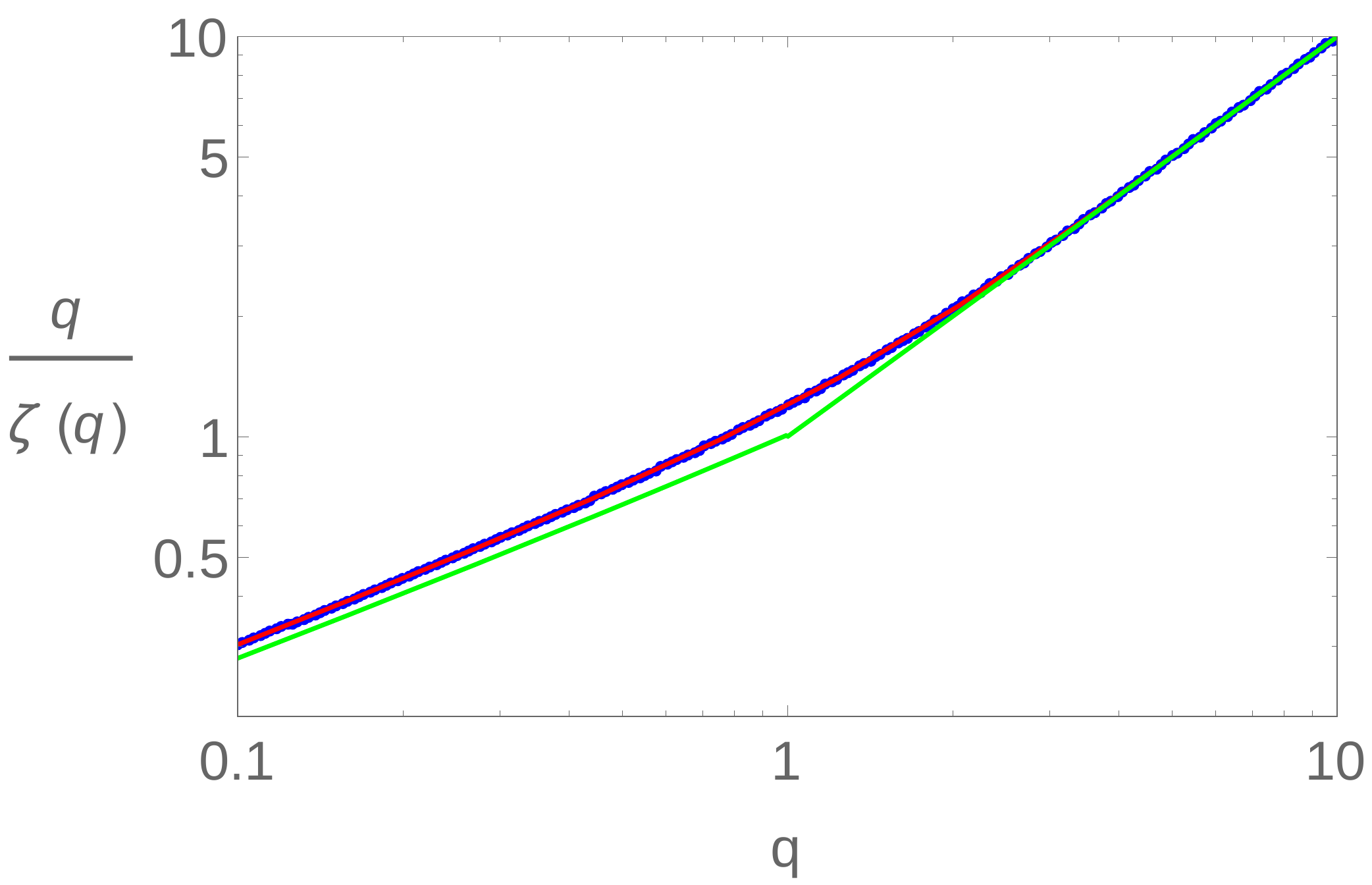}
\caption{}
\end{subfigure}\\
\begin{center}
\begin{subfigure}[]{0.53\textwidth}
\centering
\includegraphics[scale=0.36]{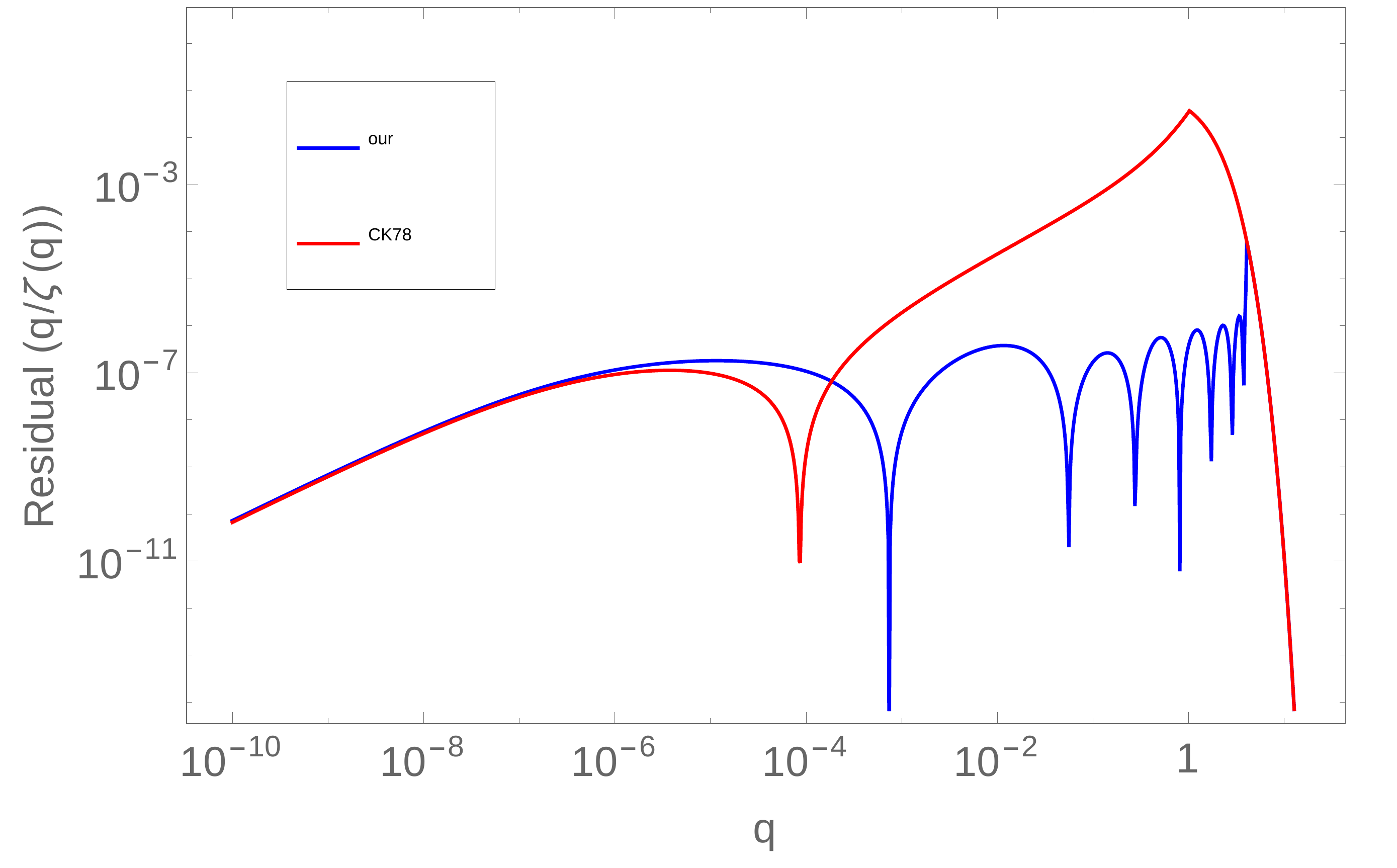}
\caption{}
\end{subfigure}
\end{center}
\caption{The figure (a) shows $q/\zeta(q)$ as a function of $q$ over all ranges under various approximations. The blue line corresponds to $\zeta(q)$ summed up to 10,000 terms. The red thin line corresponds to our approximation of $\zeta(q)$ given by Equation (\ref{azeta}). The green line shows the results obtained by \citet{1978ApJ...226.1087C} and is given by Equation (\ref{ckzeta}).  Asymptotically, the blue line follows the red thin line. The figure (b) shows $q/\zeta(q)$ that for $q$ close to unity; our approximated formula gives a good fit to $\zeta(q)$. The figure (c) shows the residual of $q/\zeta(q)$ for our approximated equation (blue) and the \citet{1978ApJ...226.1087C} approximation (red). }
\label{zeta}
\end{figure}

We have compared the $\zeta(q)$ obtained for a summation of  10,000 terms in Equation (\ref{xlc}) with $\zeta_{CK}(q)$, and it does not fit very well for $q$ close to unity, as shown in Figure \ref{zeta}. Thus we have used a better approximation to $\zeta(q)$ given by

\begin{equation}
\zeta(q) \approx \left\{
\begin{array}{ll}
1&q \geq 4 \\
\displaystyle{\frac{q}{0.86 q^{0.5} + 0.384 q - 0.379 q^{1.5} + 0.427 q^2 - 0.095 q^{2.5}}}&q<4  
\end{array}
\right. 
\label{azeta}
\end{equation}

which follows Equation (\ref{ckzeta}) for $q \ll 1$ and gives a better fit to $\zeta(q)$ for $q$ close to unity. The residual for Equation (\ref{ckzeta}) is higher than the residual for Equation (\ref{azeta}), as shown in Figure \ref{zeta}. 

\begin{figure}
\begin{center}
\includegraphics[scale=0.45]{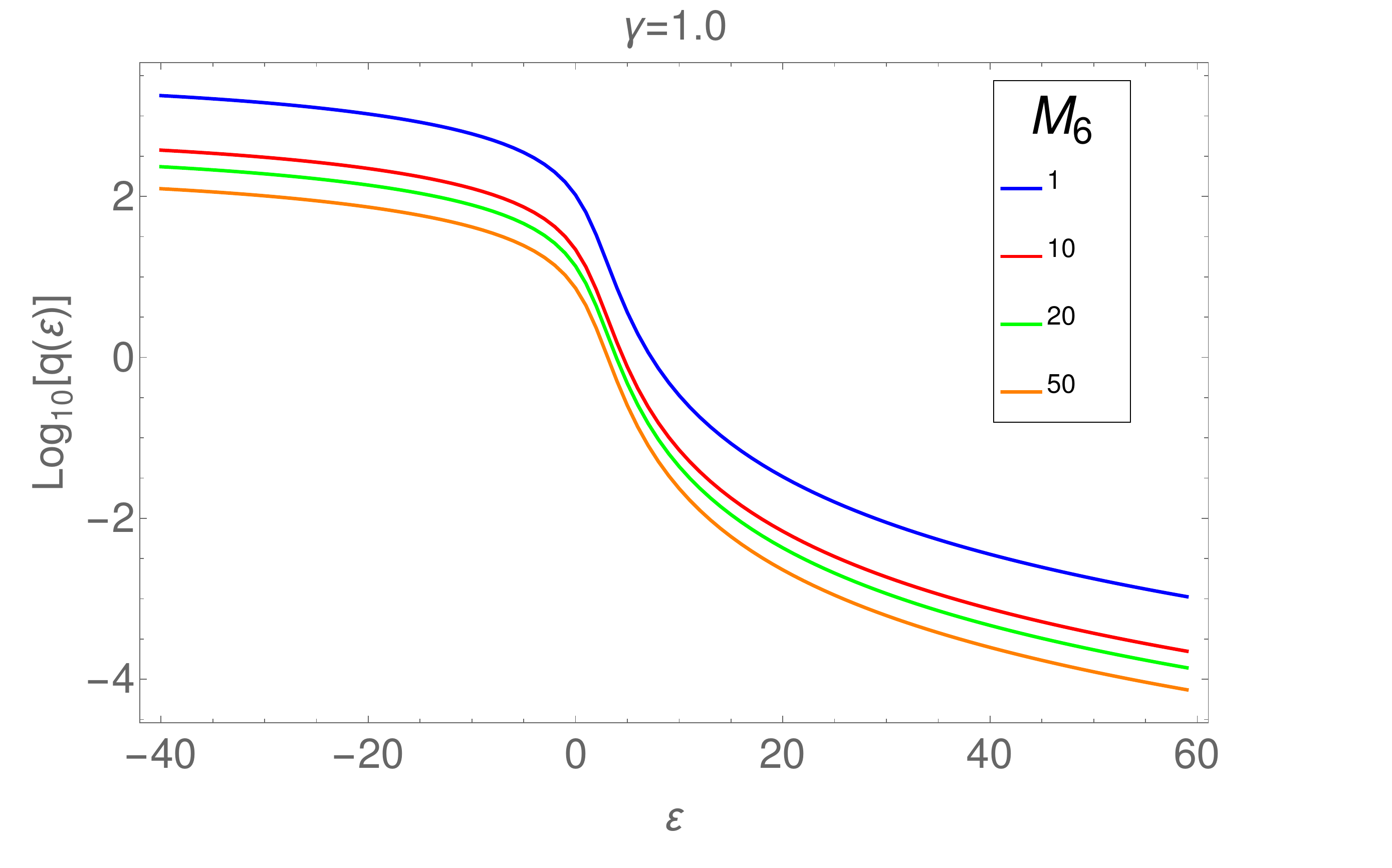}
\end{center}
\caption{ The function $q(\mathcal{E})$ given by Equation (\ref{difpar}) is shown  as a function of $\mathcal{E}$ for $\gamma=1.0$ and $m=1$; $q(\mathcal{E})$ decreases with $\mathcal{E}$ which implies that the high energy orbits have smaller diffusion angle.}
\label{qe}
\end{figure}

\begin{figure}
\begin{subfigure}[]{\textwidth}
\centering
\includegraphics[scale=0.42]{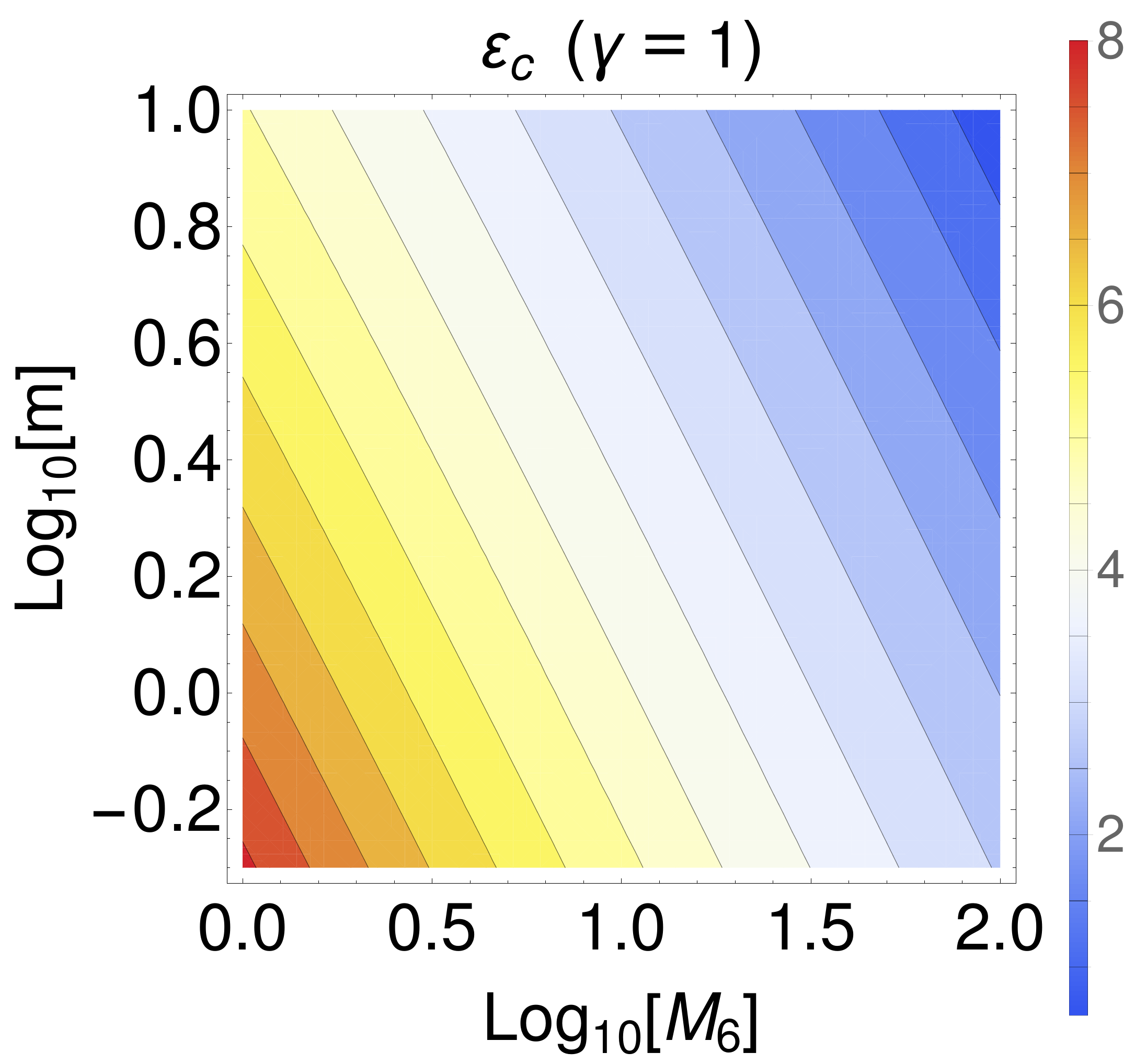}
\caption{}
\end{subfigure}
\quad \\
\begin{subfigure}[]{\textwidth}
\centering
\includegraphics[scale=0.38]{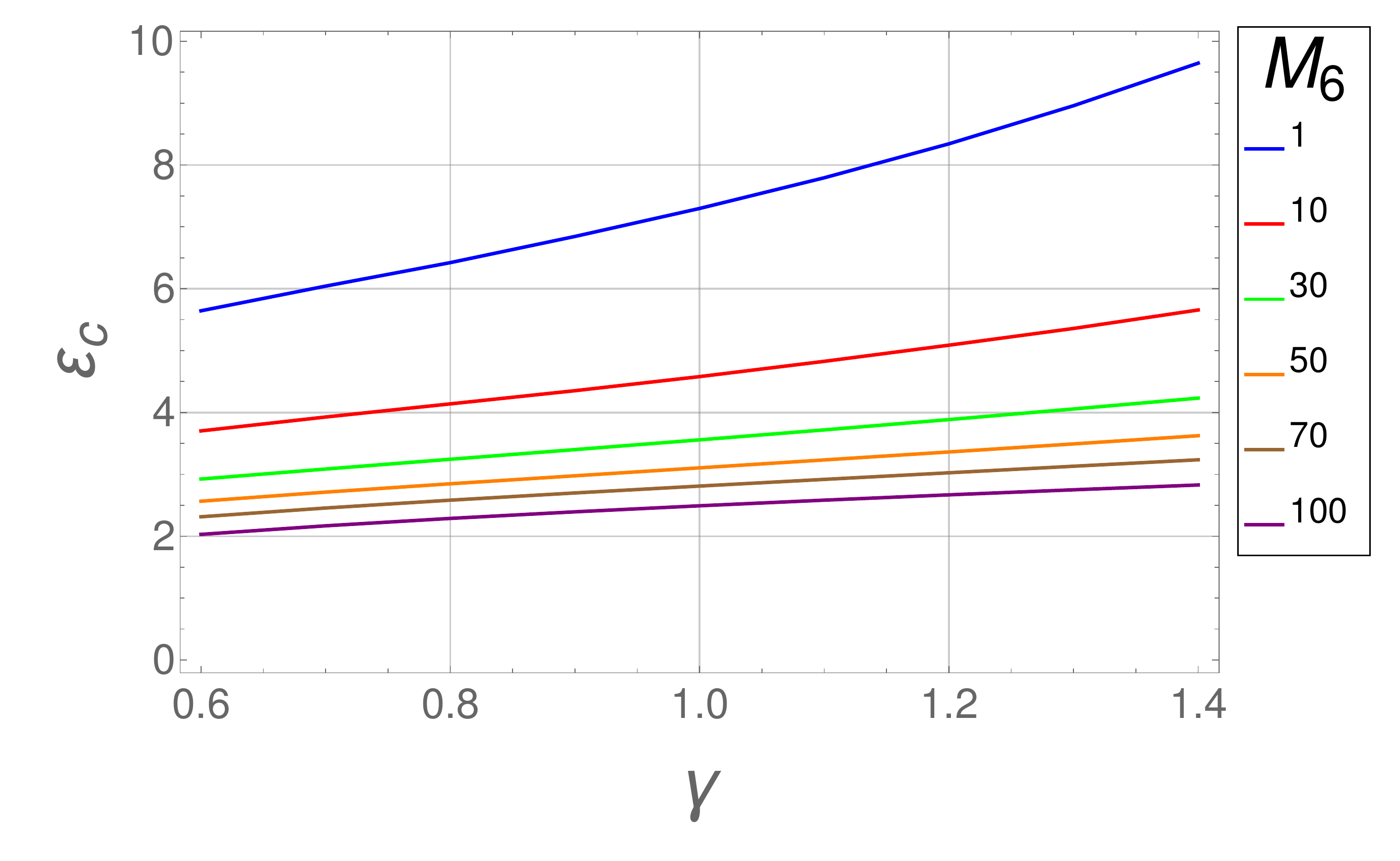}
\caption{}
\end{subfigure}
\caption{ The top panel (a) shows $\mathcal{E}_{c}$ for which $q(\mathcal{E})=1$ as function of $M_{6}$ and $m$ for $\gamma=1$. The bottom panel (b) shows the $\mathcal{E}_{c}$ as a function of $\gamma$ for a star of unit solar mass and radii that increases with $\gamma$.}
\label{ecrit}
\end{figure}

The function $q(\mathcal{E})$  given by 

\begin{equation}
q(\mathcal{E})=\frac{\left<D(\mathcal{E})\right>}{j_{lc}}=\frac{\left<D(\mathcal{E})\right>J^2_{c}(\mathcal{E})}{J^2_{lc}(\mathcal{E},r_{t})}=\frac{\left<D(\mathcal{E})\right>J^2_{c}(\mathcal{E})}{2r^2_{t}(\Phi(r_{t})-\mathcal{E}\sigma^2)}
\label{difpar}
\end{equation}

can be interpreted as the ratio of the orbital period to the timescale for diffusional refilling of the loss cone. The regime $q(\mathcal{E}) > 1$ defines the pinhole or full loss cone in which stellar encounters replenish loss cone orbits much more rapidly than they are depleted, whereas $q(\mathcal{E})<1$ defines the diffusive or empty loss cone regime.  The Figure \ref{qe} shows $q(\mathcal{E})$ plotted as a function of $\mathcal{E}$ for $\gamma=1.0$. The function $q(\mathcal{E})$ decreases with  $\mathcal{E}$ which implies that the high energy orbits have smaller diffusion angle. The smaller the diffusion angle, the higher the diffusion time and thus the lower the feeding rate to the loss cone. The critical energy $\mathcal{E}_{c}$ defined by $\displaystyle q(\mathcal{E}_{c})=1$ decreases with $M_{\bullet}$, and $m$ is shown in Figure \ref{ecrit} for $\gamma=1$. The $\mathcal{E}_{c}$ is the energy from which the majority of the loss cone flux originates \citep{1977ApJ...211..244L}. With an increase in $M_{\bullet}$, the relaxation time of the galaxy increases; thus the diffusion timescale increases \citep{1976MNRAS.176..633F} and $q$ decreases, which results in a decrease in $\mathcal{E}_{c}$. As $r_{t}(M_{\bullet},\hspace{1mm}m)\propto m^{n-1/3}$, it increases with $m$ for $n=0.8$; and $J_{lc}$ increases and so $q$ decreases, which results in a decrease of $\mathcal{E}_{c}$. As $\gamma$ increases, the diffusion timescale decreases due to an increase in the number of scatterers, and thus $q$ increases and thus, $\mathcal{E}_{c}$ increase. 

Using Equation (\ref{flr}) and the mass function of stars in the galaxy $\xi(m)$ given by Equation (\ref{smf}), the loss cone feeding rate is given by \citep{2013CQGra..30x4005M}

\begin{equation}
\frac{\diff^2 \dot{N}_{t}}{\diff E \hspace{1mm} \diff y }=4\pi^2 \int \, \diff m \hspace{1mm} \xi (m) J^2_{c}(E)\left<D(E)\right>\mathcal{F}(\chi=1,\hspace{1mm}y)
\end{equation}

The corresponding feeding rate in terms of $E$ and $j$   is given in \citet{2015ApJ...804...52M}.

The Jacobian of the transformation from $\{E,\hspace{1mm}y\}$ space to dimensionless variables $\{ \mathcal{E},\hspace{1mm}\ell^2=(J/J_{lc}(\mathcal{E},r_{t}))^2=j(J_{c}(\mathcal{E})/J_{lc}(\mathcal{E},r_{t}))^2\}$  is  given by 
\begin{equation}
\displaystyle 
\diff E\hspace{1mm}\diff y= {\rm Det}\left [\frac{\partial(E,\hspace{1mm}y)}{\partial (\mathcal{E}, \hspace{1mm} \ell^2)} \right ] \diff  \mathcal{E} \hspace{1mm}\diff \ell^2, ~~~~{\rm where}~~~
\frac{\partial(E,\hspace{1mm}y)}{\partial (\mathcal{E}, \hspace{1mm} \ell^2)} =  \frac{\partial(E,\hspace{1mm}y)}{\partial (E ,\hspace{1mm} J^2)} \cdot  \frac{\partial(E,\hspace{1mm}J^2)}{\partial (\mathcal{E}, \hspace{1mm} \ell^2)}
\end{equation}
and the following result is obtained by calculating the product of the determinants of the two Jacobians in the above equation as 
\begin{equation}
\diff E\hspace{1mm}\diff y= \sigma^2\frac{J^2_{lc}(\mathcal{E},r_{t})}{\left<D(\mathcal{E})\right>J^2_{c}(\mathcal{E})}\diff \mathcal{E} \hspace{1mm}\diff \ell^2.
\end{equation}

Then, the feeding rate  is given by

\begin{equation}
\frac{\diff^2 \dot{N}_{t}}{\diff \mathcal{E} \hspace{1mm} \diff \ell^2\hspace{1mm} \diff m}=4 \pi^2\hspace{1mm}\sigma^2 \xi (m) \hspace{1mm} J^2_{lc}(\mathcal{E})\hspace{1mm} \mathcal{F}(\chi=1,\hspace{1mm}\ell)
\label{tdnth}
\end{equation}

\begin{figure}
\centering
\includegraphics[scale=0.38]{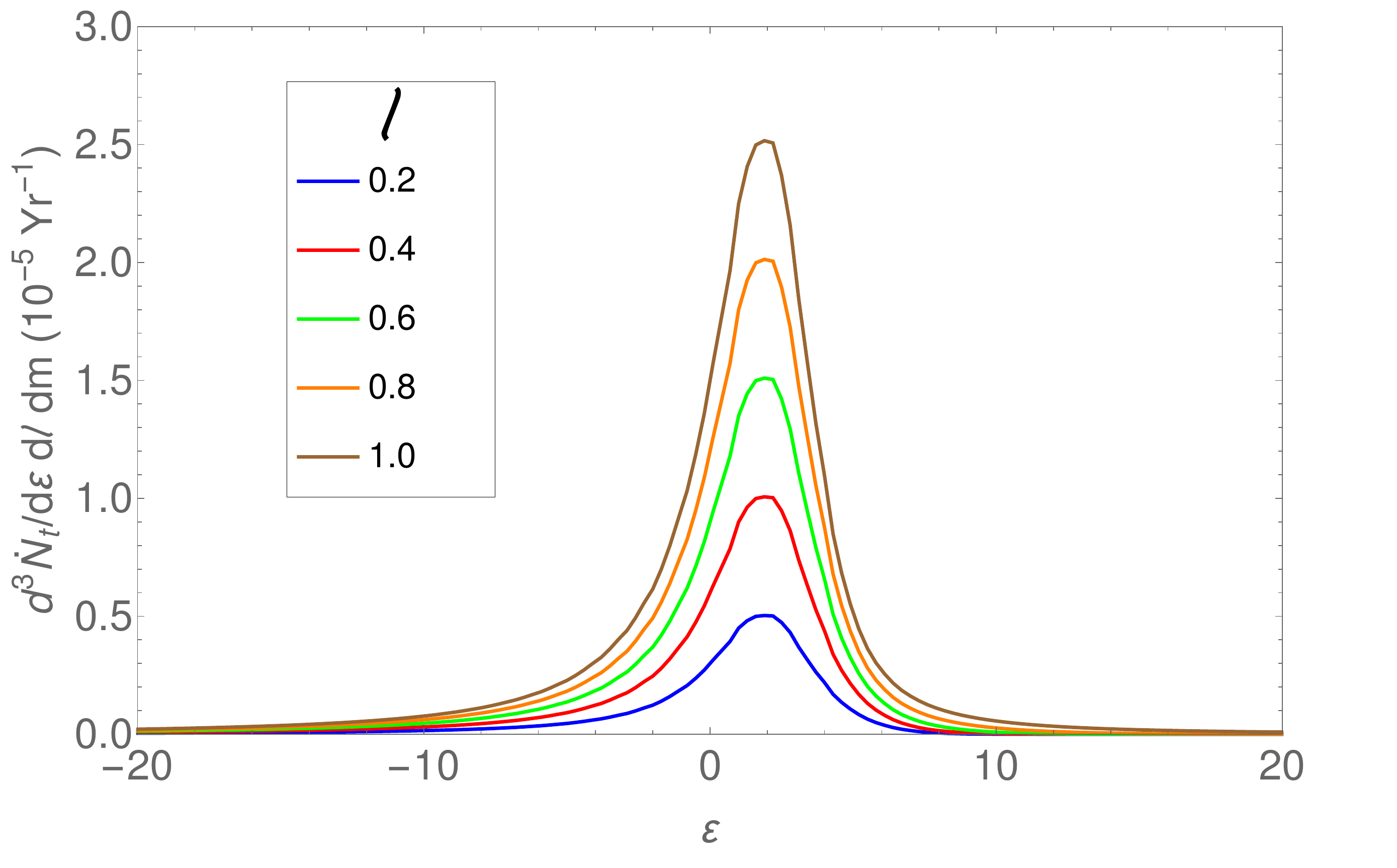}
\caption{ The theoretical capture rate $\dot{N}_{t}$ (Equation (\ref{tdnth})) is shown as a function of $\mathcal{E}$ for various $\ell$ and for $M_{6}=1$ and $\gamma=0.8$. The capture rates for high energy orbits are small and increase with $\ell$ due to an increase in $N(\mathcal{E},\hspace{1mm}\ell)$ (see Equation (\ref{nel})).}
\label{dnml}
\end{figure}

Figure \ref{dnml} shows the plot of $\dot{N}_{t}$ as a function of $\mathcal{E}$ for various $\ell$ for $M_{6}=1$ and $\gamma=0.8$. The capture rates decreases with $\mathcal{E}$ because of the decrease in diffusion coefficient $D(\mathcal{E})$ and increases with $\ell$ due to the increase in $N(\mathcal{E},\hspace{1mm}\ell)$ (see Equation (\ref{nel})).

Beacuse $J_{lc}(r_t)=\sqrt{2r^2_{t}(\Phi(r_{t})-E)}$ (see Section 2), this implies that $E<\Phi(r_t)$ and the maximum value of $E=E_{m}=\Phi(r_t)$. Because $r_{t}(M_{\bullet},\hspace{1mm}m)/r_{h}\sim 10^{-6}-10^{-5}$, and for  $r \ll r_{h}$, the potential is dominated by the BH potential as shown in the Figure \ref{pot}, which  implies that $E_{m}(M_{\bullet},\hspace{1mm}m)=\Phi(r_{t}(M_{\bullet},\hspace{1mm}m))=GM_{\bullet}/r_{t}(M_{\bullet},\hspace{1mm}m)$. The orbital motion of a star at the turning point of the orbit $r_{x}$, is given by

\begin{equation}
E=\Phi(r_{x})-\frac{J^2}{2r^2_{x}}
\label{ea}
\end{equation}

where $\Phi(r)=\displaystyle{\Phi_{\bullet}(r)+\Phi_{\star}(r)}$, $\Phi_{\bullet}(r)=GM_{\bullet}/r$ and $\Phi_{\star}(r)$ is given by Equation (\ref{poten}). In terms of dimensionless variables, $\ell=J/J_{lc}$ and $\bar{e}=E/E_{m}$, where $E_{m}=GM_{\bullet}/r_t$, such that $\bar{e}=(r_t/r_h)\mathcal{E}$, and the Equation (\ref{ea}) using Equation (\ref{psis}) is given by 

\begin{equation}
\frac{\bar{e}}{s_t}=\frac{s_x-\ell^2 s_t+\frac{2}{2-\gamma}\left[s^2_x(1-s^{2-\gamma}_x)-\ell^2 s^2_t (1-s^{2-\gamma}_t)\right]}{s^2_x-\ell^2 s^2_t}
\label{sst}
\end{equation}

where $s_x=r_x/r_h$, $s_t=r_t/r_h$. Since $\bar{e}$ is a monotonically decreasing function of $s_x$, and both the pericenter and apocenter of the orbit should lie below $r_h$, the minimum value of $\bar{e}$ is at $s_x=1$ for $r_x=r_h$; taking $\bar{e}(s_x=1)=\bar{e}_h$, the Equation (\ref{sst}) reduces to

\begin{equation}
\frac{\bar{e}_h}{s_t}=\frac{1-\ell^2 s_t-\frac{2}{2-\gamma}\ell^2 s^2_t (1-s^{2-\gamma}_t)}{1-\ell^2 s^2_t}.
\end{equation}

Now $s_t=r_{t}/r_{h} \sim 10^{-5}$--$10^{-6}$,  $\bar{e}_{h}\simeq r_{t}/r_{h}$. Beacuse $J_{lc}(r_x)=\sqrt{2r^2_{x}(\Phi(r_{x})-E)}$, this implies that $E<\Phi(r_x)$ and the maximum value of $E=\Phi(r_x)\simeq \Phi(r_t)$, which corresponds to $\bar{e}\simeq 1$ and thus  $\bar{e}\lesssim 1$.  The total potential is dominated by the BH near $r_t$, as shown in the Figure \ref{pot}. After ignoring the second term in the numerator in the righthand side of Equation (\ref{sst}), which is a factor $s_t=r_{t}/r_{h} \simeq 10^{-5}-10^{-6}\ll 1$ lower than the first,  the Equation (\ref{sst}) reduces to 

\begin{equation}
\bar{e}=\frac{x_x-\ell^2}{x_x^2-\ell^2},
\label{peri}
\end{equation}

where $x_x=s_x/s_t$. If $x_p$ is lower of the two roots of $x_x$, it is given by

\begin{equation}
x_p=\frac{1}{2\bar{e}}\left(1-\sqrt{1-4\bar{e}(1-\bar{e})\ell^2}\right).
\label{pper}
\end{equation}

For a star to be tidally disrupted, $x_p<1$ which results in 

\begin{equation}
(1-\bar{e})(1-\ell^2)>0
\label{eqal}
\end{equation}

and $x_p>0$ results in

\begin{equation}
\ell^2\bar{e}(1-\bar{e})>0.
\label{eqal1}
\end{equation}

The Equation (\ref{eqal1}) restricts the range of $\bar{e}$ to $\bar{e}<1$ and thus Equation (\ref{eqal}) implies that $\ell<1$. Thus, the applicable ranges are $\bar{e}_h<\bar{e}<1$ and $0<\ell<1$. We derived the turning points $s_x$ by solving Equation (\ref{sst}) subject to the constraint $r_p < r_t$ and $r_t< r_a<r_h$, and verified the range for $\bar{e}$ and $\ell$ derived above.

While we have ignored relativistic effects in the analysis above, we plan to include them in the future.

The lifetime of a main sequence star is $t_{MS}=t_{\odot}m^{-2.5}$ where $t_{\odot}=10^{10}~{\rm yr}$ is the lifetime of a solar type star and the radial period is given by

\begin{equation}
T_r=\int^{r_a}_{r_p} \frac{1}{v_r} \diff r.
\end{equation}

In terms of $s=r/r_h$ and using Equation (\ref{psis}), $T_r$ is given by

\begin{equation}
T_r(\bar{e},\hspace{1mm}M_{\bullet},\hspace{1mm}m)=\frac{r_h}{\sqrt{2}\sigma}\int^{s_a}_{s_p} \frac{s\sqrt{s_t}}{\sqrt{s_t \left(s-\ell^2 s_t+\frac{2}{2-\gamma}\left[s^2(1-s^{2-\gamma})-\ell^2 s^2_t (1-s^{2-\gamma}_t)\right]\right)-\bar{e}(s^2-\ell^2 s^2_t)}} \, \diff s,
\label{redv}
\end{equation}

where $s_a$ and $s_p$ are the dimensionless apocenter and pericenter that are obtained by solving Equation (\ref{sst}). We find numerically that the radial period $T_r$ is approximated by

\begin{equation}
T_r(\bar{e},\hspace{1mm}M_{\bullet},\hspace{1mm}m) \simeq \frac{\pi}{2\sqrt{2}}\frac{r_h}{\sigma}\left\{
\begin{array}{ll}
0.57 e^{[0.27(1.47-\frac{\bar{e}}{s_t})]}& \bar{e}<1.47s_t \\
\left(\frac{\bar{e}}{s_t}\right)^{-\frac{3}{2}}& \bar{e} \geq 1.47s_t 
\end{array}
\right. 
\label{radv}
\end{equation}

Because the BH potential dominates at high energy, the corresponding orbits are Keplerian and the radial period $\propto \bar{e}^{(-3/2)}$. Using the $M_{\bullet}-\sigma$ relation given by Equation (\ref{msigma}) and $r_h=GM_{\bullet}/\sigma^2$, the radial period in Keplerian regime is $T_r\propto M^{0.38}_{\bullet}\mathcal{E}^{-3/2}$, where $\mathcal{E}=\bar{e}/s_t$. The stars on the loss cone orbits are captured in the radial period timescale and the number of stars in the loss cone orbit is $N_{lc}(\mathcal{E})\diff \mathcal{E}$ \citep{2013CQGra..30x4005M}. Thus the capture rate is approximately given by $\dot{N}_t=\int (N_{lc}(\mathcal{E})/T_r(\mathcal{E}))\, \diff \mathcal{E} \propto M^{-0.38}_{\bullet}$ in the regime $\bar{e}>1.47s_t$, which is consistent with the average best-fit slope of $-0.3$ over the entire range of $\bar{e}$ that was found numerically. 

A tidally captured star is on main sequence if its main sequence lifetime is $t_{MS}>T_r$, where $T_r$ is the radial period of the orbit. Considering all possible radial phases of the star in an orbit, the probability that a star of mass $m$ is tidally captured as a main sequence is given by 

\begin{equation}
f_{\star}(\bar{e},\hspace{1mm}M_{\bullet},\hspace{1mm} m)={\rm Min}\left[1,\hspace{1mm} \frac{t_{MS}(m)}{T_r(\bar{e},\hspace{1mm}M_{\bullet},\hspace{1mm}m)}\right].
\label{fstar}
\end{equation}

Using Equations (\ref{flr}), (\ref{tdnth}), and $\diff \mathcal{E}\hspace{0.5mm}\diff \ell^2 =s^{-1}_t\hspace{0.5mm}\diff \bar{e}\hspace{0.5mm} \diff \ell^2$, the capture rate is given by

\begin{equation}
\frac{\diff^2 \dot{N}_{t}}{\diff \bar{e} \hspace{1mm} \diff \ell^2 \hspace{1mm} \diff m}=4 \pi^2\hspace{2mm}s^{-1}_t\sigma^2 \hspace{2mm} \xi (m) f_{\star}(\bar{e},\hspace{1mm}M_{\bullet},\hspace{1mm} m) \hspace{1mm} J^2_{lc}(\bar{e})\hspace{2mm} \mathcal{F}(\chi=1,\hspace{1mm}\ell).
\label{tdnth1}
\end{equation}

Figure 5 of \citet{2002A&A...394..345F}  gives the maximum mass of BH as function of mass $m$ of the star that is disrupted. We observe that stars with mass $m>0.8$ are tidally disrupted in the entire range of BH mass $10^6M_{\odot}<M_{\bullet}<10^8M_{\odot}$ and for $m < 0.8$, a substantial fraction is tidally captured without disruption. Thus we take the effective star mass range to be $0.8<m<150$. The net capture rate is given by

\begin{equation}
\dot{N}_{t}(\gamma,~M_{\bullet})=4 \pi^2\hspace{2mm} \int^{150}_{0.8} \diff m \int^{1}_{\bar{e}_{h}} \diff \bar{e} \int^{1}_{0} \diff \ell^2  \hspace{2mm} \sigma^2 s^{-1}_t(M_{\bullet},~ m) \xi (m) f_{\star}(\bar{e},\hspace{1mm}M_{\bullet},\hspace{1mm} m) \hspace{1mm} J^2_{lc}(\bar{e})\hspace{2mm} \mathcal{F}(\chi=1,\hspace{1mm}\ell).
\label{nnet}
\end{equation}  

\begin{figure}
\begin{subfigure}{\textwidth}
\centering
\includegraphics[scale=0.49]{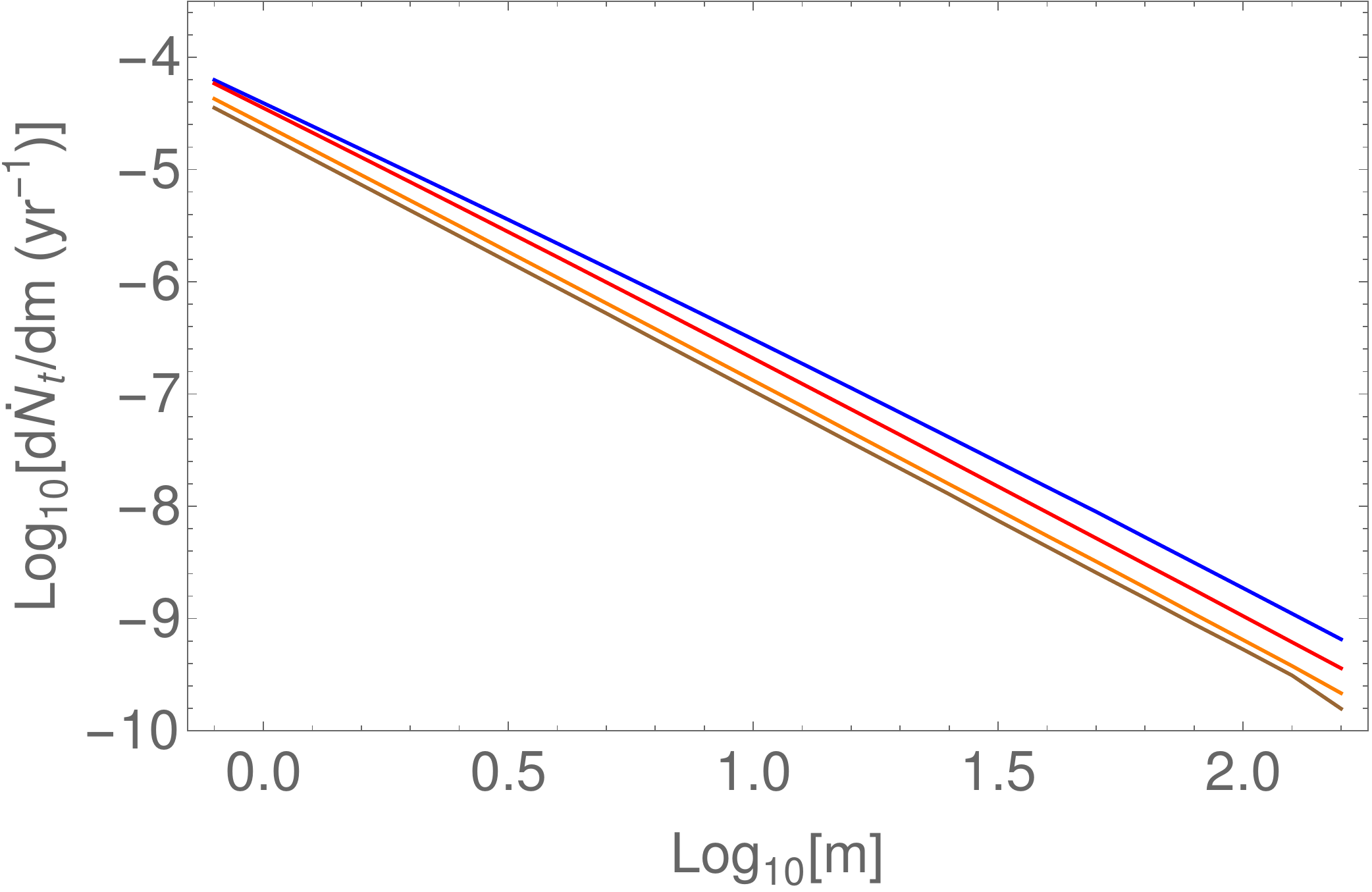}
\caption{}
\end{subfigure}
\begin{subfigure}{\textwidth}
\centering
\includegraphics[scale=0.36]{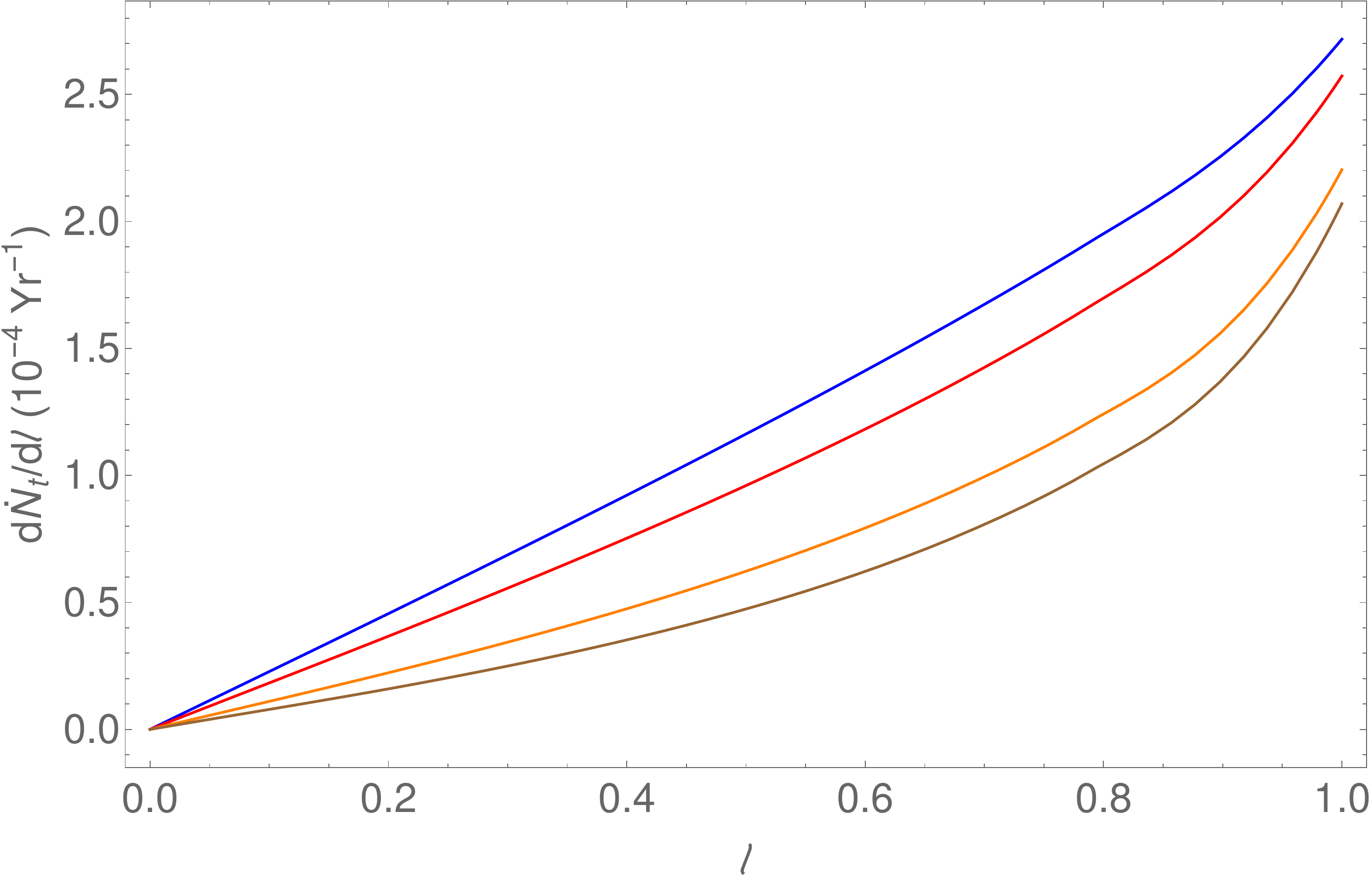}
\caption{}
\end{subfigure}
\caption{ For $\gamma=1.0$ and $M_6$= 1 (blue), 10 (red), 50 (orange), and 100 (brown). Figure (a) shows $\diff \dot{N}_{t}/\diff m$ obtained using Equation (\ref{flr}) and integrating Equation (\ref{tdnth1}) over $\bar{e}$ and $\ell$ decreases with $m$ as $\xi(m)$ decreases with $m$. Figure (b) shows $\diff \dot{N}_{t}/\diff \ell$ obtained using Equation (\ref{flr}) and integrating Equation (\ref{tdnth1}) over $\bar{e}$ and $\ell$. }
\label{dnl}
\end{figure}

We solved Equation (\ref{flr}) to obtain $\mathcal{F}(1,~\ell)$ and used it in Equation (\ref{tdnth1}) to calculate the capture rate. The integration of Equation (\ref{tdnth1}) over the energy range $\bar{e}_h<\bar{e}<1$ and angular momentum range $0<\ell<1$ results in a capture rate per unit mass $\diff \dot{N}_{t}/\diff m$, which is a decreasing function of $m$ as $\xi(m)$ decreases with $m$ and is shown in Figure~\ref{dnl}a for various $M_{\bullet}=10^6 M_{\odot}M_6$ and $\gamma=1.0$. Similarly, integrating Equation (\ref{tdnth1}) over energy $\bar{e}_h<\bar{e}<1$ and  test star mass $0.8<m<150$ results in $\diff \dot{N}_{t}/\diff \ell$, which is an increasing function $\ell$ as shown in Figure~\ref{dnl}b.

\begin{figure}
\begin{subfigure}{\textwidth}
\centering
\includegraphics[scale=0.41]{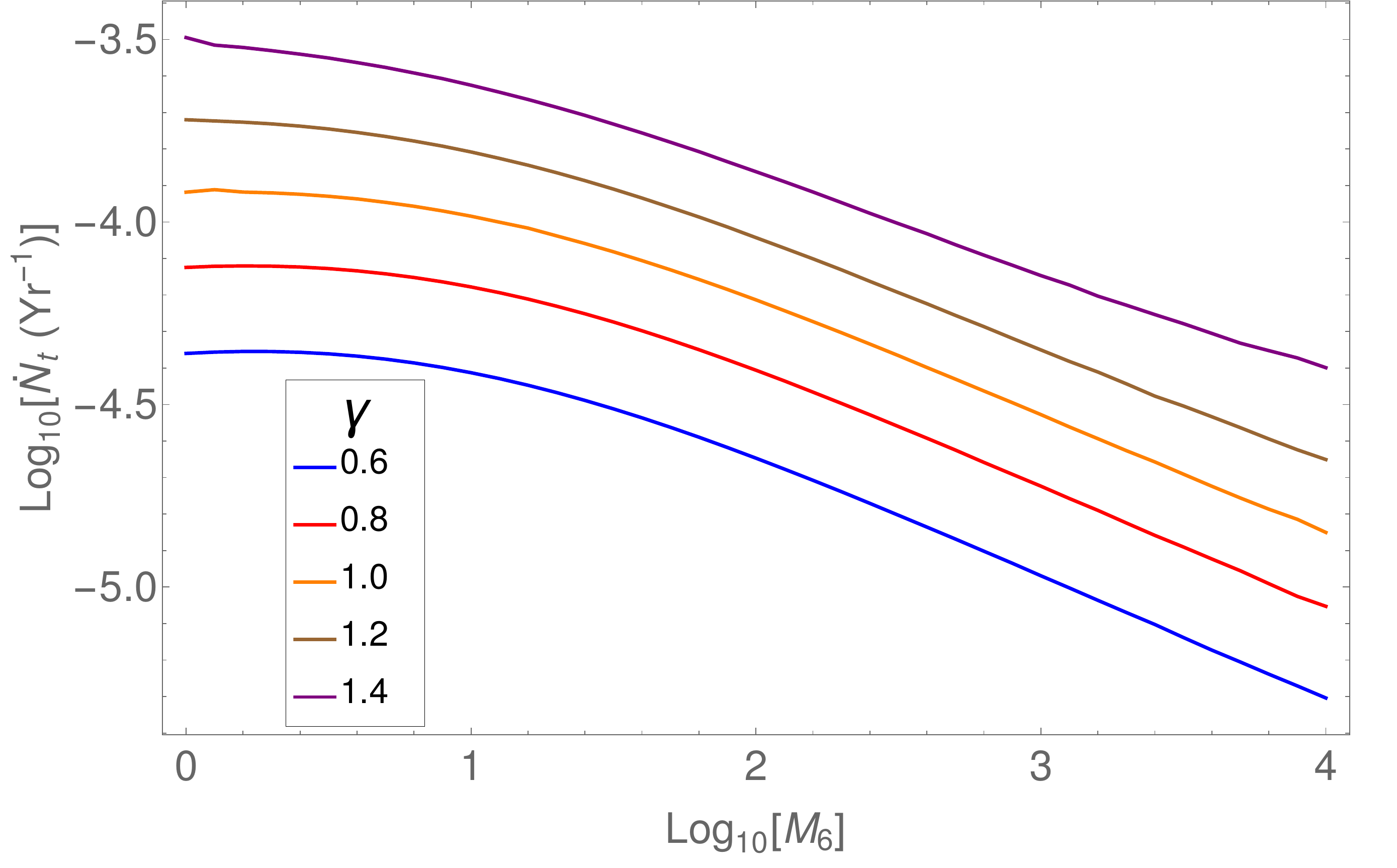}
\caption{}
\end{subfigure}
\begin{subfigure}{\textwidth}
\centering
\includegraphics[scale=0.55]{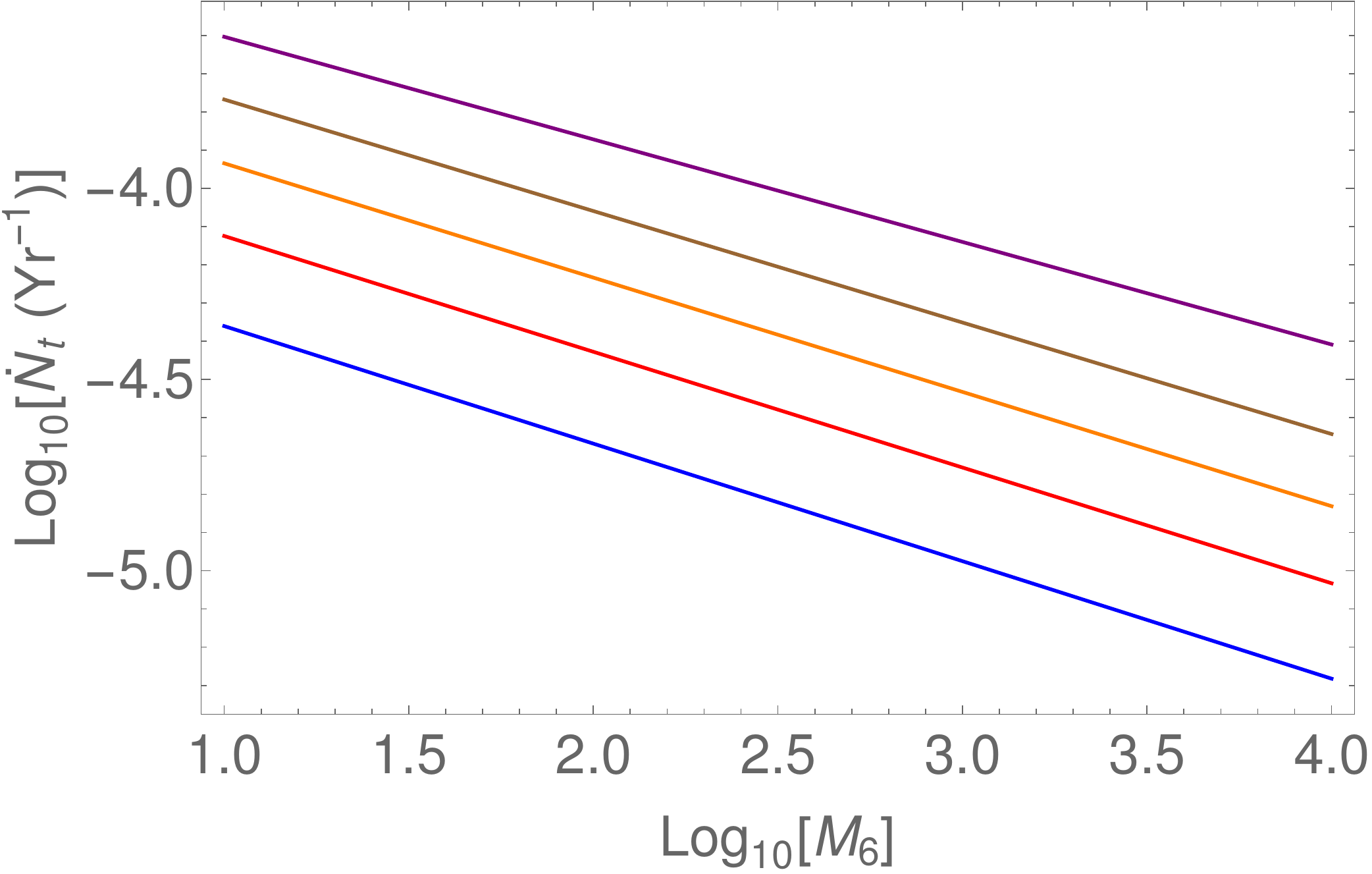}
\caption{}
\end{subfigure}
\caption{ Panel (a) shows the net $\dot{N}_{t}$ obtained using Equations (\ref{flr}) and (\ref{nnet}) as a function of $M_{6}$ for various $\gamma$. Panel (b) shows $\dot{N}_{t}$ as a function of $M_{6}$ for $M_{6}>10$ and $\gamma$=0.6 (blue), 0.8 (red), 1.0 (orange), 1.2 (brown), and 1.4 (purple), and it follows that $\dot{N}_{t}\propto M^{-\beta}_{6}$ where $\beta=0.3\pm0.01$. }
\label{nthlc}
\end{figure}

\begin{figure}
\centering
\includegraphics[scale=0.57]{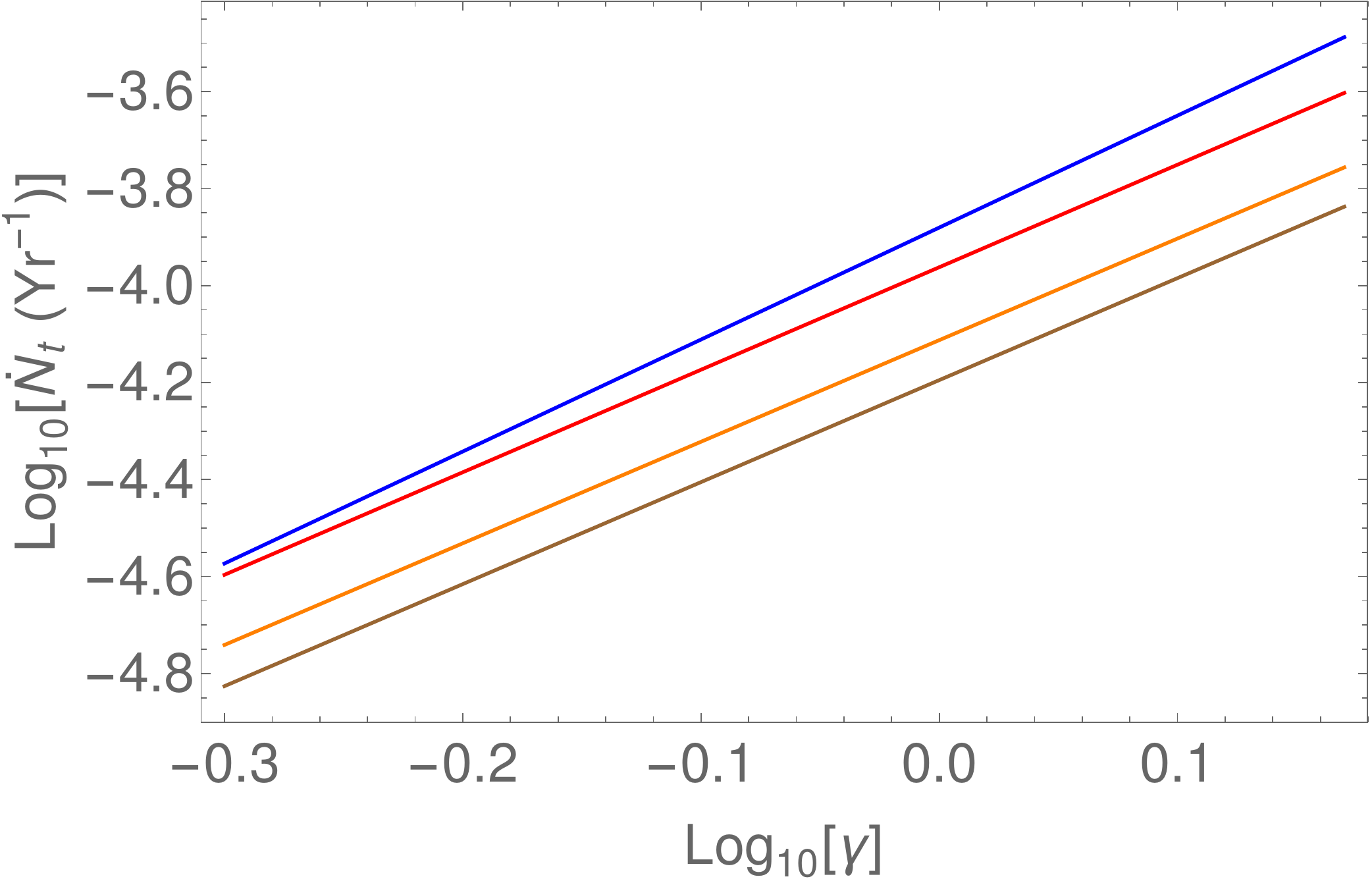}
\caption{ The figure shows the $\dot{N}_{t}$ obtained using Equations (\ref{flr}) and (\ref{nnet}) as a function of $\gamma$ for various $M_{6}$=1 (blue), 10 (red), 50 (orange), and 100 (brown), and it follows  $\dot{N}_{t}\propto \gamma^{p}$ with the best-fit value of $p\sim 2.1$. The $\dot{N}_{t}$ increases with $\gamma$ due to an increase in the density of the central stellar population.}
\label{ntg}
\end{figure}

The net $\dot{N}_{t}$ obtained using Equations (\ref{flr}) and (\ref{nnet}) increases with $\gamma$ and decreases with $M_{6}$ as shown in Figure \ref{nthlc}. For $M_{6}\geq 10$, the $\dot{N}_{t}\propto M^{\beta}_{6}$ where $\beta=0.3\pm 0.01$, as shown in the Figure \ref{nthlc}b for various $\gamma$. The increase with $\gamma$ is nonlinear as shown in Figure \ref{ntg} and for $\dot{N}_{t}\propto \gamma^{p}$, the best-fit value of $p\sim 2.1$. The galaxies with larger $\gamma$ posses higher rates because their denser central stellar populations naturally feature shorter relaxation times and faster rates of energy and angular momentum diffusion.

\section{Physics of tidal disruption}
\label{ptd}
The classical description of a TDE was outlined by \citet{1988Natur.333..523R}. In this picture, a star on parabolic orbit is tidally captured and disrupted at the pericenter and the distribution of mass of disrupted debris with respect to specific binding energy $\diff M/\diff E_{d}$ is roughly flat, where $E_{d}$ is the energy of the disrupted debris. For stars on a parabolic orbit, \citet{2009MNRAS.392..332L}  found that $\diff M/\diff E_{d}$ depends on the properties of the star and adiabatic index $\Gamma$.  Using Equation (\ref{pper}), the pericenter is given by

\begin{equation}
r_{p}(\bar{e},\hspace{1mm}\ell,\hspace{1mm}M_{\bullet},\hspace{1mm}m)=\frac{r_t}{2\bar{e}}\left(1-\sqrt{1-4\bar{e}(1-\bar{e})\ell^2}\right)=r_{t}\left(\frac{2\ell^2 (1-\bar{e})}{1+\sqrt{1-4\ell^2 \bar{e}(1-\bar{e})}}\right).
\end{equation}

Equivalently

\begin{equation}
r_p(E,\hspace{1mm}J,\hspace{1mm}M_{\bullet})=\frac{GM_{\bullet}}{2E}\left[1-\sqrt{1-\frac{2E J^2}{G^2 M^2_{\bullet}}}\right]=\frac{J^2}{G M_{\bullet}}\left[1+\sqrt{1-\frac{2E J^2}{G^2 M^2_{\bullet}}}\right]^{-1}.
\end{equation}

The stars on the loss cone orbits are captured within the dynamical time $t_{d}=(r_p^3/GM_{\bullet})^{0.5}$ with tidal acceleration $a_{t}=GM_{\bullet}\Delta R/r_p^3$ where $\Delta R$ is the debris distance from the star center at the moment of breakup. Then, the energy of the disrupted debris is given by

\begin{equation}
E_d(\bar{e},\hspace{1mm}\ell,\hspace{1mm}M_{\bullet},\hspace{1mm}m,\hspace{1mm}\Delta R)\approx \bar{e}E_{\boldsymbol{m}}(M_{\bullet},\hspace{1mm}m)-\frac{2kGM_{\bullet}\Delta R}{r^2_p(\bar{e},\hspace{1mm}\ell,\hspace{1mm}M_{\bullet},\hspace{1mm}m)} 
\label{ed}
\end{equation}

where $\Delta R \in \{-R_{\star},\hspace{1mm}R_{\star}\}$, the negative sign corresponds to the region toward the BH and $k$ is the spin-up factor due to tidal torque by the SMBH on a star given by \citep{2001ApJ...549..948A}

\begin{equation}
k = \left\{
\begin{array}{ll}
1& {\rm non\hspace{1mm} spin\hspace{1mm} up\hspace{1mm} (no\hspace{1mm} change\hspace{1mm} in\hspace{1mm} angular\hspace{1mm} velocity)}\nonumber\\
3& {\rm spin\hspace{1mm} up\hspace{1mm} to\hspace{1mm} break\hspace{1mm} up\hspace{1mm} angular\hspace{1mm} velocity} \nonumber 
\end{array}
\right. 
\end{equation}

The maximum distance from the center of a star to the point where the debris is bound to the BH at the moment of disruption is obtained by setting $E_d=0$ in Equation (\ref{ed}) and is given by

\begin{equation}
R_{l}(\bar{e},\hspace{1mm}\ell,\hspace{1mm}M_{\bullet},\hspace{1mm}m)=\frac{r^2_p(\bar{e},\hspace{1mm}\ell,\hspace{1mm}M_{\bullet},\hspace{1mm}m)\bar{e}}{2kr_{t}(M_{\bullet},\hspace{1mm}m)}.
\label{rlim}
\end{equation}

\begin{figure}
\begin{center}
\includegraphics[scale=0.8]{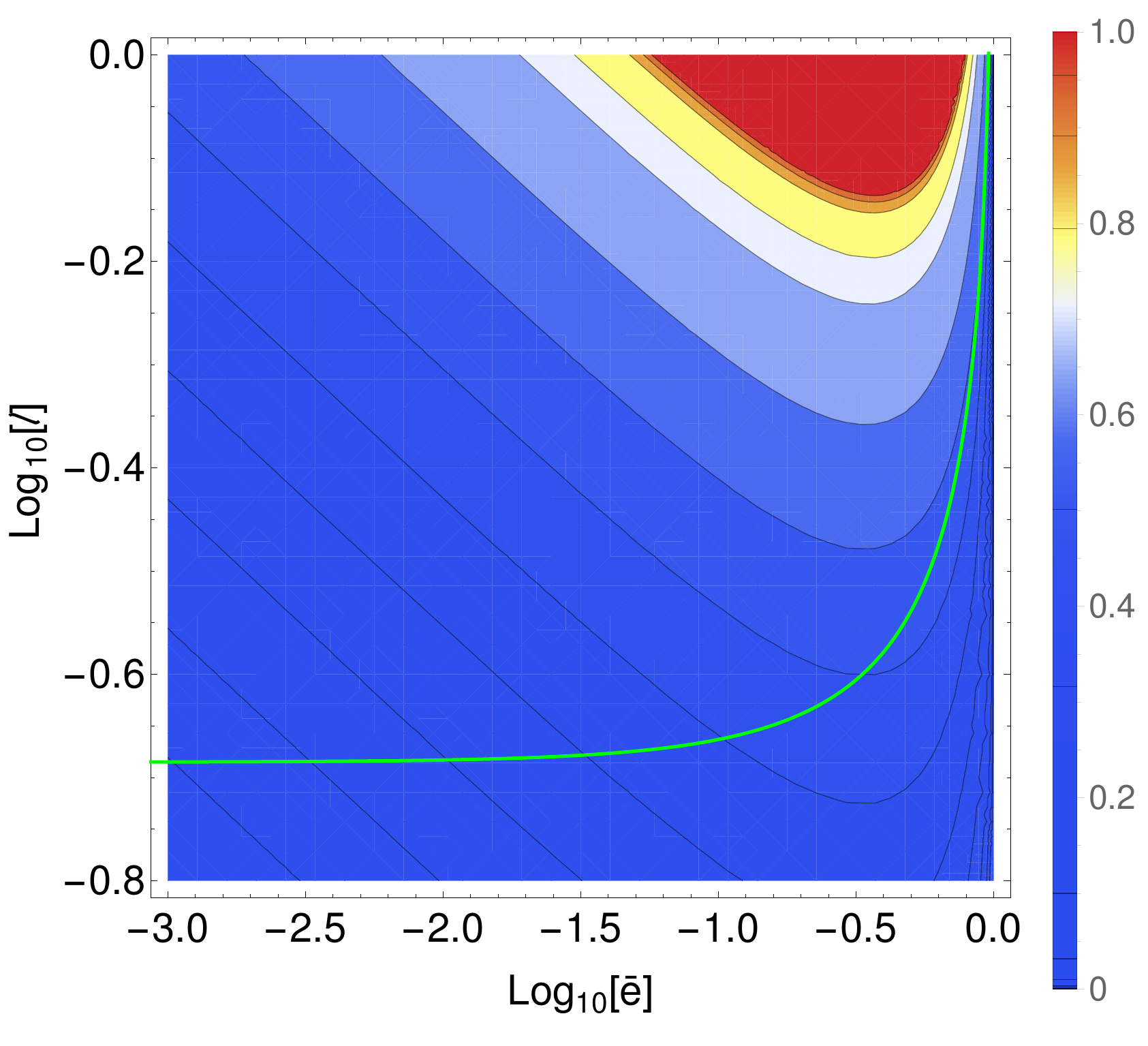}
\caption{A contour plot of the maximum distance from the star center $x_{l}(\bar{e},\hspace{1mm}\ell,M_{\bullet},\hspace{1mm}m)=R_l/R_{\star}$ (Equation (\ref{rlim})) to the point where the debris is bound to the black hole for $M_{6}=1$ and $m=1$. The green line corresponds to $r_{p}=R_{s}$ and for $r_{p}>R_{s}$, the contours lie above the green line. }   
\label{rplim}
\end{center}
\end{figure}

Figure \ref{rplim} shows the contour plot of $x_{l}\equiv x_{l}(\bar{e},\hspace{1mm}\ell,M_{\bullet},\hspace{1mm}m)=R_{l}(\bar{e},\hspace{1mm}\ell,M_{\bullet},\hspace{1mm}m)/R_{\star}$ for $M_{6}=1$ and $m=1$. The value of $r_{p}(\bar{e},\hspace{1mm}\ell,\hspace{1mm}M_{\bullet},\hspace{1mm}m)$ is less than Schwarzschild radius $R_{s}(M_{\bullet})$ for $\ell \leq 0.2$; whereas $x_{l}(\bar{e},\hspace{1mm}\ell,M_{\bullet},\hspace{1mm}m)$ increases with $\bar{e}$ and the increase with $\ell$ is significant for high energy orbits. With the increase in the value of $x_{l}(\bar{e},\hspace{1mm}\ell,M_{\bullet},\hspace{1mm}m)$, the mass of the debris bound to the BH increases and for $x_{l}(\bar{e},\hspace{1mm}\ell,M_{\bullet},\hspace{1mm}m)=1$, the entire debris is bound to the BH.

The time taken for the most tightly bound debris to return its pericenter after disruption is given by 

\begin{equation}
t_{m}(\bar{e},\hspace{1mm}\ell,\hspace{1mm}M_{\bullet},\hspace{1mm}m)=\frac{2\pi GM_{\bullet}}{[2E_d(\bar{e},\hspace{1mm}\ell,\hspace{1mm}M_{\bullet},\hspace{1mm}m,\hspace{1mm}-R_{\star})]^{3/2}}.
\end{equation}

\begin{figure}
\begin{center}
\includegraphics[scale=0.6]{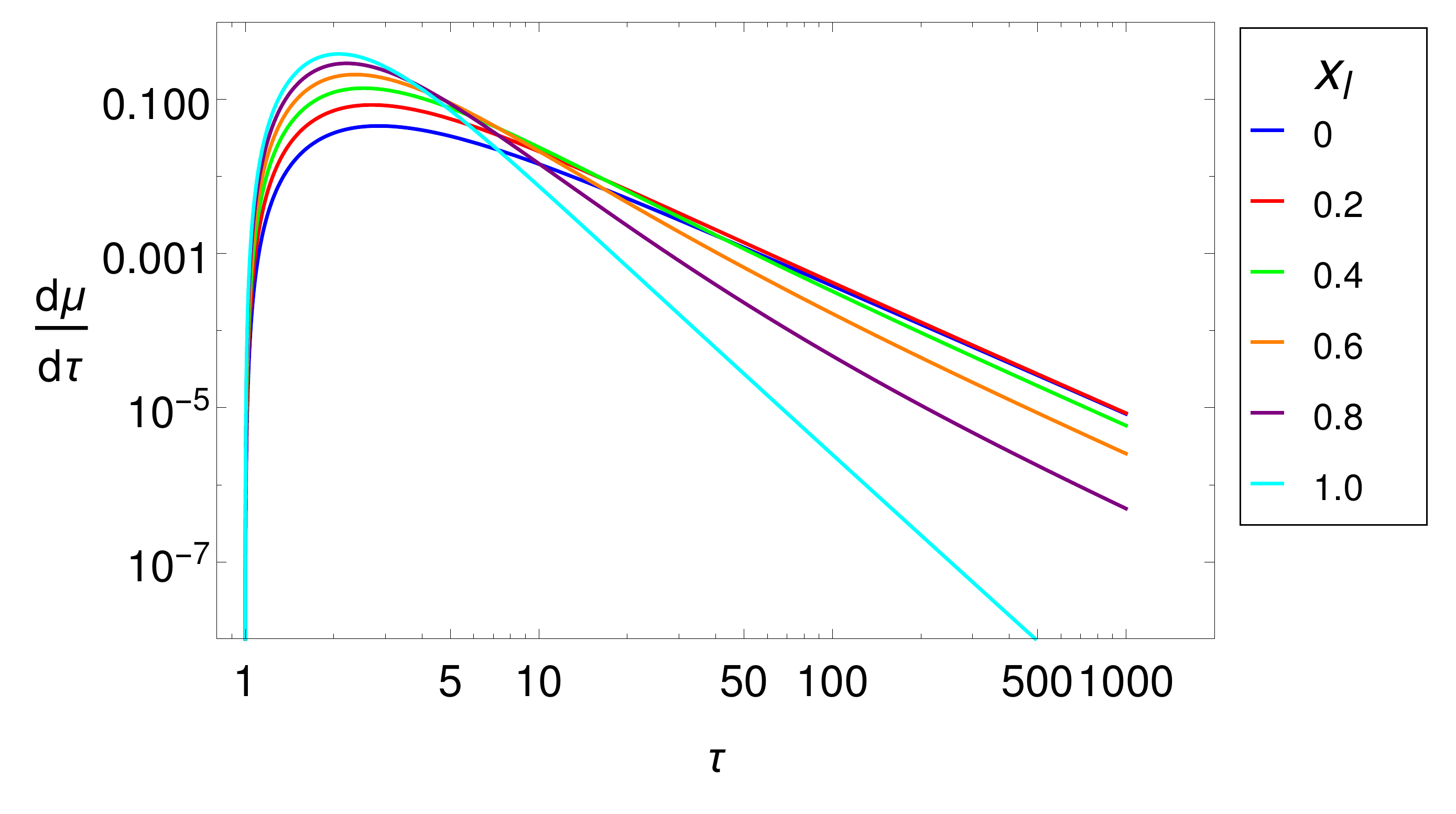}
\caption{The dimensionless mass accretion rate given by Equation (\ref{lod}), as a function of dimensionless time $\tau$ for various $x_{l}$. The peak accretion rate increases with $x_{l}$ whereas the time for peak accretion decreases with $x_{l}$.}   
\label{masst}
\end{center}
\end{figure}

As the bound material falls back to its pericenter, it loses its energy and angular momentum, thus accreting into the SMBH and giving rise to the flare \citep{1989IAUS..136..543P}. The in-fall mass accretion rate at time $t$ after disruption for the debris following Keplerian orbits is given by

\begin{equation}
\frac{\diff M}{\diff t}=\frac{\diff M}{\diff E_d} \frac{\diff E_d}{\diff a}\frac{\diff a}{\diff t}= \frac{1}{3} (2 \pi G M_{\bullet})^\frac{2}{3}\frac{\diff M}{\diff E_d} t^{-\frac{5}{3}}
\label{acc}
\end{equation}

where $a$ is the semimajor axis of the debris with orbital energy $E_d(\bar{e},\hspace{1mm}\ell,\hspace{1mm}M_{\bullet},\hspace{1mm}m,\hspace{1mm}\Delta R)$. The term $\displaystyle{\diff M/ \diff E_d}$ is the energy distribution within the bound matter and depends on the internal structure of the star \citep{1989IAUS..136..543P,2009MNRAS.392..332L}. We now write it in terms of dimensionless quantities using Equations (\ref{ed}) and (\ref{rlim}), and modify the dimensionless quantities given in \citet{2009MNRAS.392..332L} by including the dependence on $r_p(\bar{e},\hspace{1mm}\ell,\hspace{1mm}M_{\bullet},\hspace{1mm}m)$ through $R_{l}(\bar{e},\hspace{1mm}\ell,\hspace{1mm}M_{\bullet},\hspace{1mm}m)$ and express

\begin{equation}
\varepsilon =\frac{x_{l}-x}{x_{l}+1},  \hspace{10mm}   x =x_{l}-\tau^{-2/3}(1+x_{l})
\end{equation}

\begin{equation}
\frac{\diff \mu}{\diff \varepsilon} = (1+x_{l}) \frac{\diff \mu}{\diff x},\hspace{10mm} \frac{\diff \mu}{\diff \tau}=\frac{2}{3} \frac{\diff \mu}{\diff \varepsilon} \tau^\frac{-5}{3}
\label{lod}
\end{equation}

where $\displaystyle{\varepsilon(\bar{e},\hspace{1mm}\ell,\hspace{1mm}M_{\bullet},\hspace{1mm}m)= E_d(\bar{e},\hspace{1mm}\ell,\hspace{1mm}M_{\bullet},\hspace{1mm}m,\hspace{1mm}\Delta R)/ E_{d}(\bar{e},\hspace{1mm}\ell,\hspace{1mm} M_{\bullet},\hspace{1mm}m,\hspace{1mm}-R_{\star})}$, $x =\Delta R/R_{\star}$, $\displaystyle{\mu=M/M_{\star}}$ and $\tau(\bar{e},\hspace{1mm}\ell,\hspace{1mm}M_{\bullet},\hspace{1mm}m) \equiv t/t_{m}(\bar{e},\hspace{1mm}\ell,\hspace{1mm}M_{\bullet},\hspace{1mm}m)$ where $E_{dm}=E_{d}(\bar{e},\hspace{1mm}\ell,\hspace{1mm}M_{\bullet},\hspace{1mm}m,\hspace{1mm}-R_{\star})$ is the energy of inner-most tightly bound debris. The term $\diff \mu$

\begin{equation}
\frac{\diff \mu}{\diff x} = \frac{3}{2} b \int_{x}^1 \theta^u (x') x' \,\diff x'
\end{equation}

where $b$ is the ratio of central density $\rho_c$ to mean density $\overline{\rho_{\star}}=3M_{\star}/4\pi R_{\star}^3$ and $\theta$ is the solution of Lane--Emden equation for the given polytrope $u$ related to the density by $\rho=\rho_c \theta^{u}$. The total mass accretion rate is given by $ \dot{M}(\bar{e},\hspace{1mm}\ell,\hspace{1mm} M_{\bullet},\hspace{1mm}m,\hspace{1mm}t)=(M_{\star}/t_m)(d \mu/d \tau)$ which depends on the orbital parameters through $x_l$ and $t_m$. We simulated the mass fallback rate for $u=1.5$, which corresponds to $\Gamma=5/3$. Figure \ref{masst} shows the plot of $\diff \mu/\diff \tau$ for various values of $x_{l}$. With an increase in $\bar{e}$, the $x_{l}$ increases and the orbital period of the debris decreases, which implies that the mass in-fall rate increases. Thus, the peak accretion rate increases with $x_{l}$. Next, we examine the
conditions for formation of an accretion disk.

\section{Formation of an accretion disk}
\label{fad}

The debris of the disrupted star follows a Keplerian orbit around the BH. This debris experiences stream--stream collision either due to an incoming stream that intersects with the outflowing stream at the pericenter \citep{1994ApJ...422..508K} or due to relativistic precession at the pericenter \citep{2013MNRAS.434..909H}. This stream--stream collision results in a shock breakout which circularizes the debris and forms an accretion disk \citep{2009ApJ...697L..77R}. 

Even though the circularization timescale (time required for the debris to circularize to form an accretion disk) is not accurately known, it is roughly given by $t_{c}\approx n_{\rm orb}t_{m}$, where $n_{\rm orb}$ is the minimum number of  orbits required for circularization \citep{1999ApJ...514..180U}. As the debris falls toward the pericenter, it is accreted with an accretion rate given by Equation (\ref{acc}).  The formation of an accretion disk depends on $t_{c}$ and the accretion timescale $t_{a}$, which we define as the time required to consume 99\% (at the 3$\sigma$ level) of bound debris. If this timescale is less than $t_{c}$, the matter is accreted before the disk is formed. We approximated $\diff \mu/\diff x$ for convenience with a Gaussian function because it depends on the solution of Lane--Emden equation, which is symmetric about the center of the star and is given by $\diff \mu/\diff x \simeq 1.192 e^{-4.321x^2}$. The total mass consumed in dimensionless time $\tau_{a}=t_{a}/t_{m}$ is given by

\begin{equation}
\Delta \mu=\int^{\tau_{a}}_{1} \frac{\diff \mu}{\diff \tau} \, \diff \tau
\end{equation}

where $\diff \mu/\diff \tau$ is given by Equation (\ref{lod}). If $f_{r}$ is the fraction of debris bound to the BH, then in time $\tau_{a}$, the mass accreted by the BH is $\sim$ 0.99$f_{r}$. Then, the accretion timescale $t_{a}$ is given by

\begin{equation}
t_{a}(\bar{e},\hspace{1mm}\ell,\hspace{1mm} M_{\bullet},\hspace{1mm}m)= t_{m}\left(\frac{1+x_{l}}{x_{l}+\frac{1}{2.0787}{\rm Erf}^{-1}[0.997-1.962f_r]}\right)^{\frac{3}{2}}
\label{acctime}
\end{equation} 

where

\begin{equation}
f_{r}\equiv f_{r}(\bar{e},\hspace{1mm}\ell,\hspace{1mm} M_{\bullet},\hspace{1mm}m)=\int^{x_{l}}_{-1} \frac{\diff \mu}{\diff x} \, \diff x 
\end{equation}

The orbital period of debris will vary due to energy. The energy gradient between the bound debris will fill out a ring and the initial spatial distance between the bound debris will determine the ring formation timescale $t_{r}$. Let the dispersion in the energy around the initial energy $E$ be $\Delta E$. A ring is formed in the timescale $t_{r}=2\pi /\Delta \Omega$, where $\Delta \Omega$ is the dispersion in the orbital frequency \citep{2001ragt.meet...15H}. The orbital frequency $\Omega=(2E)^{3/2}/(GM_{\bullet})$ and dispersion $\Delta \Omega \propto E^{1/2}\Delta E$ and $\Delta E$ are given by 

\begin{equation}
\Delta E(\bar{e},\hspace{1mm}\ell,\hspace{1mm}M_{\bullet},\hspace{1mm}m)= \frac{2kGM_{\bullet}}{r^2_{p}(\bar{e},\hspace{1mm}\ell,\hspace{1mm}M_{\bullet},\hspace{1mm}m)}({\rm Min[}R_{l}(\bar{e},\hspace{1mm}\ell,\hspace{1mm}M_{\bullet},\hspace{1mm}m),\hspace{1mm} R_{\star} {\rm ]}+R_{\star})
\end{equation}

Then, $t_{r}$ is given by

\begin{equation}
t_{r}(\bar{e},\hspace{1mm}\ell,\hspace{1mm}M_{\bullet},\hspace{1mm}m)= \frac{\pi}{3\sqrt{2}k}\frac{r^2_p(\bar{e},\hspace{1mm}\ell,\hspace{1mm}M_{\bullet},\hspace{1mm}m)}{\bar{e}^{1/2}E^{1/2}_{\boldsymbol{m}}(M_{\bullet},\hspace{1mm}m)}(R_{\star}+{\rm Min}[R_{l}(\bar{e},\hspace{1mm}\ell,\hspace{1mm}M_{\bullet},\hspace{1mm}m),\hspace{1mm}R_{\star}])^{-1}
\label{trring}
\end{equation}  

\begin{figure}
\begin{subfigure}{\textwidth}
\centering
\includegraphics[scale=0.35]{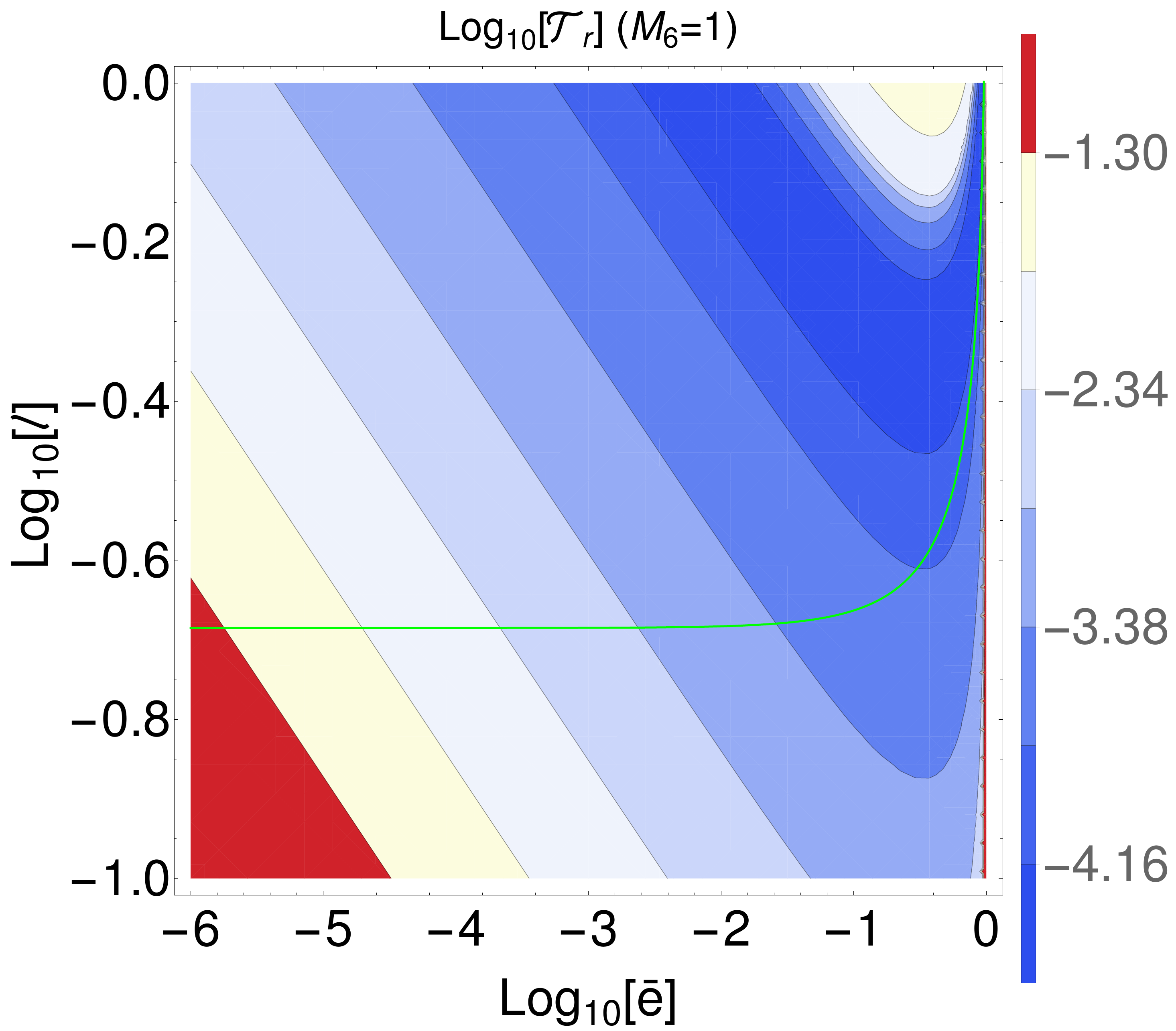}
\caption{}
\end{subfigure}
\begin{subfigure}{\textwidth}
\centering
\includegraphics[scale=0.45]{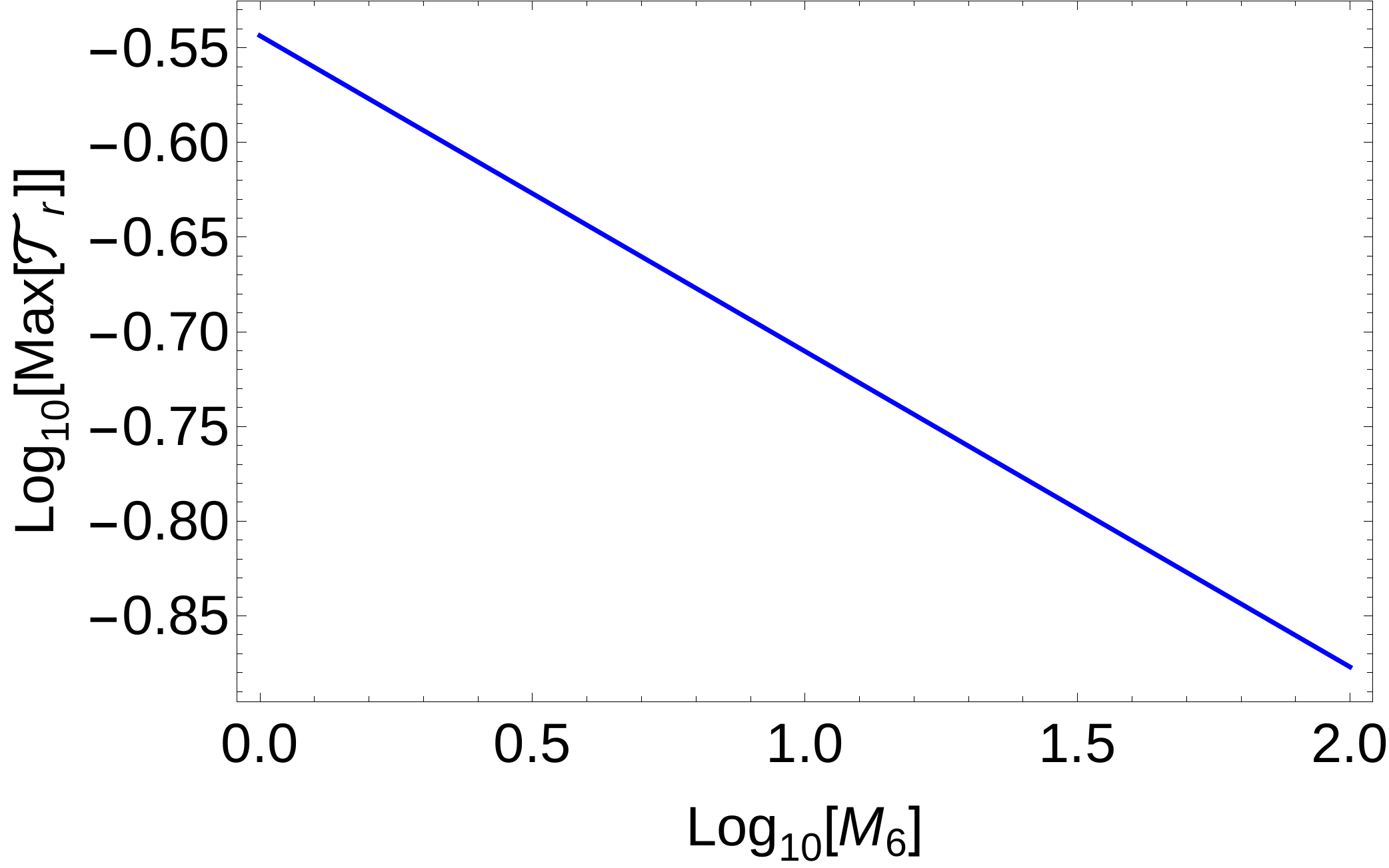}
\caption{}
\end{subfigure}
\caption{The top panel (a) shows a contour plot of the ratio $\mathcal{T}_{r}(\bar{e},\hspace{1mm}\ell,\hspace{1mm}M_{6},\hspace{1mm}m)$ (Equation (\ref{taur})) for $M_{6}=1$ and $m=1$. The green line corresponds to $r_{p}=R_{s}$. For $r_{p}>R_{S}$ which lies above green line, $\mathcal{T}_{r}(\bar{e},\hspace{1mm}\ell)<1$ and thus an accretion disk is formed. The bottom panel (b) shows the Max[$\mathcal{T}_{r}(\bar{e},\hspace{1mm}\ell)$] as a function of $M_{6}$ obtained in the range $10^{-6} \hspace{0.5mm} \leq \hspace{0.5mm} \bar{e}  \hspace{0.5mm}\leq  \hspace{0.5mm} 1$ and $0\hspace{0.5mm}\leq \hspace{0.5mm} \ell \hspace{0.5mm} \leq \hspace{0.5mm} 1$.  }
\label{diskform}
\end{figure}

The ratio $\mathcal{T}_{r}(\bar{e},\hspace{1mm}\ell,\hspace{1mm}M_{\bullet},\hspace{1mm}m)=t_{r}(\bar{e},\hspace{1mm}\ell,\hspace{1mm}M_{\bullet},\hspace{1mm}m)/t_{a}(\bar{e},\hspace{1mm}\ell,\hspace{1mm}M_{\bullet},\hspace{1mm}m)$ is given by

\begin{equation}
\mathcal{T}_{r}(\bar{e},\hspace{1mm}\ell,\hspace{1mm}M_{\bullet},\hspace{1mm}m)=\frac{1}{3\sqrt{2}}\frac{x_{l}}{1+{\rm Min[}1,\hspace{1mm}x_{l}]}\left(\frac{x_{l}+\frac{1}{2.0787}{\rm Erf}^{-1}[0.997-1.962f_r{\rm ]}}{x_{l}}\right)^{\frac{3}{2}}
\label{taur}
\end{equation}

and an accretion disk is formed if $\mathcal{T}_{r}(\bar{e},\hspace{1mm}\ell,\hspace{1mm}M_{\bullet},\hspace{1mm}m)<1$. In Figure \ref{diskform}, the top panel (a) shows the contour plot of $\mathcal{T}_{r}(\bar{e},\hspace{1mm}\ell,\hspace{1mm}M_{\bullet},\hspace{1mm}m)$ for $M_{6}$=1, $m=1$, and for $r_{p}(\bar{e},\hspace{1mm}\ell,\hspace{1mm}M_{6},\hspace{1mm}m)>R_{s}$, $\mathcal{T}_{r}<1$. The bottom panel (b) shows that Max[$\mathcal{T}_{r}(\hspace{1mm} 10^{-6} \hspace{0.5mm} \leq \hspace{0.5mm} \bar{e}  \hspace{0.5mm}\leq  \hspace{0.5mm} 1, \hspace{1mm} 0\hspace{0.5mm}\leq \hspace{0.5mm} \ell \hspace{0.5mm} \leq \hspace{0.5mm} 1,1\hspace{0.5mm} \leq \hspace{0.5mm} M_{6} \hspace{0.5mm} \leq \hspace{0.5mm} 100)$]$<$ 1, which implies that the bound debris will form an accretion disk. 

The radiation timescale of the disk is given by

\begin{equation}
t_{R}(\bar{e},\hspace{1mm}\ell,\hspace{1mm}M_{\bullet},\hspace{1mm}m)=\frac{f_r M_{\star}c^2}{\eta \dot{M}(\bar{e},\hspace{1mm}\ell,\hspace{1mm}M_{\bullet},\hspace{1mm}m)}\hspace{1mm} c^2
\end{equation} 

where $c$ is the light speed, $\dot{M}$ is the accretion rate, and $\eta$ is the radiative efficiency of the disk.  The viscous timescale $t_{v}$ of the disk formed is given by

\begin{equation}
t_{v}=\int^{r_{c}}_{r_{\rm in}} \frac{1}{V_r} \, \diff r
\label{visct}
\end{equation}

where $V_r$ is the radial inflow velocity of matter in the disk, $r_{\rm in}$ is the inner radius of disk, and $r_c(\bar{e},\hspace{1mm}\ell,\hspace{1mm}M_{\bullet},\hspace{1mm}m) = 2r_p(1-r_{p}\bar{e}/r_{t})$, where $r_c(\bar{e},\hspace{1mm}\ell,\hspace{1mm}M_{\bullet},\hspace{1mm}m)$ is the circularization radius. The radial inflow velocity is given by \citep{1973A&A....24..337S,2009MNRAS.400.2070S}

\begin{equation}
V_r=-\frac{3}{2}\frac{\nu}{r}\frac{1}{f}
\label{vr}
\end{equation}

where $\displaystyle{f=1-\sqrt{\frac{r_{\rm in}}{r}}}$ and $\nu$ is the viscosity of the medium given by $\nu=\alpha c_s H$. The parameter $\alpha$ is taken to be 0.1 and $c_s=H\sqrt{GM_{\bullet}/r^3}$ where $H$ is the disk scale height. \citet{2009MNRAS.400.2070S} have calculated the disk scale height for a slim disk (see Section \ref{SE}) to be

\begin{equation}
\frac{H}{r}(\bar{e},\hspace{1mm}\ell,\hspace{1mm}M_{\bullet},\hspace{1mm}m,\hspace{1mm}r)=\frac{3}{4}f \left(\frac{\dot{M}}{\eta \dot{M}_E}\right) \left(\frac{r}{R_s}\right)^{-1}\left\{\frac{1}{2}+\left \{\frac{1}{4}+\frac{3f}{2} \left (\frac{\dot{M}}{\eta \dot{M_{E}}}\right)^2 \left(\frac{r}{R_s}\right)^{-2}\right \}^\frac{1}{2}\right\}^{-1}
\end{equation}

where $\dot{M}_E$ is the Eddington mass accretion rate and $\dot{M}$ is taken to be the time averaged accretion rate. Using Equations (\ref{visct}) and (\ref{vr}), the viscous timescale is given by

\begin{equation}
t_{v}(\bar{e},\hspace{1mm}\ell,\hspace{1mm}M_{\bullet},\hspace{1mm}m)=\frac{2}{3\alpha}\frac{1}{\sqrt{GM_{\bullet}}}\int^{r_{c}(\bar{e},\hspace{1mm}\ell,\hspace{1mm}M_{\bullet},\hspace{1mm}m)}_{r_{\rm in}} \left(\frac{H(\bar{e},\hspace{1mm}\ell,\hspace{1mm}M_{\bullet},\hspace{1mm}m,\hspace{1mm}r)}{r}\right)^{-2}\sqrt{r} f \, \diff r
\end{equation}

The ratio $\mathcal{T}_{v}(\bar{e},\hspace{1mm}\ell,\hspace{1mm}M_{\bullet},\hspace{1mm}m)=t_{v}(\bar{e},\hspace{1mm}\ell,\hspace{1mm}M_{\bullet},\hspace{1mm}m)/t_{R}(\bar{e},\hspace{1mm}\ell,\hspace{1mm}M_{\bullet},\hspace{1mm}m)$ is given by

\begin{figure}
\begin{subfigure}{\textwidth}
\centering
\includegraphics[scale=0.36]{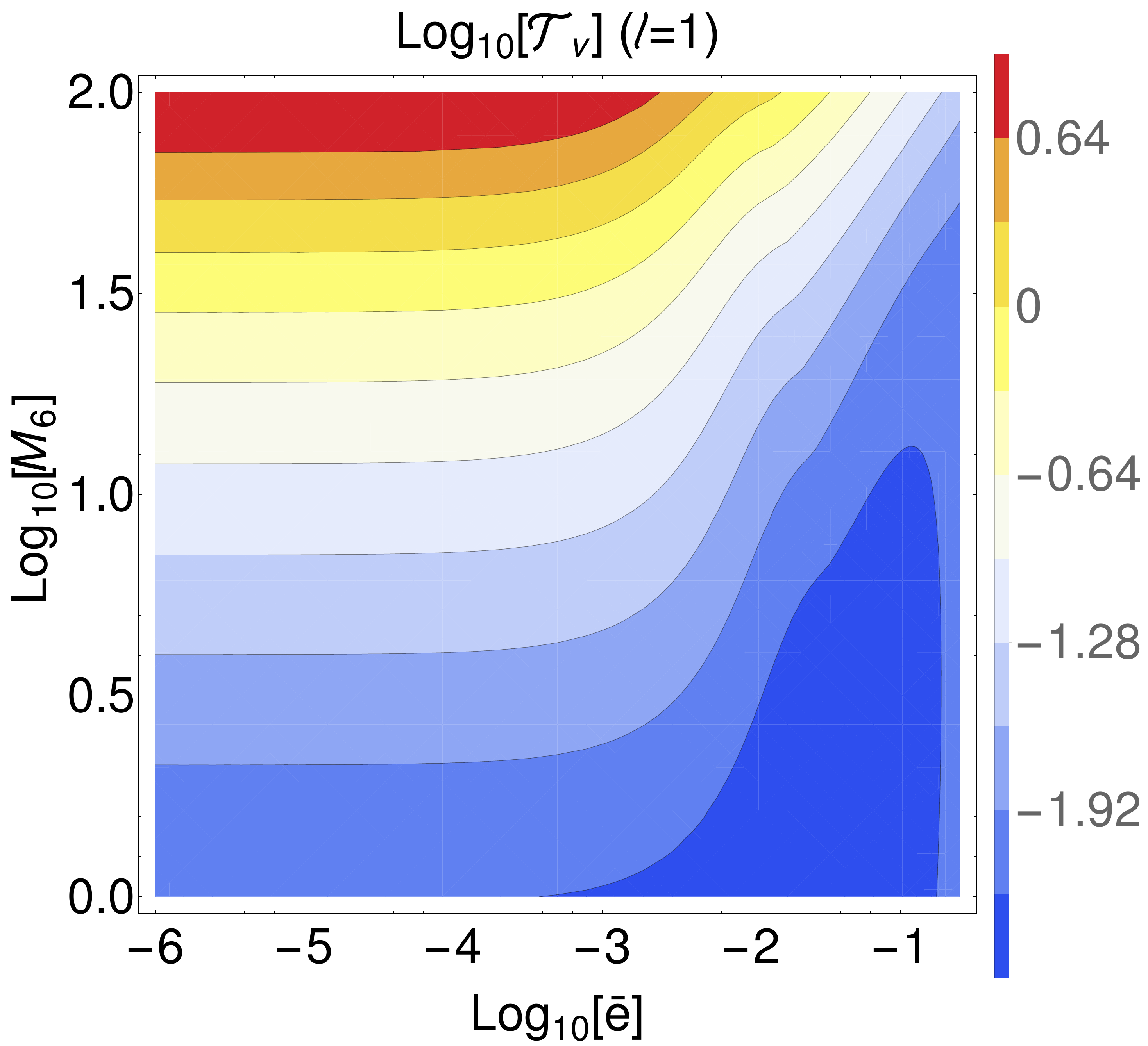}
\caption{}
\end{subfigure}
\begin{subfigure}{\textwidth}
\centering
\includegraphics[scale=0.36]{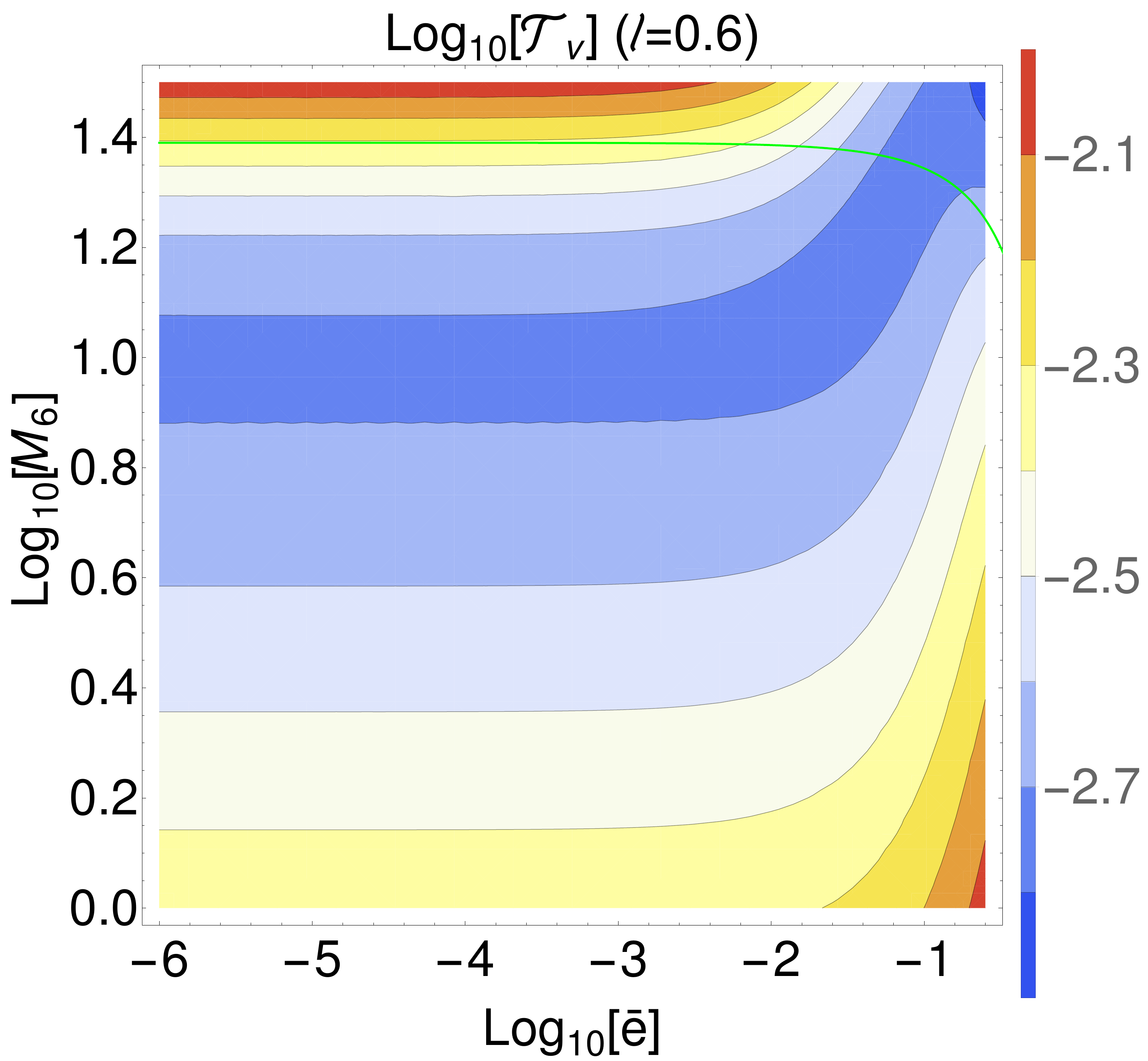}
\caption{}
\end{subfigure}
\caption{The contour plot of $\mathcal{T}_{v}(\bar{e},\hspace{1mm}\ell,\hspace{1mm}M_{\bullet},\hspace{1mm}m)$ (Equation (\ref{tauv})) is shown for $\ell=1$ (top) and $\ell=0.6$ (bottom) for $m=1$. The green line corresponds to $r_{p}=R_{s}$. For $M_{6}\leq \hspace{1mm} 31.6$, the accretion disk formed is a slim disk.}
\label{tvtr}
\end{figure}

{\footnotesize
\begin{equation}
\mathcal{T}_{v}(\bar{e},\hspace{1mm}\ell,\hspace{1mm}M_{\bullet},\hspace{1mm}m)=\frac{4}{9\alpha}\frac{1}{\sqrt{GM_{\bullet}}}\frac{\eta^3}{f_{r}M_{\star}}\frac{\dot{M}^2_{E}}{\dot{M}}\int^{r_{c}}_{r_{\rm in}}\left(\frac{r}{R_{s}}\right)^{2}\frac{\sqrt{r}}{f}\left(\frac{1}{2}+\left \{\frac{1}{4}+\frac{3f}{2} \left (\frac{\dot{M}}{\eta \dot{M}_E}\right)^2 \left(\frac{r}{R_s}\right)^{-2}\right \}^\frac{1}{2}\right)^{2}\, \diff r
\label{tauv}
\end{equation}
}
where radiative efficiency is typically $\eta=0.1$ and an accretion disk formed is a slim disk if $\mathcal{T}_{v}<1$. The Figure \ref{tvtr} shows the contour plot of $\mathcal{T}_{v}(\bar{e},\hspace{1mm}\ell,\hspace{1mm}M_{\bullet},\hspace{1mm}m)$ for $m=1$ and $\ell$=1 and 0.6. We conclude that the accretion disk formed is a slim disk for $M_{6}\leq 31.6$. For higher mass SMBHs, a thin disk forms from the disrupted debris of a star on low energy orbit and $\ell\hspace{0.5mm} \sim\hspace{0.5mm} 1$, and a thick disk for a star on high energy orbit.

\section{Accretion disk phase}
\label{adp}
\begin{figure}
\begin{center}
\includegraphics[scale=1]{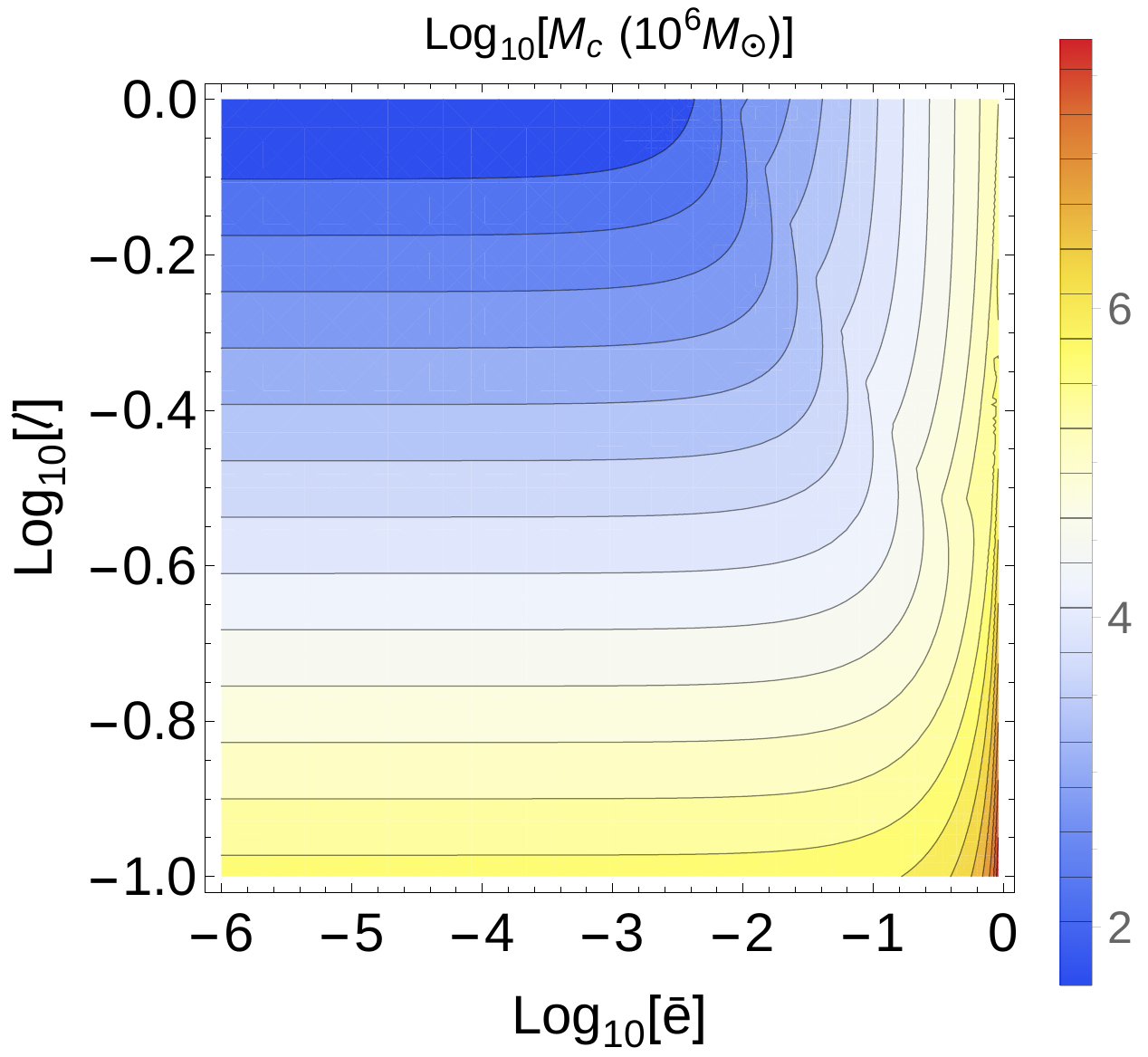}
\caption{ A contour plot of $M_{c}(\bar{e},\hspace{1mm}\ell,\hspace{1mm}m)$ is shown for the disruption of a star of solar mass. The peak $\dot{M}$  increases with a decrease in $\ell$ and an increase in $\bar{e}$, and thus, the $M_{c}$ increases with decreasing in $\ell$ and increasing in $\bar{e}$. }
\label{mcr}
\end{center}
\end{figure}

The Eddington mass accretion rate is given by 

\begin{equation}
\dot{M}_E=\frac{4\pi G M_{\bullet}}{\eta \kappa c}
\end{equation}

where $\kappa$ is the opacity of the medium taken to be Thompson opacity and $\eta$ is the radiative efficiency. For a given tidally disrupted star, the accretion disk formed has a super Eddington phase if $M_{c}(\bar{e},\hspace{1mm}\ell,\hspace{1mm}m)> M_{\bullet}$ where $M_{c}(\bar{e},\hspace{1mm}\ell,\hspace{1mm}m)$ is the critical BH mass. We have numerically equated the peak accretion rate $\dot{M}_{p}$ and $\dot{M}_E$, and obtained the $M_{c}(\bar{e},\hspace{1mm}\ell,\hspace{1mm}m)$ as shown in Figure \ref{mcr} for $m=1$. The $M_{c}(\bar{e},\hspace{1mm}\ell,\hspace{1mm}m)$ decreases with $\ell$ and increases with $\bar{e}$. For a given $\ell$, an increase in $\bar{e}$ increases $E_{d}$ and thus the orbital period of the disrupted debris decreases, which results in an increase in the $\dot{M}$ and hence the peak accretion rate $\dot{M}_{p}$. For a given $\bar{e}$, the pericenter $r_{p}$ increases with $\ell$, which results in a decrease in $E_{d}$ and thus $\dot{M}_{p}$ decreases. For a given $\bar{e}$ and $\ell$, $r_{p}(\bar{e},\ell,m)$ increases with $m$, which results in a decrease in $E_{d}$. The decrease in $E_{d}$ implies an increase in fallback time and a decrease in peak accretion rate $\dot{M}_{p}$ which results in a decrease in $M_{c}(\bar{e},\ell,m)$.

\subsection{Super Eddington Phase}
\label{SE}

For $M_{\bullet} \leq M_{c}(\bar{e},\hspace{1mm}\ell,\hspace{1mm}m)$, the radiation produced by viscous stress in the rotating disk is trapped by electron scattering and the disk is radiatively inefficient. The time for the photon to diffuse out of the gas is longer than both the inflow time in the disk and the dynamical time for the outflow. Thus, the disk is radiation pressure dominated and the opacity is given by the electron scattering. We assume that the opacity due to the Thompson scattering and in our flux calculation adopt the work of \cite{2009MNRAS.400.2070S}. The strong radiative pressure induces an outflowing wind. At the launch radius $r_L(\bar{e},\hspace{1mm}\ell,\hspace{1mm}M_{\bullet},\hspace{1mm}m)= r_c(\bar{e},\hspace{1mm}\ell,\hspace{1mm}M_{\bullet},\hspace{1mm}m)= 2r_p(1-r_{p}\bar{e}/r_{t})$, the internal energy is converted into the kinetic energy of the outflows, and the material leaves the disk with the temperature at the launch radius as determined by $aT_L^4\approx (1/2)\rho(r_L) v_w^2$ \citep{2011MNRAS.410..359L}, where $a$ is the radiation constant.  The outflow geometry is assumed to be spherical. The photons are trapped up to the radius where $\rho \kappa r \sim 1$ and the radius of photosphere $r_{\rm ph}\sim (\rho \kappa)^{-1}$ is given by

\begin{equation}
r_{\rm ph}(\bar{e},\hspace{1mm}\ell,\hspace{1mm}M_{\bullet},\hspace{1mm}m,\hspace{1mm}t)= \frac{f_{\rm out}\dot{M}\kappa}{4\pi v_{w}},
\label{rph}
\end{equation}

where $M_{\bullet}=M_6 10^6 M_{\odot}$, $f_{\rm out}=(\dot{M}_{\rm out}/\dot{M})$ and $v_w=f_{v}\sqrt{G M_{\bullet}/r_L}$ is the velocity of outflowing wind where $f_{v}$ is taken to be unity. The outflowing wind is assumed to expand adiabatically so that the density is $\rho(r) \propto T^3(r)$, where $T$ is the temperature of the out flowing wind. Using this scaling relation, the temperature at the photosphere is $T_{\rm ph}=T_L (\rho (r_{\rm ph})/\rho (r_L))^{1/3}$ and using Equation (\ref{rph}), $T_{\rm ph}$ is given by

\begin{equation}
T_{\rm ph}(\bar{e},\hspace{1mm}\ell,\hspace{1mm}M_{\bullet},\hspace{1mm}m,\hspace{1mm}t)=(4\pi)^{\frac{5}{12}}\left(\frac{1}{2a}\right)^{\frac{1}{4}}\kappa^{-\frac{2}{3}}f_{\rm out}^{-\frac{5}{12}}f_{v}^{\frac{11}{12}}\dot{M}^{-\frac{5}{12}}r_{L}^{-\frac{7}{24}}(GM_{\bullet})^{\frac{11}{24}}.
\end{equation} 

\cite{2011MNRAS.413.1623D} have calculated the fraction of outflowing material from the super Eddington slim disk with $\dot{M}/\dot{M}_E$=1, 5, 10, and 20 respectively. We approximated their result by the following relation \citep{2011MNRAS.410..359L} 

\begin{equation}
f_{\rm out}(\bar{e},\hspace{1mm}\ell,\hspace{1mm}M_{\bullet},\hspace{1mm}m,\hspace{1mm}t)=\frac{2}{\pi} \arctan \left[\frac{1}{4.5}\left(\frac{\dot{M}}{\dot{M}_E}-1\right)\right].
\end{equation}

Thus, the luminosity from the outflowing wind is given by $L^{\rm out}_{\nu}(\bar{e},\hspace{1mm}\ell,\hspace{1mm}M_{\bullet},\hspace{1mm}m,\hspace{1mm}t)=4\pi r_{\rm ph}^2 B_{\nu}(T_{\rm ph})$ and $B_{\nu}(T_{\rm ph})$ is the intensity obtained assuming the outflowing wind as a black body.

In the super Eddington phase, the time for the photon to diffuse out of the disk is longer than the viscous time, so  that the disk that is formed is thick and advective, whereas in the case when $\dot{M} \leq \dot{M}_E$, the disk is thin and cools by radiative diffusion. \citet{2009MNRAS.400.2070S}  considered a slim disk model by introducing an additional advection term the in energy conservation equation, where the effective temperature profile of the disk as a function of radius is given by

\begin{equation}
\sigma_{\rm SB} T_{e}^4(\bar{e},\hspace{1mm}\ell,\hspace{1mm}M_{\bullet},\hspace{1mm}m,\hspace{1mm}r,\hspace{1mm}t)=8.54 \times 10^{17} \hspace{1mm} \frac{ M_6^{-1} \left(\frac{\eta}{0.1}\right)^{-1} \left(\frac{r}{R_{s}}\right)^{-3}\left(\frac{\dot{M}}{\dot{M}_E}\right)f}{\frac{1}{2}+\left \{\frac{1}{4}+\frac{3f}{2} \left (\frac{\dot{M}}{\eta \dot{M}_{E}}\right)^2 \left(\frac{r}{R_s}\right)^{-2}\right \}^\frac{1}{2}} \hspace{3mm} {\rm W\hspace{0.5mm}m^{-2}}
\label{se}
\end{equation} 

where $\displaystyle{f=1-\sqrt{\frac{r_{\rm in}}{r}}}$, $\sigma_{\rm SB}$ is Stefan--Boltzmann constant, $r_{\rm in}$ is the inner radius of the disk and $R_s$ is the Schwarzschild radius BH.

\subsection{Sub-Eddington Phase} 
\label{sep}

The disk is sub-Eddington for the BH mass $M_{\bullet}>M_{c}(\bar{e},\hspace{1mm}\ell,\hspace{1mm}m)$ and $\dot{M}<\dot{M}_E$. We then consider the disk as the radiative thin disk whose the temperature profile is given by 

\begin{equation}
\sigma_{\rm SB} T_{e}^4(\bar{e},\hspace{1mm}\ell,\hspace{1mm}M_{\bullet},\hspace{1mm}m,r,\hspace{1mm}t)=8.54 \times 10^{17}  M_6^{-1} \left(\frac{\eta}{0.1}\right)^{-1} \left(\frac{r}{R_{s}}\right)^{-3}\left(\frac{\dot{M}}{\dot{M}_E}\right)f \hspace{3mm} {\rm W\hspace{0.5mm}m^{-2}}
\label{sub}
\end{equation}

We see that Equation (\ref{se}) is the modified temperature profile of the thin disk and follows the thin disk for $\dot{M}<\dot{M}_E$. Since the super-Eddington phase exists only for a certain duration, we assume that  Equation (\ref{se}) is the temperature profile for the entire duration as it approaches the thin disk model for $\dot{M}< \dot{M}_E$. Assuming the disk to be a black body, the intensity of the disk is given by $B_{\nu}(T_{e}(\bar{e},\hspace{1mm}\ell,\hspace{1mm}M_{\bullet},\hspace{1mm}m,r,\hspace{1mm}t))$ and thus the disk luminosity is given by

\begin{equation}
L^{\rm Disk}_{\nu}(\bar{e},\hspace{1mm}\ell,\hspace{1mm}M_{\bullet},\hspace{1mm}m,\hspace{1mm}t)= \int_{r_{\rm in}}^{r_{c}(\bar{e},\hspace{1mm}\ell,\hspace{1mm}M_{\bullet},\hspace{1mm}m)} B_{\nu}(T_{e}(\bar{e},\hspace{1mm}\ell,\hspace{1mm}M_{\bullet},\hspace{1mm}m,r,\hspace{1mm}t)) 2\pi r \,\diff r,
\label{lum}
\end{equation}

The total luminosity can be written as

\begin{equation}
L_{\nu}(\bar{e},\hspace{1mm}\ell,\hspace{1mm}M_{\bullet},\hspace{1mm}m,\hspace{1mm}t)=\left\{
\begin{array}{ll}
L^{\rm Disk}_{\nu}(\bar{e},\hspace{1mm}\ell,\hspace{1mm}M_{\bullet},\hspace{1mm}m,\hspace{1mm}t)\hspace{1mm}+ L^{\rm out}_{\nu}(\bar{e},\hspace{1mm}\ell,\hspace{1mm}M_{\bullet},\hspace{1mm}m,\hspace{1mm}t) & M_{\bullet}<M_{c}(\bar{e},\hspace{1mm}\ell,\hspace{1mm}m) \nonumber\\
L^{\rm Disk}_{\nu}(\bar{e},\hspace{1mm}\ell,\hspace{1mm}M_{\bullet},\hspace{1mm}m,\hspace{1mm}t)\hspace{1mm} & M_{\bullet}\geq M_{c}(\bar{e},\hspace{1mm}\ell,\hspace{1mm}m) \nonumber 
\end{array}
\right. 
\end{equation}

If $\nu_{l}$ and $\nu_{h}$ are the minimum and maximum frequency of the spectral band, then the luminosity of the emitted radiation in the given spectral band in the rest frame of the galaxy is given by

\begin{equation}
L_{e}(\bar{e},\hspace{1mm}\ell,\hspace{1mm}M_{\bullet},\hspace{1mm}m,\hspace{1mm}z,\hspace{1mm}t)= \int_{\nu_{l}(1+z)}^{\nu_{h}(1+z)} L_{\nu}(\bar{e},\hspace{1mm}\ell,\hspace{1mm}M_{\bullet},\hspace{1mm}m,\hspace{1mm}t) \,\diff \nu
\label{ltot}
\end{equation} 

The observed flux $\displaystyle{f_{\rm obs}(\bar{e},\hspace{1mm}\ell,\hspace{1mm}M_{\bullet},\hspace{1mm}m,\hspace{1mm}z,\hspace{1mm}t)=L_{e}(\bar{e},\hspace{1mm}\ell,\hspace{1mm}M_{\bullet},\hspace{1mm}m,\hspace{1mm}z,\hspace{1mm}t)/(4\pi {\rm d}_L^2(z))}$, where $z$ is the redshift and $d_L$ is the luminosity distance, and the radiation is observed only if 

\begin{equation}
f_{l}< \hspace{0.5mm} f_{\rm obs}(\bar{e},\hspace{1mm}\ell,\hspace{1mm}M_{\bullet},\hspace{1mm}m,\hspace{1mm}z,\hspace{1mm}t)=\hspace{0.5mm}\frac{L_{e}(\bar{e},\hspace{1mm}\ell,\hspace{1mm}M_{\bullet},\hspace{1mm}m,\hspace{1mm}z,\hspace{1mm}t)}{4\pi d_L^2(z)}
\label{cond}
\end{equation}

where $f_{l}$ is the sensitivity of the detector. The Equation (\ref{cond}) is utilized to generate a digital signal $A(t)$ such that

\begin{equation}
A(t)=\left\{
\begin{array}{ll}
1 & {\rm if \hspace{1mm} Equation \hspace{1mm}(\ref{cond}) \hspace{1mm} holds \hspace{1mm} true } \\\\
0 & {\rm if \hspace{1mm} Equation \hspace{1mm}(\ref{cond}) \hspace{1mm} does\hspace{1mm} not \hspace{1mm} hold \hspace{1mm} true} 
\end{array}
\right. 
\label{A}
\end{equation}

The width of the digital signal gives the duration of the flare detection used in the event rate calculation (see Section \ref{pro}). 

\begin{figure}
\begin{subfigure}{\textwidth}
\centering
\includegraphics[scale=0.40]{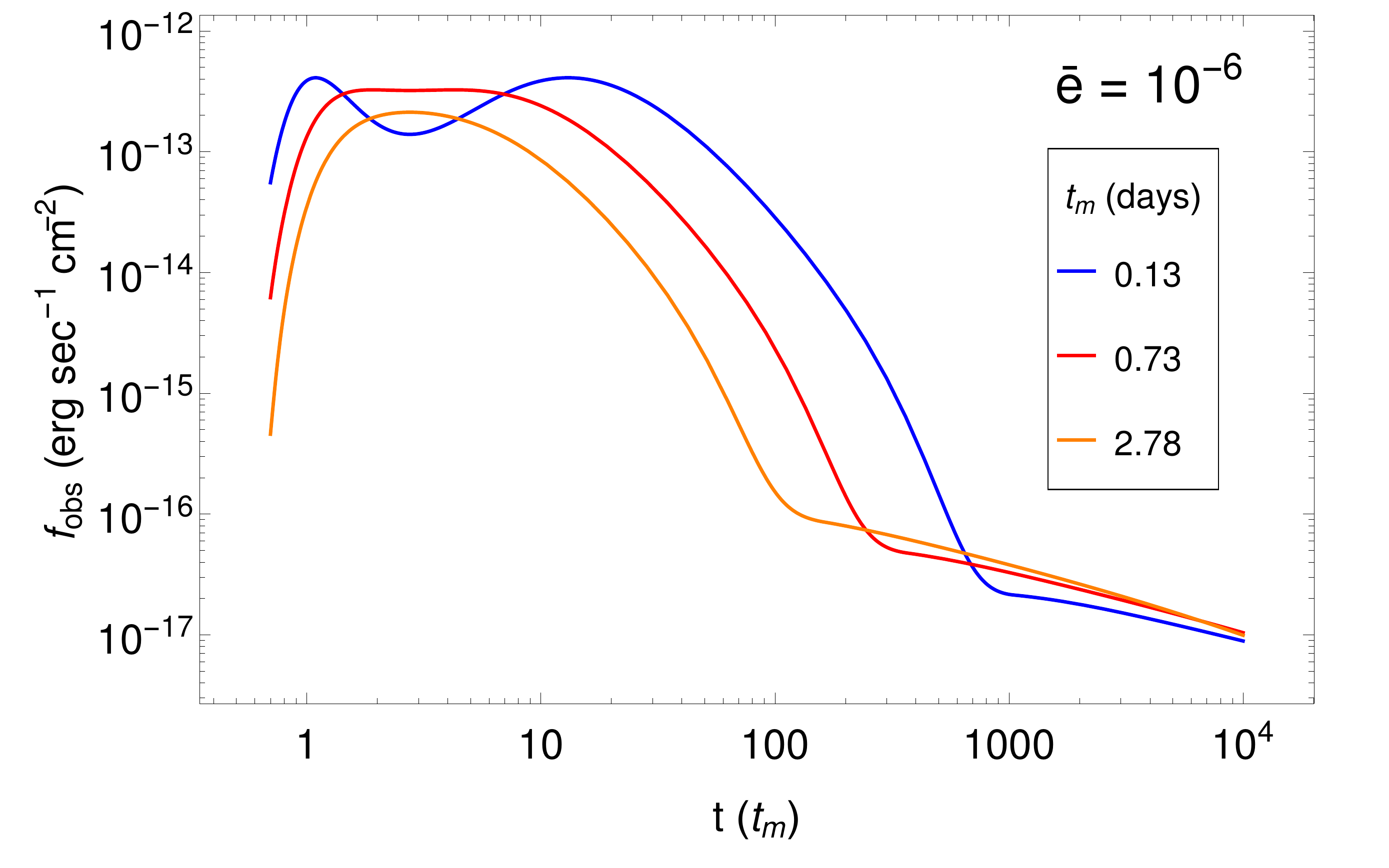}
\caption{}
\end{subfigure}
\begin{subfigure}{\textwidth}
\centering
\includegraphics[scale=0.40]{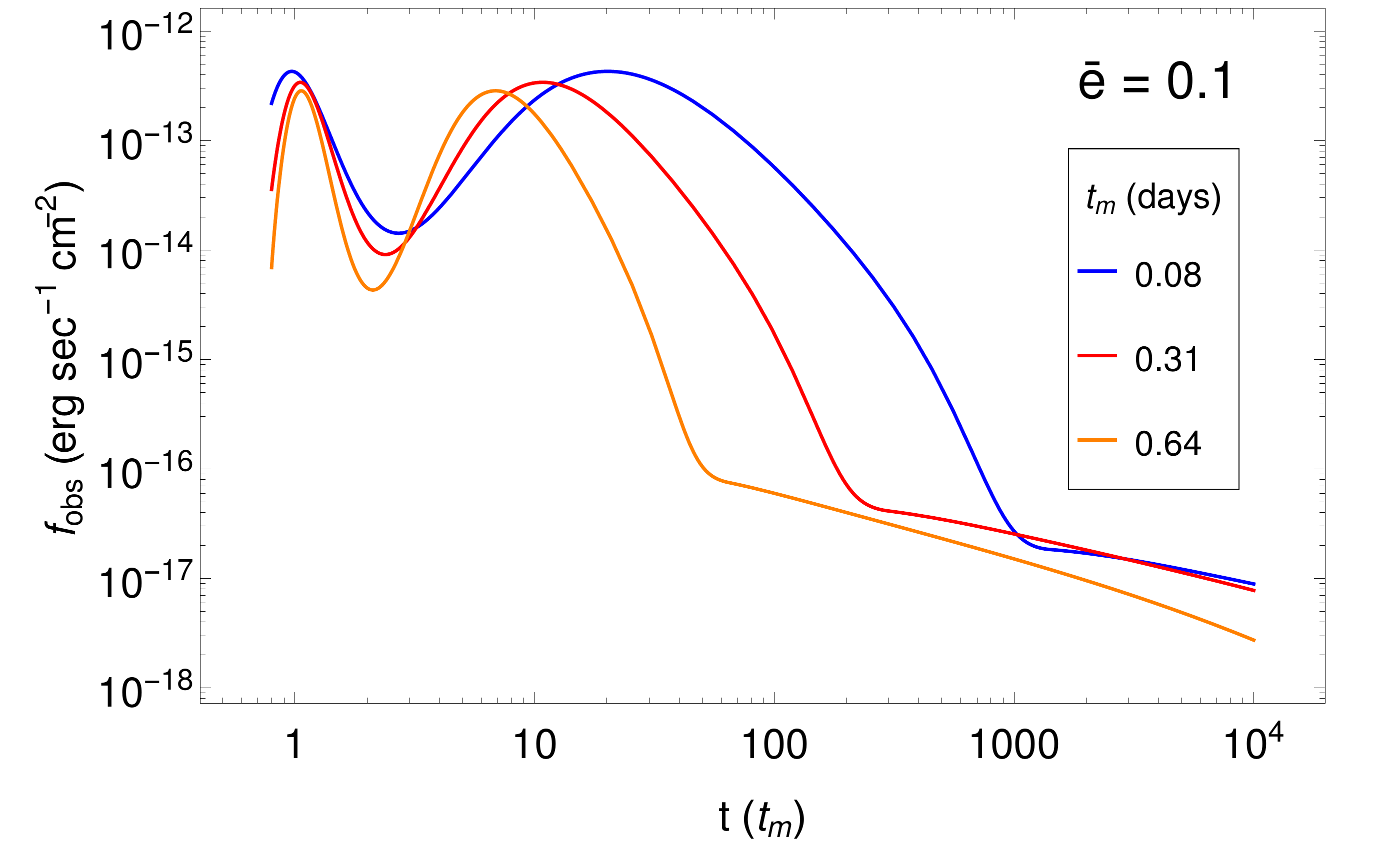}
\caption{}
\end{subfigure}
\caption{The observed flux $f_{\rm obs}$ (Equation (\ref{cond})) in the optical g band for $M_{6}=1$, $m=1$, redshift $z=0.1$, and $\ell$= 0.6 (blue), 0.8 (red), 1.0 (orange). The peak flux decreases with $\ell$ and the light curve profile gets widened with a decrease in $\ell$. The initial dip in the flux is due to the outflowing wind. The time is scaled with $t_m$ (in days), which is the orbital period of inner-most bound debris that decreases with $\ell$ due to the increase in the energy of the disrupted debris.}
\label{opt}   
\end{figure}

As an example, the observed flux in optical g band is shown in Figure \ref{opt} for $M_{6}=1$, $z=0.1$, and $\ell$= 0.6 (blue), 0.8 (red), 1.0 (orange). For $M_{6}=1$, both the outflowing wind and disk contribute to the observed flux and the flux from wind dominates in the initial time. We can observe a dip in flux due to the outflowing wind whose $r_{\rm ph}\propto \dot{M}$ and $T_{\rm ph}\propto \dot{M}^{-5/12}$ and the occurrence time of dip is nearly at the time of peak accretion rate $\dot{M}_{p}$ (see Figure \ref{masst}). With an increase in $\bar{e}$, the $\dot{M}_{p}$ increases, which results in an increase in $r_{\rm ph}$ and a decrease in $T_{\rm ph}$ and thus a decrease in the intensity of radiation $B(T_{\rm ph})$. The flux is due to out flowing wind $\propto r^2_{\rm ph}B(T_{\rm ph})$ and decreases with $\dot{M}_{p}$ if the decline in $B(T_{\rm ph})$ is higher than the rise in $r_{\rm ph}$. \citet{1999ApJ...514..180U} predicted that the minimum value of $n_{\rm orb}$ for the disrupted debris to get circularized is $\sim 2$--$3$ and thus we have utilized the $f_{\rm obs}$ starting from the time $\tau = 3$ to generate a digitized signal and calculate the duration of the detection.  

\section{Event rate calculation}
\label{erc}

For any transient survey, the net detectable TDE rate depends on the number density of non-active galaxies, the theoretical capture rate per galaxy (see Section \ref{tcr}), the luminosity distance of galaxies, the sensitivity of the detector, and the duration of flare detection. In this section, we will carry out the detailed calculation of each quantity separately and then combine them in Section \ref{pro}.

\subsection{Number density of non active galaxies}

The number density of quasars is a function of redshift and luminosity, where quasars emitting radiation of low intensity ($L < 10^{40} \hspace{1mm} \rm{erg \hspace{1mm} sec^{-1}}$) are non-active galaxies as compared to the normal quasars ($L \gg 10^{45}-10^{46} \hspace{1mm} \rm{erg \hspace{1mm} sec^{-1}}$). According to the \citet{1982MNRAS.200..115S} argument, if quasars were powered by accretion onto a SMBH, then such SMBH must exist in our local universe as ``dead'' quasars or non-active galaxies. The number density of galaxies (quasars) can be obtained by using the quasar luminosity function (QLF; \citet{2007ApJ...654..731H}),

\begin{equation}
\frac{\diff \psi}{\diff \log L}= \frac{\psi_{\ast}}{(\frac{L}{L_{\ast}})^{\gamma_1}+(\frac{L}{L_{\ast}})^{\gamma_2}},
\end{equation}

where $\displaystyle{\psi_{\ast},\hspace{1mm} L_{\ast},\hspace{1mm}  \gamma_1,\hspace{1mm} \gamma_2}$ are a function of redshift $z$. As TDEs are the main observational signatures of quiescent galaxies, we need to determine the number density of quiescent galaxies at any redshift $z$. \cite{2007ASPC..373..667C} used the QLF to obtain the duty cycle $\displaystyle{\delta(z) = 10^{-3} \left(z/0.1 \right)^{2.5}}$ where $\delta(z)$ is defined as the ratio of the number of active galaxies to the total number of galaxies. Thus, the BH mass function of quiescent galaxies is given by

\begin{equation}
\frac{\diff \psi}{\diff M_{\bullet}}=(1-\delta(z)) \frac{\diff \psi}{\diff \log L} \left|\frac{\diff \log L}{\diff M_{\bullet}}\right|
\end{equation}

where $L$ is the luminosity of the quasars, which is taken to be $L=\eta L_{E}=\eta 4\pi GM_{\bullet}c/\kappa$ where $L_{E}$ is the Eddington luminosity, $\eta $ is taken to be 0.1, and $\kappa$ is the opacity due to Thompson scattering. This gives the number density of quiescent galaxies as a function of BH mass $M_{\bullet}$ and redshift $z$ as

\begin{equation}
\frac{\diff \psi}{\diff M_{\bullet}}(M_{\bullet},\hspace{1mm} z)=(1-\delta(z))\frac{1}{M_{\bullet}} \frac{\diff \psi}{\diff \log L} 
\label{numden}
\end{equation}

\subsection{Luminosity Distance}
\label{LD}

We assume $\Lambda$CDM cosmology with $\Omega_m=0.315, \Omega_{\Lambda}=0.685$, $H_o=67.3 \hspace{1mm} \rm{Km\hspace{1mm}sec^{-1}\hspace{1mm}Mpc^{-1}}$ \citep{2013arXiv1303.5089P}. The luminosity distance as a function of redshift $z$ is  given as

\begin{equation}
{\rm d}_{L}(z)=(1+z) \frac{c}{H_o}  \int_0^z  \frac{1}{((1+z')^{3} \Omega_m+ \Omega_{\Lambda})^{0.5}} \,\diff z'. 
\end{equation}
\\
Consider now a small volume of the universe at redshift $z$ with radial width $\diff z$ covering an opening angle $\omega$ on the observer's sky \citep{2014MNRAS.437..327K}. The comoving volume of the slice is

\begin{equation}
\diff V_c=\omega d_H^3 \frac{I^2(z)}{W(z)} \diff z
\label{dvol}
\end{equation}
\\
where $\omega=4\pi f_{s},\hspace{1mm} d_H=\displaystyle{c/H_o},\hspace{1mm} W(z)=((1+z)^{3} \Omega_m+ \Omega_{\Lambda})^{0.5}$,

\begin{equation}
I(z)= \int_0^z  \frac{1}{((1+z')^{3} \Omega_m+ \Omega_{\Lambda})^{0.5}} \,\diff z'
\end{equation}

and $f_{s}$ is the fraction of sky observed.

\subsection{Probability of flare detection}
\label{pro}

We generate the spectrum in the form of digital signal using Equation (\ref{A}), and the width of the digital signal provides the duration of flare detection $\delta t_{f}(\bar{e},\hspace{1mm}\ell,\hspace{1mm}M_{\bullet},\hspace{1mm}m,\hspace{1mm}z)$. If $\displaystyle{t_{\rm cad} \hspace{1mm} {\rm and} \hspace{1mm} t_{\rm int}}$ are the cadence and integration time of the detector, then the probability of detection of an event is given by 

\begin{equation}
P(\bar{e},\hspace{1mm}\ell,\hspace{1mm}M_{\bullet},\hspace{1mm}m,\hspace{1mm}z)={\rm Min}\left[1,\hspace{1mm}\frac{\delta t_{f}(\bar{e},\hspace{1mm}\ell,\hspace{1mm}M_{\bullet},\hspace{1mm}m,\hspace{1mm}z)}{t_{\rm cad}+ t_{\rm int}}\right]
\label{proba}
\end{equation}

Using Equations (\ref{flr}), (\ref{tdnth1}, (\ref{numden}, (\ref{dvol} and (\ref{proba}), the net detectable event rate by the detector is given by

\begin{equation}
 \frac{\diff^5\dot{N}_{D}(\gamma,\bar{e},\ell,M_{\bullet},m,z)}{\diff M_{\bullet}\hspace{0.2mm}\diff m \hspace{0.2mm}\diff \bar{e}\hspace{0.2mm}\diff \ell \hspace{0.2mm}\diff z}=\omega d^3_H \left(\frac{\diff \Psi}{\diff M_{\bullet}}\right) \frac{\diff^2 \dot{N}_{t}}{\diff \bar{e} \hspace{0.2mm} \diff \ell \hspace{0.2mm} \diff m}(\gamma,\bar{e},\ell,M_{\bullet},m) \frac{I^2(z)}{W(z)} P(\bar{e},\ell,M_{\bullet},m,z)
\label{net}
\end{equation}

With the given initial parameters $M_{\bullet},~m,~\bar{e},~\ell$ and $z$, we generate the light curves using Equation (\ref{ltot}) in the optical g and soft X-ray bands. The generated spectrum is compared with the sensitivity of the detector $f_l$ to generate a digital signal using Equation (\ref{A}), and the width of the digital signal gives the duration of flare detection. The range of initial parameters in the calculation are taken to be $M_6=M_{\bullet}/10^6M_{\odot}={\rm 1-100},~m={\rm 0.8-150},~\bar{e}=\bar{e}_h-1,~\ell=0-1$ and $z=0-z_s$, where $z_s(M_{\bullet},~m,~\bar{e},~\ell)$ is the detection limit of the survey. Then, using Equation (\ref{net}), we calculated the net detectable rate by integrating in steps over redshift $z,~\ell,~\bar{e},~m$, and finally over $M_{\bullet}$ such that 

\begin{equation}
\displaystyle \dot{N}_{D}= \int^{100}_{1} \diff M_6 \int^{150}_{0.8} \diff m \int^{1}_{\bar{e}_{h}} \diff \bar{e} \int^{1}_{0} \diff \ell \int^{z_{s}}_{0} \diff z  \hspace{2mm} \frac{\diff^5\dot{N}_{D}(\gamma,\bar{e},\ell,M_6,m,z)}{\diff M_6\hspace{0.2mm}\diff m\hspace{0.2mm}\diff \bar{e}\hspace{0.2mm}\diff \ell \hspace{0.2mm}\diff z}.
\label{det}
\end{equation}  

\begin{figure}
\begin{subfigure}{\textwidth}
\centering
\includegraphics[scale=0.5]{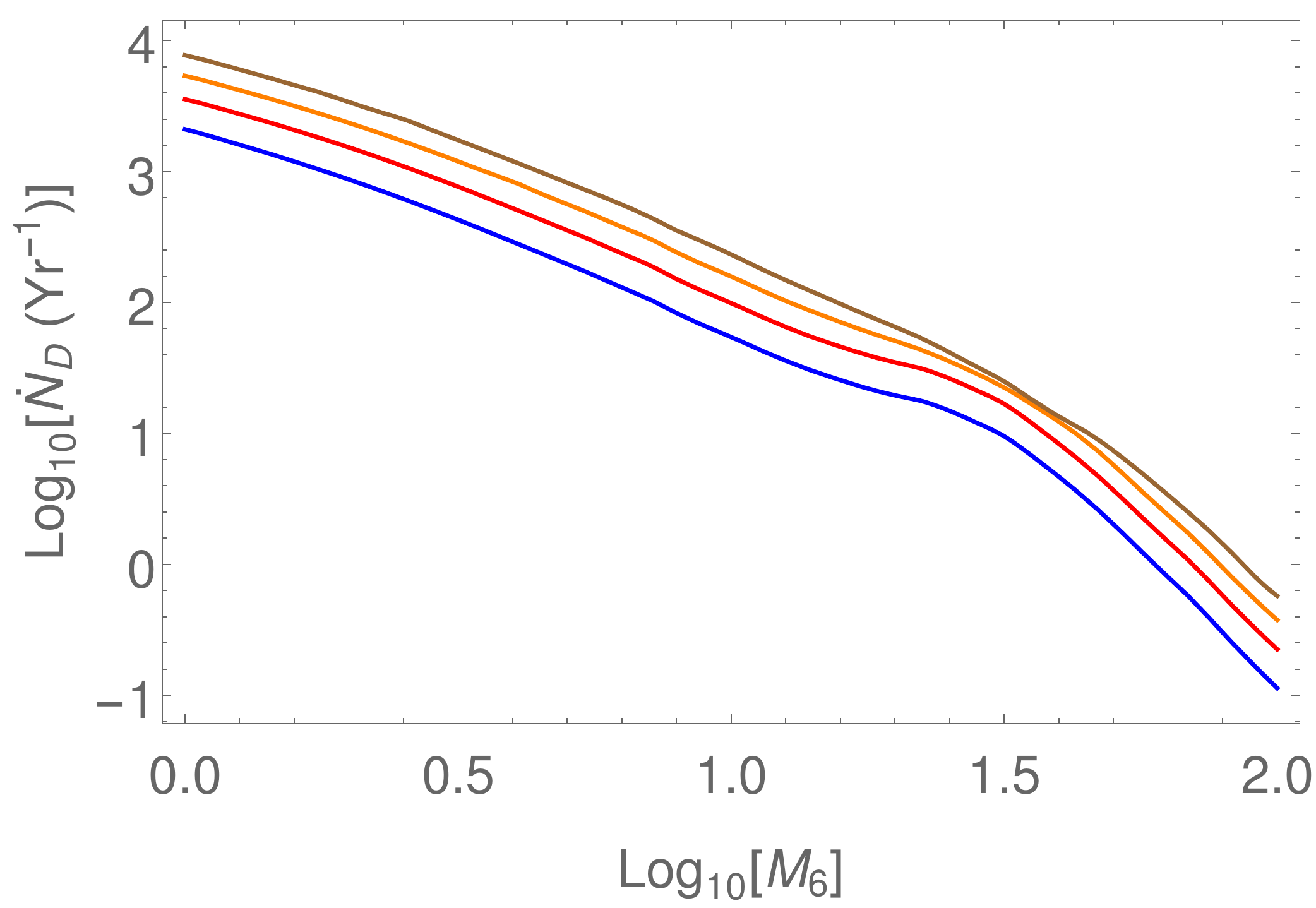}
\caption{}
\end{subfigure}
\begin{subfigure}{\textwidth}
\centering
\includegraphics[scale=0.5]{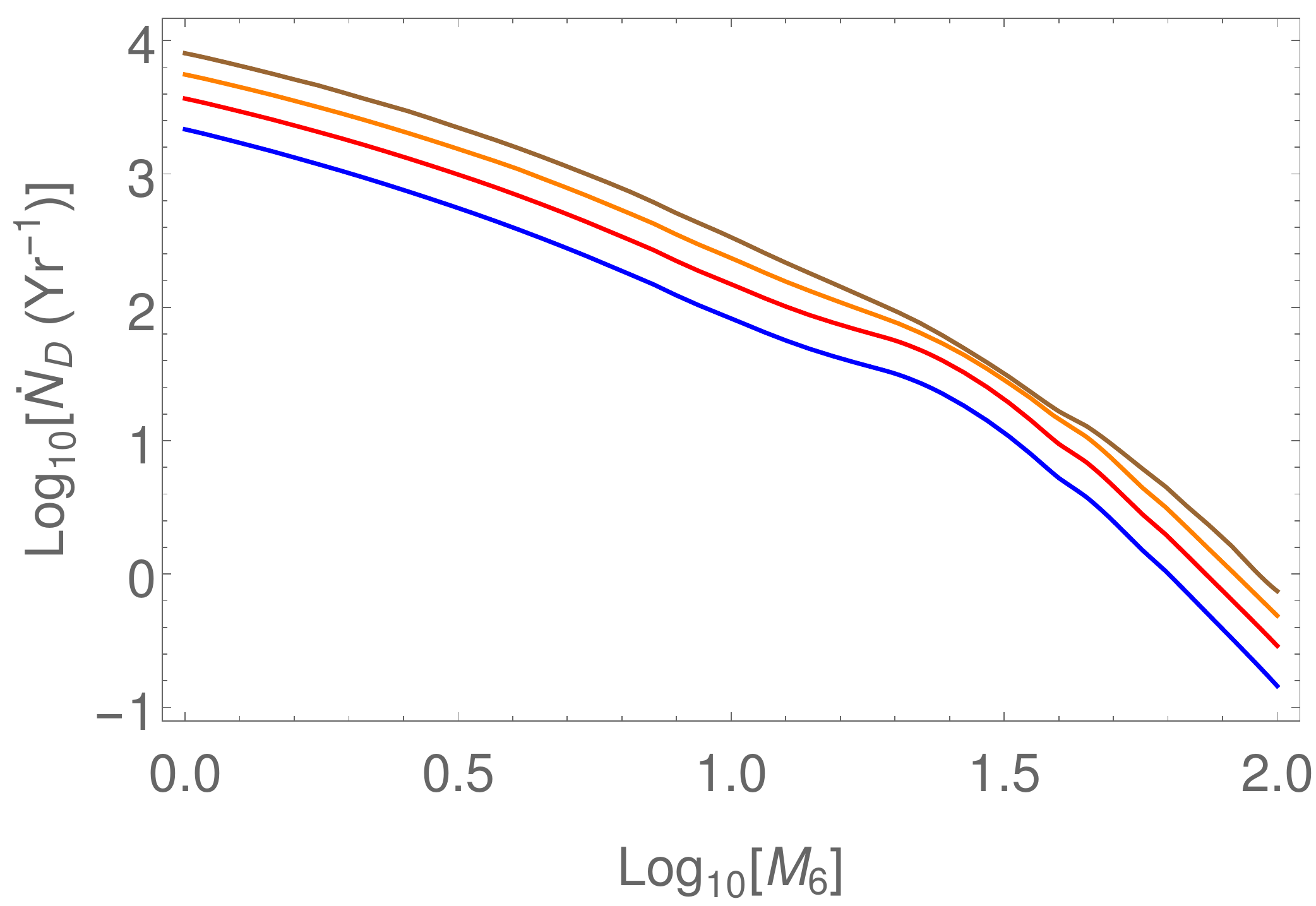}
\caption{}
\end{subfigure}
\caption{ The detectable rate, $\dot{N}_{D}$ per $M_{6}$ obtained by integrating Equation (\ref{net}) in steps over $z,~\ell,~\bar{e}~{\rm and}~m$ for various $\gamma$ for (a) LSST survey and (b) Pan-STARRS 3$\pi$ survey for $\gamma=$0.6 (blue), 0.8 (red), 1.0 (orange), and 1.2 (brown). With increase in $\gamma$, the detectable rate increases due to the increase in $\dot{N}_{t}$.} 
\label{lsmt}   
\end{figure}

\begin{figure}
\begin{center}
\includegraphics[scale=0.41]{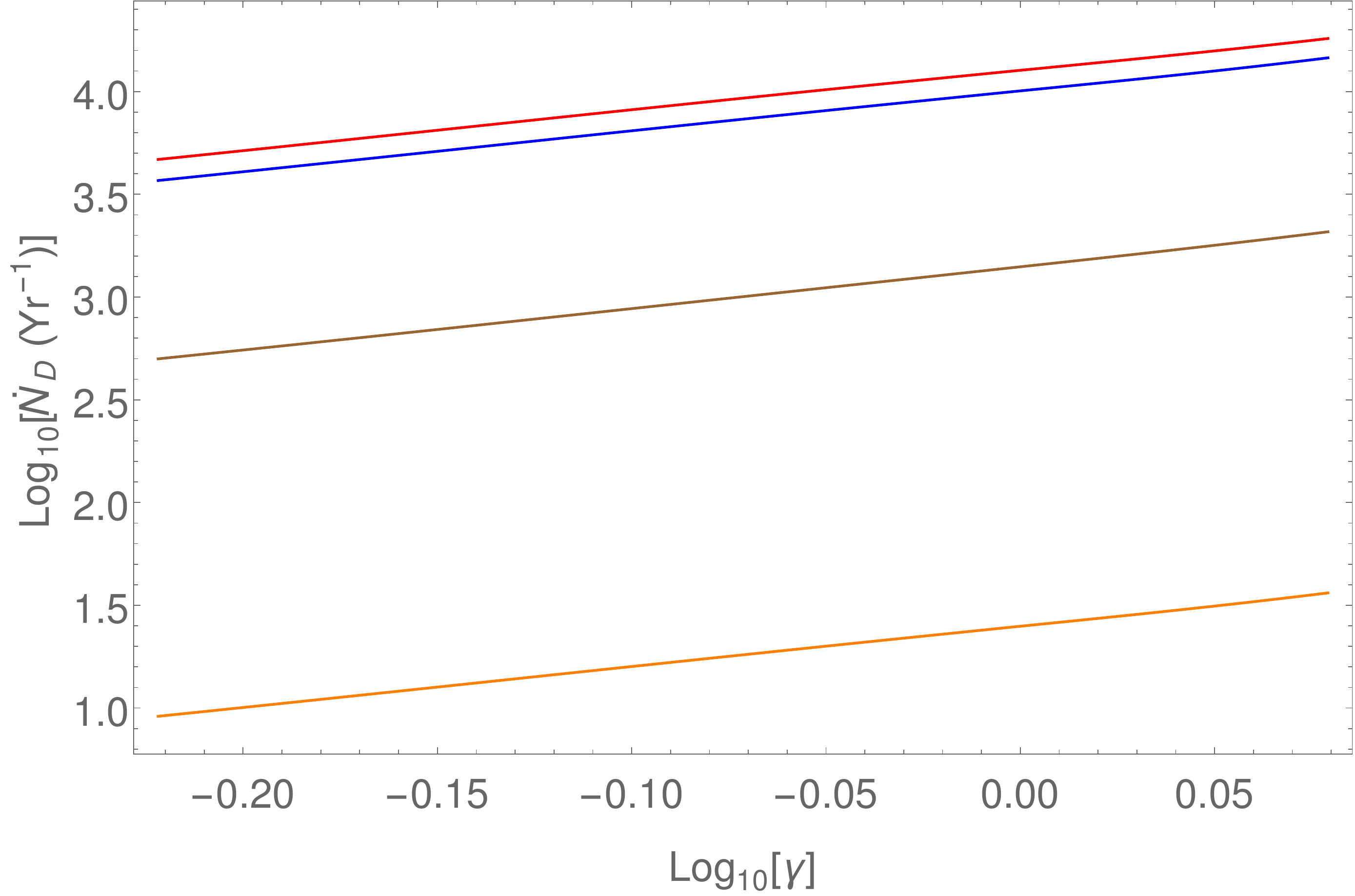}
\caption{The detectable rate, $\dot{N}_{D}$ (Equation (\ref{det})),  as a function of $\gamma$ for LSST (blue), Pan-STARRS 3$\pi$ (red), Pan-STARRS MDS (orange), and eROSITA (brown). It is seen that $\dot{N}_{D}\propto \gamma^{\Delta}$ where $\Delta$ is the slope given in Table \ref{survey}.}
\label{ndga}
\end{center}
\end{figure}

The detectable rate per $M_{6}$ integrated over $z$, $\bar{e}$, $\ell$ and $m$ for various $\gamma$ is shown in Figure \ref{lsmt} for LSST and Pan-STARRS 3$\pi$ detectors parameters. As the $\dot{N}_{t}$ and BH mass function decreases with $M_{6}$, the $\diff \dot{N}_{D}/\diff M_{6}$ decreases with $M_{6}$. The net $\dot{N}_{D}$ integrated over $\bar{e}$, $\ell$, $m$ and $M_{6}$ is plotted as a function of $\gamma$ in Figure \ref{ndga} for various missions. The resulting $\dot{N}_{D}\propto \gamma^{\Delta}$ where $\Delta$ is given in Table \ref{survey}. 

Using Equation (\ref{flr}), (\ref{tdnth}), (\ref{numden}) and (\ref{dvol}), the occurrence rate of TDE is given by 

\begin{equation}
\displaystyle \dot{N}_{o}= \int^{100}_1 \diff M_6 \int^{150}_{0.8} \diff m \int^{1}_{\bar{e}_{h}} \diff \bar{e} \int^{1}_{0} \diff \ell \int^{z_{s}}_{0} \diff z  \hspace{2mm} \omega d^3_H \left(\frac{\diff \Psi}{\diff M_6}\right)\frac{\diff^3 \dot{N}_{t}}{\diff \bar{e} \hspace{0.2mm} \diff \ell \hspace{0.2mm} \diff m} (\gamma,\bar{e},\ell,M_6,m) \frac{I^2(z)}{W(z)}
\label{ocr}
\end{equation}  

where integration limits are same as those taken for Equation (\ref{det}). We define the detection efficiency of TDE for a detector to be

\begin{equation}
\Upsilon=\frac{\dot{N}_{D}}{\dot{N}_{o}}
\label{up}
\end{equation}

The $\dot{N}_{D}$ calculated for LSST, Pan-STARRS 3$\pi$, Pan-STARRS MDS, and eROSITA mission along with their detection efficiency are given in Table \ref{survey}. 
\citet{2011ApJ...741...73V} estimated the event rates on the basis of observational studies in the optical bands for the Sloan Digital Sky survey (SDSS) and scaled the result of SDSS to the other missions using the relation $\dot{N} \propto f_{s}\hspace{0.5mm}f_{l}^{-3/2}$ \citep{2009ApJ...698.1367G}. This relation is valid only if we assume all other parameters, such as cadence and integration time of the detector, to be same and Table \ref{survey} shows the estimated rates by \citet{2011ApJ...741...73V} $\dot{N}_{obs}$ (column with c notation) and our predicted rates $\dot{N}_{D}$ (column with a notation). The value of $\gamma$ from the observed density profile is in the range $\sim$ 0.5-1.2\citep{2004ApJ...600..149W,2014arXiv1410.7772S}. Our results are in reasonable agreement with their results. \citet{2014MNRAS.437..327K} calculated the number of events $N$ detectable at any moment in the X-ray band for the eROSITA mission assuming a constant theoretical rate for $M_{6} \sim$ 1--10 and light curve profile to follow the $t^{-5/3}$ law, whereas we have followed a more rigorous calculation to obtain $\dot{N}_{D}$. Our prediction for eROSITA  does not include the limitations in \citet{2014MNRAS.437..327K}, but are more precise and in agreement with their rough estimate.

The values of $\gamma$ for which our predictions of $\dot{N}_{D}$ match with the scaled-up values in \citet{2011ApJ...741...73V} are shown as $\gamma_{s}$ in Table \ref{survey}. The only free parameter in our estimate is $\gamma$ and
this is likely to vary from source to source. Not knowing the expected distribution of $\gamma$ as a function of say redshift, we have calculated the error in our estimation of $\dot{N}_D$ by taking a fiduciary range in the observed median of $\gamma=0.7\pm 0.1$, as is shown in Table \ref{survey}.

\begin{landscape}

\begin{table}
\scriptsize
\scalebox{1.05}{
\begin{tabular}{|ccccccccccc|}
\hline
&&&&&&&&&&\\
Survey & Band & $f_{s}$  & Sensitivity/flux & Cadence time & Integration time & $\dot{N}_{D}$ (${\rm yr^{-1}}$)$^{a}$ & $\gamma_{s}^{b}$ & $\dot{N}_{obs}$ (${\rm yr^{-1}}$)$^{c}$ & $\Upsilon^{d}$ & $\Delta^{e}$ \\
&&& & (s) & (s) & $\gamma=0.7~\pm~0.1$ & & & & \\
\hline
&&&&&&&&&&\\
LSST & Optical & 0.5  & 24.5 AB mag (g band) &  2.6 $\times$ 10$^5$  & 10 & 5003 $\pm$ 1421&  0.63 & 4131 & 0.91 & 1.97\\
&&&&&&&&&&\\
Pan-STARRS (MDS) & Optical & 0.0012  & 24.8 AB mag (g band) & 3.46 $\times$ 10$^5$  & 30 & 12.3 $\pm$ 3.5 & 0.77 & 15 & 0.92  & 1.98\\
&&&&&&&&&&\\
Pan-STARRS 3$\rm{\pi}$ & Optical &  0.75  & 24 AB mag (g band) & 6.05 $\times$ 10$^5$  & 30 & 6337 $\pm$ 1800 & 0.48 & 3106 &  0.85 & 1.94\\
&&&&&&&&&&\\
eROSITA  & X-ray & 1 & 2.4$\rm{\times 10^{-14}} ({\rm erg\hspace{1mm}sec^{-1}\hspace{1mm}cm^{-2}})$ & 1.58 $\times$ 10$^7$   & 1.6 $\times$ 10$^3$  &   679.5 $\pm$ 195 &-- & -- & 0.7 & 2.06\\
&&&&&&&&&&\\
\hline
\end{tabular}
}
\caption{Mission instrument parameters and predicted rate of the surveys}
\tablecomments{The parameters of the survey are taken from (1) LSST (\cite{2009MNRAS.400.2070S} \& \url{http://www.lsst.org/lsst/overview/}), (2)  Pan-STARRS (MDS; Medium Deep survey; \citet{2011ApJ...741...73V}), (3) Pan-STARRS  3$\rm{\pi}$ (\cite{2009MNRAS.400.2070S} \& \url{http://pan-starrs.ifa.hawaii.edu/public/}), (4) eROSITA (SRG; \citet{2014MNRAS.437..327K} \& \url{http://www.mpe.mpg.de/eROSITA})\\ $^{\rm a}$ Our predicted values along with the error estimates for an assumed range of $\Delta \gamma=0.1$ around a typically observed median of $\gamma=0.7$. \\ $^{\rm b}$$\dot{N}_{D}(\gamma_{s})=\dot{N}_{D}$ estimated by \citet{2011ApJ...741...73V}. \\ $^{\rm c}$ Results from \citet{2011ApJ...741...73V}. \\ $^{\rm d}$ Detection efficiency of the detector given by Equation (\ref{up}). \\ $^{\rm e}$ Detectable rate $\dot{N}_{D} \propto \gamma^{\Delta}$.}
\label{survey}
\end{table}
\end{landscape}

\begin{figure}
\begin{center}
\includegraphics[scale=0.5]{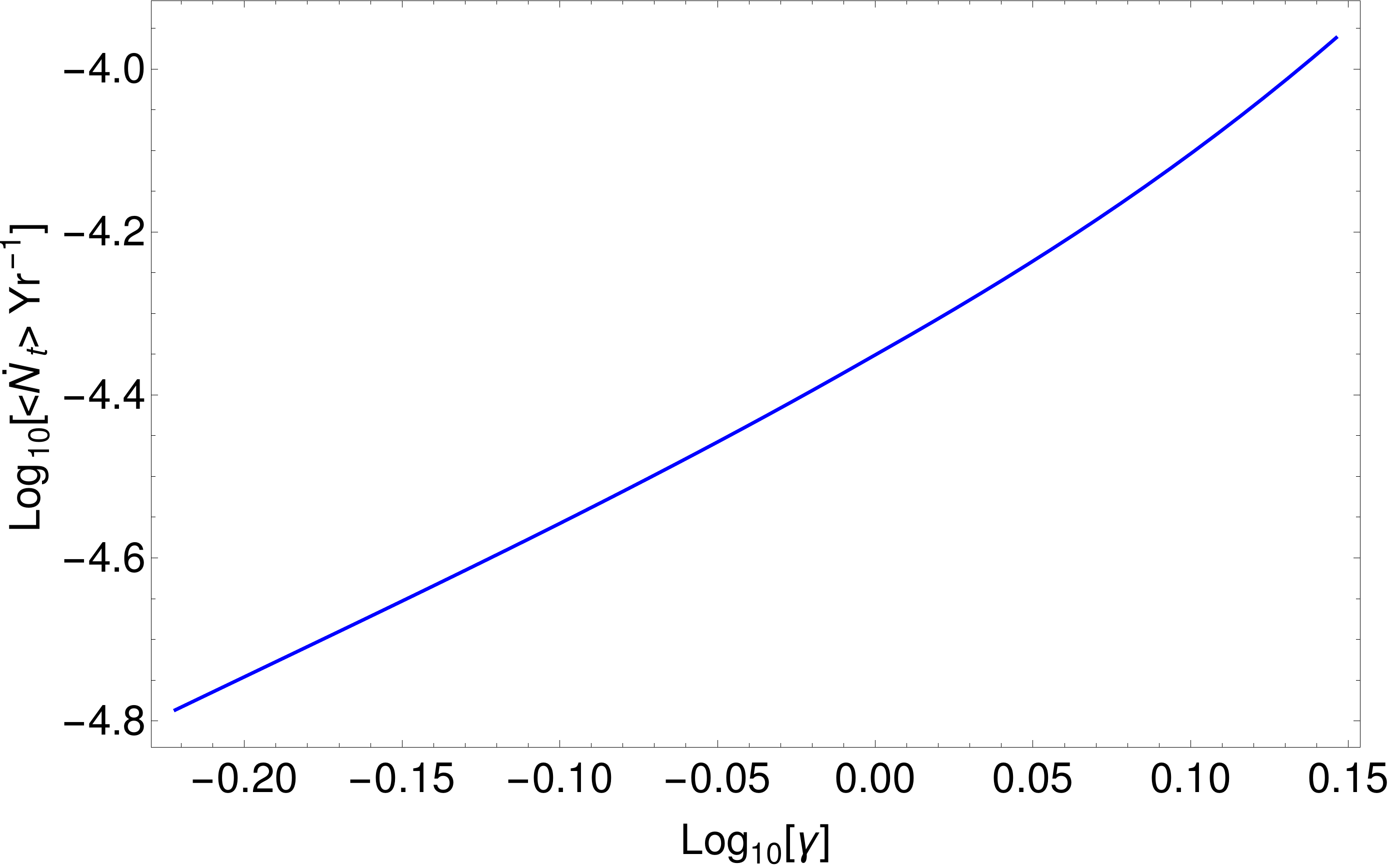}
\caption{The galaxy averaged $\dot{N}_{t}$ (Equation (\ref{gavn})) increases with $\gamma$ and for $\gamma\leq 1.2$, $\left<\dot{N}_{t}\right>\approx \gamma^2$. }
\label{avgnet}
\end{center}
\end{figure}

\section{Discussion of the results}

The star's initial orbital parameters $E$ and $J$ have significant effects on both stellar and accretion dynamics. We have seen that the effect of $J$ which has not been included previously, plays a crucial role in constructing the shape of light curve profiles.   

 We have employed a single power density model because most of the galaxies given in \citet{2004ApJ...600..149W} and \citet{2014arXiv1410.7772S} have a break radius $r_b > r_h$ and we calculated the $\dot{N}_{t} \sim 10^{-5}-10^{-4}$ Yr$^{-1}$, which shows a nonlinear dependence with $M_{\bullet}$ as shown by \citet{2004ApJ...600..149W} for single stellar mass distribution with a Nuker profile. Using Equations (\ref{flr}), (\ref{nnet}), (\ref{numden}) and (\ref{dvol}), the galaxy average $\dot{N}_{t}$ is given by

\begin{equation}
\displaystyle{\left<\dot{N}_{t}\right>(\gamma)=\frac{\int^{10^{8}M_{\odot}}_{10^{6}M_{\odot}} \diff M_{\bullet} \int^{1}_{0} \diff z \hspace{2mm} \left(\frac{\diff \psi}{\diff M_{\bullet}}\right) \frac{I^2(z)}{W(z)}\dot{N}_{t}(\gamma,\hspace{1mm} M_{\bullet})}{\int^{10^{8}M_{\odot}}_{10^{6}M_{\odot}} \diff M_{\bullet} \int^{1}_{0} \diff z \hspace{2mm} \left(\frac{\diff \psi}{\diff M_{\bullet}}\right)\frac{I^2(z)}{W(z)} }}
\label{gavn}
\end{equation}

and is shown in Figure \ref{avgnet}. For $\gamma=0.7$, $\left<\dot{N}_{t}\right>\hspace{1mm} \sim 6.8 \hspace{1mm} \times 10^{-5}$ yr$^{-1}$ which is close to the observational inferred value $\sim \hspace{0.5mm}10^{-5}$ yr$^{-1}$ and $\left<\dot{N}_{t}\right> \approx \gamma^2$ for $\gamma \leq 1.2$. The sample of galaxies taken by \citet{2014arXiv1410.7772S} have $\gamma$ varying over all ranges upto $\gamma=1.2$, which implies that their $<\dot{N}_{t}>$ is $\gamma$ independent, whereas we have calculated the $<\dot{N}_{t}>$ assuming that all galaxies to have the same $\gamma$. The discrepancy in theory and observation is smaller in our model for $\gamma \leq 1$ compared with \citet{2014arXiv1410.7772S} who have predicted $<\dot{N}_{t}> \hspace{1mm} \sim \hspace{1mm} {\rm few} \hspace{1mm} \times \hspace{1mm} 10^{-4}$ yr$^{-1}$ by taking into account the Schechter BH mass function and a Nuker profile. The $\gamma$ averaged $\left<\dot{N}_{t}\right>$ integrated over the range $0.6\leq \gamma \leq 1.2$ is $\sim 1.3 \times 10^{-4}$ yr$ ^{-1}$.

Our calculation of $\dot{N}_{t}$ is based on a simple model of two body relaxations in a spherical potential. The alternative relaxation mechanism such as resonant relaxation \citep{1996NewA....1..149R}, which dominates for high energy orbits and anomalous relaxation for highly eccentric orbits $e \rightarrow 1$ \citep{2014MNRAS.443..355H} and non-spherical potential can moderately change the values of $\dot{N}_{t}$. \citet{2015ApJ...804..128M}, taking into account the resonant relaxation theory in a potential dominated by the BH, has shown that the enhancement of angular momentum diffusion at large binding energies results in a depletion of DF at those energies, and results in a density deficit core, but the effect of resonant relaxation on $\dot{N}_{t}$ is moderate for a power law model such as the one used here; however, resonant relaxation for more general DF can be important. The assumption of isotropic velocity distribution overestimates the TDE rates if the true velocity distribution is anisotropic in a tangentially biased, and underestimates if the aniostropy is radially biased. This is because a tangentially biased distribution will have longer angular momentum relaxation time compared to a radially biased distribution. Recent $N$-body simulations by \citet{2014ApJ...792..137Z} have indicated that the presence of a loss cone will bias the orbits toward tangential anisotropy at smaller distance and radial anisotropy at a larger distance, and that the tangential anisotropy is minor near $\mathcal{E}_{c}$. From both observational and theoretical perspectives, it is unclear whether the galactic nuclei are sufficiently anisotropic (and overwhelmingly in the tangential direction) so as to reduce TDE rates to the observationally inferred values. 

 \citet{1988Natur.333..523R} and others have considered the stellar orbit to be nearly parabolic. We have included the angular momentum $J$ in the calculation and studied the effect of $J$ on accretion dynamics. We have modified the dimensionless quantities given in \citet{2009MNRAS.392..332L} and for the low eccentric orbits, which results in an increase in peak accretion rate. \citet{2009MNRAS.400.2070S} and \citet{2011MNRAS.410..359L} have calculated the spectral profile for a parabolic orbit that does not have any dip in their luminosity profile. The inclusion of $J$ induces a dip in the light curve profile, which gets deeper with increased energy. We can also see that our results in the optical band match with the result of \citet{2011MNRAS.410..359L} for $\bar{e}\ll1$. 

In general, the accretion of matter into the BH is non-steady because the mass at the outer radii are higher than the mass at inner radii. \citet{2011ApJ...736..126M} evaluated the surface density and temperature profile assuming the accretion disk to be thin and the accretion rate $\propto t^{-5/3}$. A model for a non-steady accretion mechanism that includes both super- and sub-Eddington phase is required to better understand the evolution and emission from the disk. The $\alpha$ viscosity prescription used by \citet{2009MNRAS.400.2070S} is not applicable in the super-Eddington phase due to low efficiency and high opacity of the disk, so a general viscosity prescription, such as $\nu \propto \Sigma^{d}(r)r^{e}$ where $d$ and $e$ are constants, can be used to evaluate the accretion disk, where $\Sigma(r)$ is surface density profile  \citep{2001A&A...379.1138M,2014ApJ...784...87S}.  

We have built simple analytical expressions to evaluate the condition for the formation of an accretion disk. We did not include the stream interactions and relativistic effects in the calculations, which can speed up the rate of formation of the disk through the angular momentum exchange. The hydrodynamical simulations by \citet{2009ApJ...697L..77R} have shown that the debris interactions result in the formation of an accretion disk with mass accretion rate showing deviation from \citet{2009MNRAS.392..332L} in early time and following $t^{-5/3}$ in the late stage. Very recently, \citet{2015arXiv150104635B} have performed hydrodynamical simulations for a star on a highly elliptical orbit with the resulting debris undergoing apsidal precession; they found that the higher the eccentricity and/or the deeper the encounter, the faster is the circularization. For an efficient cooling, the debris forms a thin and narrow ring of gas. For an inefficient cooling, they settle in a thick and extended torus, mostly centrifugally supported against gravity. The general relativistic hydrodynamical simulation by \citet{2015ApJ...804...85S} have shown that the accretion rate still rises sharply and then decays as a power law; however, its maximum is 10 \% smaller than the previous expectation, and timescale of the peak accretion is longer than the previously predicted values. This is due to the mass accumulation at higher radius because of angular momentum exchange at large radii. The overall conclusion is that the resulting debris will form an accretion disk. The thickness of the disk formed and the circularization timescale as a function of stellar parameters and $M_{\bullet}$ are still need to be evaluated. We think that while it is important to include relativistic effects and stream collisions, it is unlikely that it will change the conclusion, and will change only slightly change the conditions derived for the disk formation.

\citet{2009MNRAS.400.2070S} have predicted $\dot{N}_{D}$ assuming a constant capture rate, stellar orbits to be parabolic, and the flare duration to be the duration of Eddington phase, which is obtained assuming $\diff M/\diff E_{d}$ as a constant. \citet{2008ApJ...676..944G} and \citet{2011ApJ...741...73V} have used the observed detectable rate for the GALEX mission in the Near Ultra Violet (NUV) and SDSS in optical band respectively and scaled it to the other missions assuming survey parameters such as cadence and integration time to be same. \citet{2011ApJ...741...73V} have observationally estimated higher rates compared to the estimation by \citet{2008ApJ...676..944G} due to low sample size. We have performed a detailed calculation, taking in account the both stellar and accretion dynamics, and predicted the detectable rates that are in agreement with the prediction by \citet{2011ApJ...741...73V}. We have not included the filter transmission in generating the spectrum. As the filter transmission varies over the wavelength in the given spectral band, and is less than unity, the simulated flux gets reduced, which results in the reduction in the $\delta t_f$ and hence the detectable rate $\dot{N}_{D}$.

India's space mission ASTROSAT that was launched recently, has a payload SSM (Sky Scanning Monitor) to follow up the transient universe in the X-ray band by nearly scanning half the sky in about 6 hours duration for a continued same stellar pointing of the spacecraft (\url{http://astrosat.iucaa.in/?q=node/13}). The sensitivity of the instrument is $\sim 7.2 \times 10^{-10} \hspace{1mm} {\rm erg \hspace{1mm} sec^{-1} \hspace{1mm} cm^{-2}}$ with the integration time of the detector to be 10 min. With these parameters, the detectable rate for ASTROSAT is expected to be less than $\sim$ 1 yr$^{-1}$.

For the optical surveys in the g-band, namely LSST and PAN-STARRS, the TDE may not be resolved and the rates corresponding predicted could be an over estimate by a factor of a few.

\section{Summary and conclusions}

We studied in detail, the model of the TDE, taking into account the both stellar dynamical and gas dynamical inputs.
  The overall system parameters include BH mass $M_{\bullet}$, specific orbital energy $E$ and angular momentum $J$, star mass $M_{\star}$ and radius $R_{\star}$ and pericenter of the star orbit $r_{p}(E,\hspace{1mm}J,\hspace{1mm}M_{\bullet})$. We solved the steady state FP equation using the standard loss cone theory for the galactic density profile $\rho (r) \propto r^{-\gamma}$ and stellar mass function $\xi(m) $, where $m=M_{\star}/M_{\odot}$ and obtained the feeding rate of stars to the BH $\dot{N}(E,\hspace{1mm}J,\hspace{1mm}m,\hspace{1mm}\gamma)$
that it is an increasing function of $J$ and $\gamma$, but a decreasing function of $E$ and $m$. Because the stars evolve along their orbits toward the BH, we compared the lifetime of main sequence star to the radial period of its orbit and calculated the probability $f_{\star}$ for a star to be captured as a main sequence given by Equation (\ref{fstar}). Using this we model the in-fall rate of the disrupted debris, $\dot{M}(E,\hspace{1mm}J,\hspace{1mm}m,\hspace{1mm}t)$, and discuss the conditions for the formation of an accretion disk considering accretion, viscous, ring formation, and radiation timescales. We find that the accretion disk is almost always formed for the fiduciary range of the physical parameters. By equating the peak of $\dot{M}(E,\hspace{1mm}J,\hspace{1mm}m,\hspace{1mm}t)$ to the Eddington rate, we derive the critical black mass $M_c(E, J,\hspace{1mm}m)$. We simulated the light curve profiles in relevant optical g band and soft X-rays for both super- and sub-Eddington accretion disks as a function of $\dot{M}(E,\hspace{1mm}J,\hspace{1mm}m,\hspace{1mm}t)$, taking typical stellar system parameters. Specifically, we have found
the following key results:

\begin{enumerate}

\item In Section \ref{lct}, we have approximated the radial period of an orbit with Equation (\ref{radv}) and the $M_{\bullet}-\sigma$ relation. The radial period of an orbit in Keplerian potential is $T_r\propto M^{-0.38}_{\bullet}\mathcal{E}^{-3/2}$. The capture rate $\dot{N}_t=\int (N_{lc}(\mathcal{E})/T_r)\, \diff \mathcal{E}\propto M^{-0.38}_{\bullet}$.

\item The applicable ranges of dimensionless energy $\bar{e}$ and angular momentum $\ell$ are given by \{$\bar{e}_h<\bar{e}<1$, $0<\ell<1$\} where $\bar{e}_h=s_t=r_t/r_h$.

\item We solved the steady state FP equation in Section \ref{lct} and obtained the capture rate using Equation (\ref{nnet}). We found that the capture rate $\dot{N}_{t}$  does not show a power law dependence with $M_{\bullet}$ and it increases with $\gamma$. Even though the increase in $\dot{N}_{t}$ with $\gamma$ is non linear, an approximate fit gives $\dot{N}_{t} \propto \gamma^{p}$, where $p\sim 2.1$ (see Figure \ref{ntg}). For $M_{6}>10$, $\dot{N}_{t}\propto M^{-\beta}_{6}$ and $\beta$ $\sim 0.3\pm 0.01$ (see Figure \ref{nthlc}).

\item In Section \ref{ptd}, we show that the fractional radius from the star center $x_{l}(\bar{e},\hspace{1mm}\ell,\hspace{1mm}M_{\bullet},\hspace{1mm}m)$ to the point where the debris is bound to the BH increases with $\bar{e}$ and $\ell$ (see Figure \ref{rplim}). The increase with $\ell$ is significant only for high energy orbits. The peak accretion rate increases with $x_{l}$. The decline to later $t^{-5/3}$ law is steeper if the energy of the initial orbit is higher, as shown in Figure \ref{masst} .   

\item In Section \ref{adp}, by equating $\dot{M}_{p}$ and $\dot{M}_E$, the critical BH mass $M_{c}(\bar{e},\hspace{1mm}\ell,\hspace{1mm}m)$ is found to increase with $\bar{e}$ and decrease with $\ell$ and $m$ (see Figure \ref{mcr}). With the decrease in $\ell$, the $r_{p}$ decreases and thus $\dot{M}$ increases, which results in an increase in $\dot{M}_{p}$. For higher $\bar{e}$ and lower $\ell$, $M_{c}(\bar{e},\hspace{1mm}\ell,\hspace{1mm}m)$ can exceed the BH mass limit for TDE to occur (i.e, $\sim$ 3 $\times \hspace{1mm} 10^{8} M_{\odot}$).

\item In Section \ref{fad}, we found that  Max[$\mathcal{T}_{r}(\hspace{1mm} 10^{-6} \hspace{0.5mm} \leq \hspace{0.5mm} \bar{e}  \hspace{0.5mm}\leq  \hspace{0.5mm} 1, \hspace{1mm} 0\hspace{0.5mm}\leq \hspace{0.5mm} \ell \hspace{0.5mm} \leq \hspace{0.5mm} 1,1\hspace{0.5mm} \leq \hspace{0.5mm} M_{6} \hspace{0.5mm} \leq \hspace{0.5mm} 100$)]$<$ 1, which implies that the debris will form an accretion disk (see Figure \ref{diskform}). The ratio $\mathcal{T}_{v}<1$ for $M_{6}\leq 31.6$ which implies that the accretion disk formed is a slim disk. The $\mathcal{T}_{v}$ increases with $M_{6}$ and decreases with $\bar{e}$ as shown in Figure \ref{tvtr}. The higher mass SMBHs form a thin disk from the disrupted debris of a star on low energy orbit and $\ell\hspace{0.5mm} \sim\hspace{0.5mm} 1$ and a thick disk for a star on a high energy orbit.

\item In Section \ref{sep}, we derive the observed flux as a function of $\bar{e}$ and $\ell$. Figure \ref{opt} shows the observed fluxes $f_{\rm obs}$ in the optical g band and the peak observed flux increases with a decrease in $\ell$. The decline of the light curve profile to the later stage gets steeper with increasing $\ell$.   

\item In Section \ref{erc}, the net detectable rate $\dot{N}_{D}$ is calculated for the various missions observing in optical and X-ray bands. Using standard cosmological parameters and mission instrument details, we predict the detectable tidal disruption rates for $\gamma=0.7$ for LSST to be $\sim$ 5003 ${\rm yr^{-1}}$; Pan-STARRS in the optical g band performing in either all sky survey (ASS) mode or the deep imaging survey (DIS) mode were predicted to be $\sim$ 6337 yr$^{-1}$ for operation in 3 $\pi$ mode and $\sim$ 12.3 yr$^{-1}$ in the MDS mode, which are in reasonable agreement with scaled-up values based on Sloan Digital Sky Survey (SDSS) detection. Our prediction for eROSITA in the soft X-ray band is about $\sim$ 679.5 yr$^{-1}$, which is consistent with \citet{2014MNRAS.437..327K}. The values of $\gamma$ for which our predictions of $\dot{N}_{D}$ match with the scaled-up values in \citet{2011ApJ...741...73V} are shown as $\gamma_{s}$ in Table \ref{survey}. We have also estimated the error in $\dot{N}_D$ for an error in fit to $\gamma$ which is taken to be 0.1 and is also shown in Table \ref{survey}.

\item
Our results are in reasonable agreement with the scaled-up values from the SDSS observations \citep{2011ApJ...741...73V}, as given in Table \ref{survey} along with the detection efficiency of the detector $\Upsilon$. The $\Upsilon$ is lowest for the eROSITA mission due to the high cadence of half year and is highest for Pan-STARRS MDS due to very high sensitivity. The $\dot{N}_{D}\propto \gamma^{\Delta}$, where $\Delta \sim 1.95$ in optical band is shown in Figure \ref{ndga}.

\end{enumerate}

   We can use the TDE rate as a proxy to estimate the BH mass distribution as a function of redshift and relate it to the occupation fraction of SMBHs in galaxies \citep{2014arXiv1410.7772S}. The TDE can influence jet activity from BH systems in two ways. The stellar debris can quench the jet \citep{2008ApJ...680L..13G} for BH masses below $3 \times 10^8$ $M_{\odot}$ or the Poynting flux from a spinning hole can
be boosted by the ingestion of a star. There have been several observations (e.g, \citep{2011Natur.476..421B}) indicating
radio transients after a TDE and these have been studied by simulations \citep{2014MNRAS.437.2744T}. In the future, we plan to include more details of the underlying physics, such as resonant relaxation, non-spherical systems, relativistic gas dynamics, and viscosity prescription, to improve our predictions. 

 We thank the anonymous referee for insightful and helpful comments that has improved the paper significantly.  We also thank Andrea Merloni and M. C. Ramadevi for useful discussions.
\bibliography{reference}
\end{document}